\documentclass[preprint,authoryear,3p,a4paper]{elsarticle}
\usepackage{graphicx}
\usepackage{url,amssymb,subfigure}

\journal{J. Comp. Phys.}
\newcommand{\pder}[2]{\ensuremath{\frac{\partial #1}{\partial #2}}}

\begin{document}

\begin{frontmatter}

\title{Smoothed Particle Hydrodynamics and Magnetohydrodynamics}
\tnotetext[t1]{This review and associated material germinated as lectures and tutorials given as part of the ASTROSIM summer school on computational astrophysics held during July 2010 in Toru\'n, Poland. A video of the original lectures can be viewed online at \url{http://supercomputing.astri.umk.pl/}. The paper also presents an updated version of much of the otherwise unpublished material in my PhD thesis \citep{price04}.}
\author{Daniel J. Price}
\ead{daniel.price@monash.edu}
\ead[url]{http://users.monash.edu.au/$\sim$dprice/SPH/}
 \address{Centre for Stellar and Planetary Astrophysics \& School of Mathematical Sciences, Monash University, Melbourne Vic. 3800, Australia}

\begin{abstract}
 This paper presents an overview and introduction to Smoothed Particle Hydrodynamics and Magnetohydrodynamics in theory and in practice. Firstly, we give a basic grounding in the fundamentals of SPH, showing how the equations of motion and energy can be self-consistently derived from the density estimate. We then show how to interpret these equations using the basic SPH interpolation formulae and highlight the subtle difference in approach between SPH and other particle methods.  In doing so, we also critique several `urban myths' regarding SPH, in particular the idea that one can simply increase the `neighbour number' more slowly than the total number of particles in order to obtain convergence. We also discuss the origin of numerical instabilities such as the pairing and tensile instabilities. Finally, we give practical advice on how to resolve three of the main issues with SPMHD: removing the tensile instability, formulating dissipative terms for MHD shocks and enforcing the divergence constraint on the particles, and we give the current status of developments in this area. Accompanying the paper is the first public release of the \textsc{ndspmhd} SPH code, a 1, 2 and 3 dimensional code designed as a testbed for SPH/SPMHD algorithms that can be used to test many of the ideas and used to run all of the numerical examples contained in the paper.

\end{abstract}


\end{frontmatter}


\section{Introduction}
\label{sec:intro}
 Smoothed Particle Hydrodynamics (SPH), originally formulated by \citet{lucy77} and \citet{gm77}, is by now very widely used for many diverse applications in astrophysics, geophysics, engineering and in the film and computer games industry. Whilst numerous excellent reviews already exist \citep[e.g.][]{monaghan92,monaghan05,price04,rosswog09}, there remain -- particularly in the astrophysical domain -- some widespread misconceptions about its use, and more importantly, its fundamental basis.
 
  Our aim in this -- mainly pedagogical -- review is therefore not to provide a comprehensive survey of SPH applications, nor the implementation of particular physical models, but to re-address the fundamentals about why and how the method works, and to give practical guidance on how to formulate general SPH algorithms and avoid some of the common pitfalls in using SPH. Since such an understanding is critical to the development of robust and accurate methods for Magnetohydrodynamics (MHD) in SPH (hereafter referred to as ``Smoothed Particle Magnetohydrodynamics'', SPMHD), this will lead us naturally on to review the background and current status in this area -- particularly relevant given the importance of MHD in most, if not all, astrophysical problems. Whilst the paper is written with an astrophysical flavour in mind (given the topical issue of JCP for which it is written), the principles are general and thus are applicable in any of the areas in which SPH is applied.
  
   Finally, alongside this article I have released a public version of my \textsc{ndspmhd} SPH/SPMHD code, along with a set of easy-to-follow numerical exercises -- consisting of setup and input files for the code and step-by-step instructions for each problem in 1, 2 and 3 dimensions -- the problems themselves having been chosen to illustrate many of the theoretical points in this paper. Indeed, $\textsc{ndspmhd}$ has been used to compute all of the test problems and examples shown. The hope is that this will become a useful resource\footnote{\textsc{ndspmhd} is available from \url{http://users.monash.edu.au/~dprice/SPH/}. Note that we do not advocate the use of \textsc{ndspmhd} as a ``performance'' code in 3D, since it is not designed for this purpose and excellent parallel 3D codes already exist (such as the GADGET code by \citealt{springel05}). Rather it is meant as a testbed for algorithmic experimentation and understanding.} not only for advanced researchers but also for students embarking on an SPH-based research topic.

\section{The foundations of SPH: Calculating density}
\label{sec:calculatingdensity}
 The usual introductory lines on SPH refer to it as a ``Lagrangian particle method for solving the equations of hydrodynamics''. However, SPH starts with a basis much more fundamental than that, as the answer to the following question: 
\begin{quote}
 How does one compute the density from an arbitrary distribution of point mass particles?
\end{quote}
 This problem arises in many areas other than hydrodynamics, for example in obtaining the solution to Poisson's equation for the gravitational field $\nabla^{2} \Phi = 4\pi G \rho ({\bf r})$ when a (continuous) density field is represented by a collection of point masses.

\begin{figure}[t]
\begin{center}
   \includegraphics[width=0.2\columnwidth]{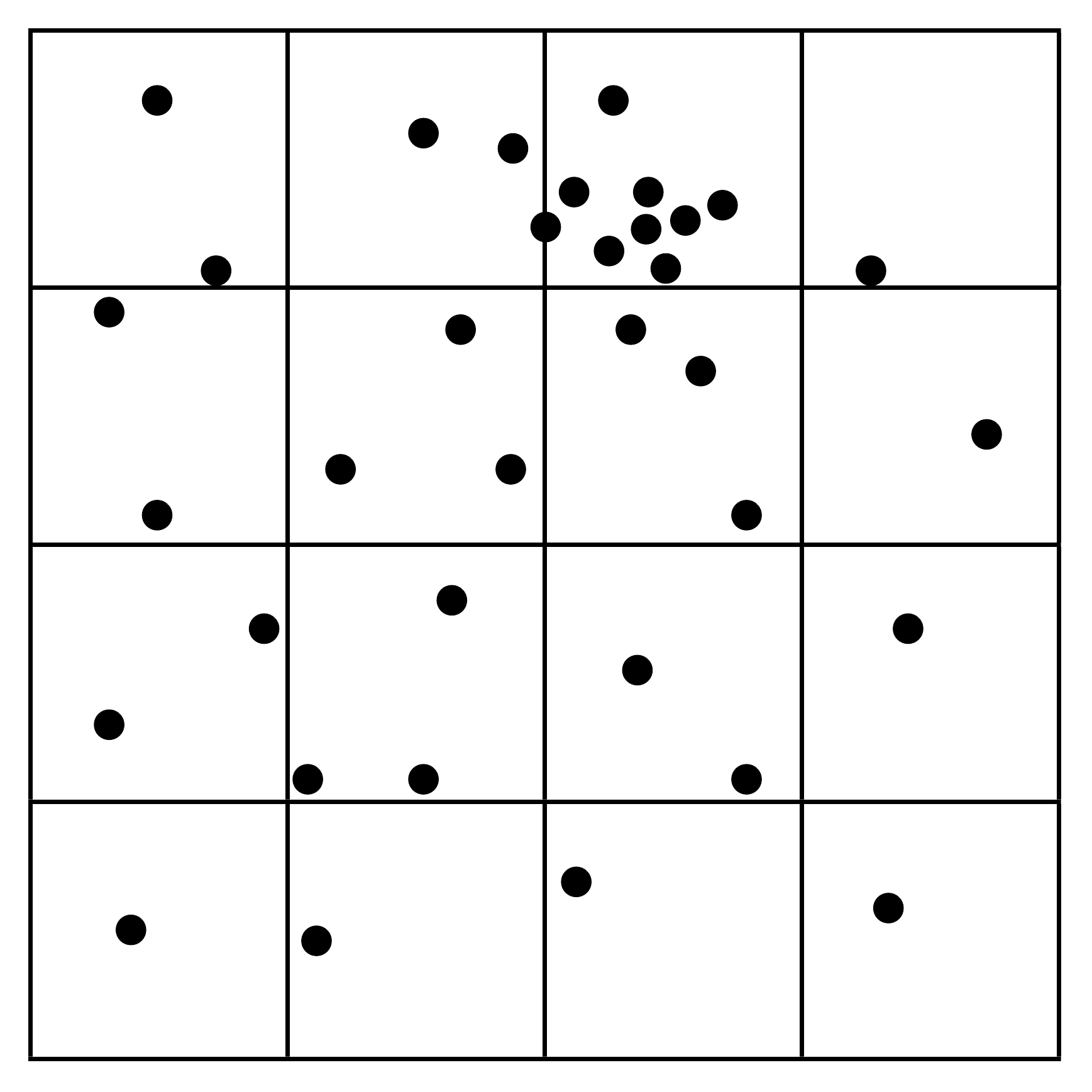}
   \hspace{0.05\columnwidth}
   \includegraphics[width=0.2\columnwidth]{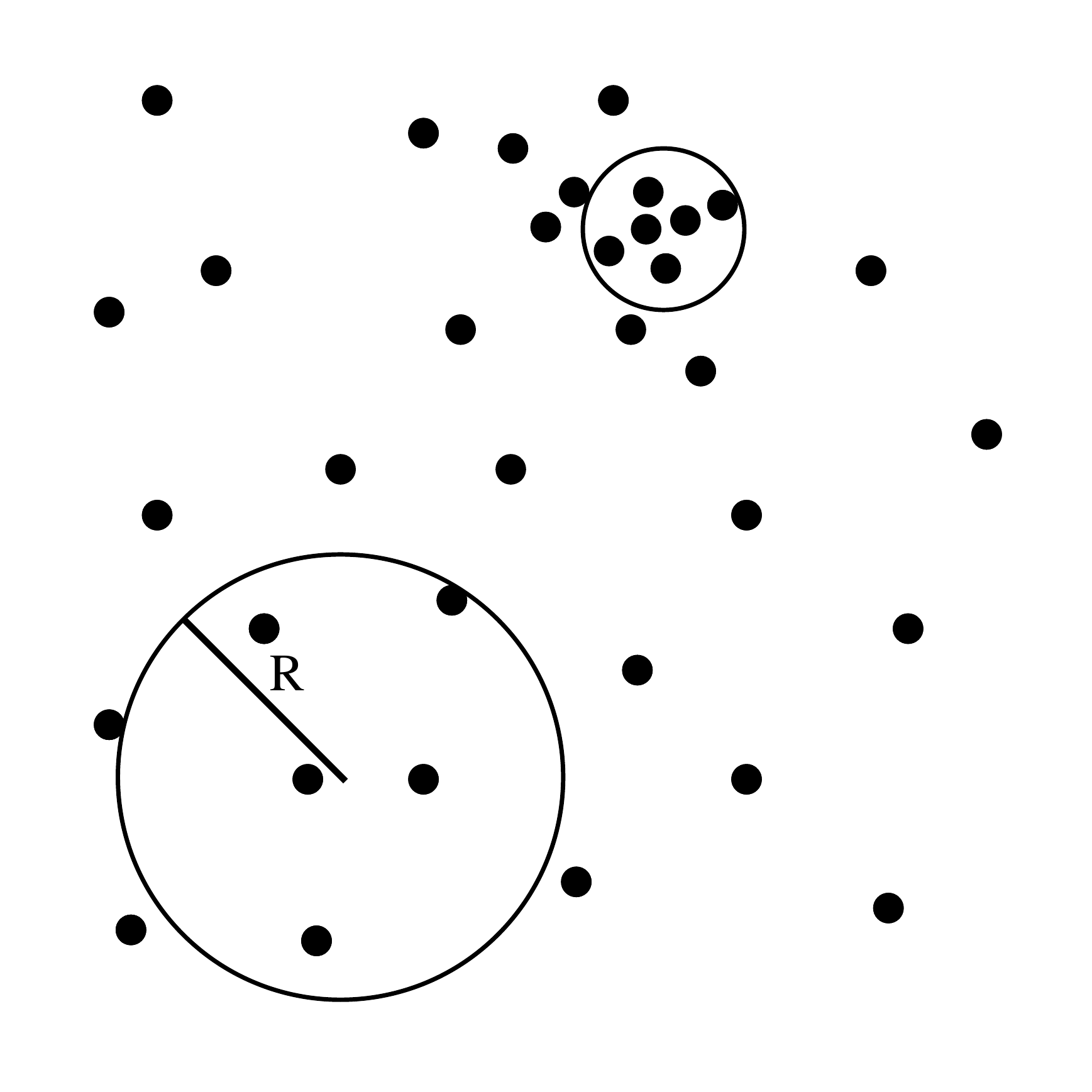} 
   \hspace{0.05\columnwidth}
   \includegraphics[width=0.2\columnwidth]{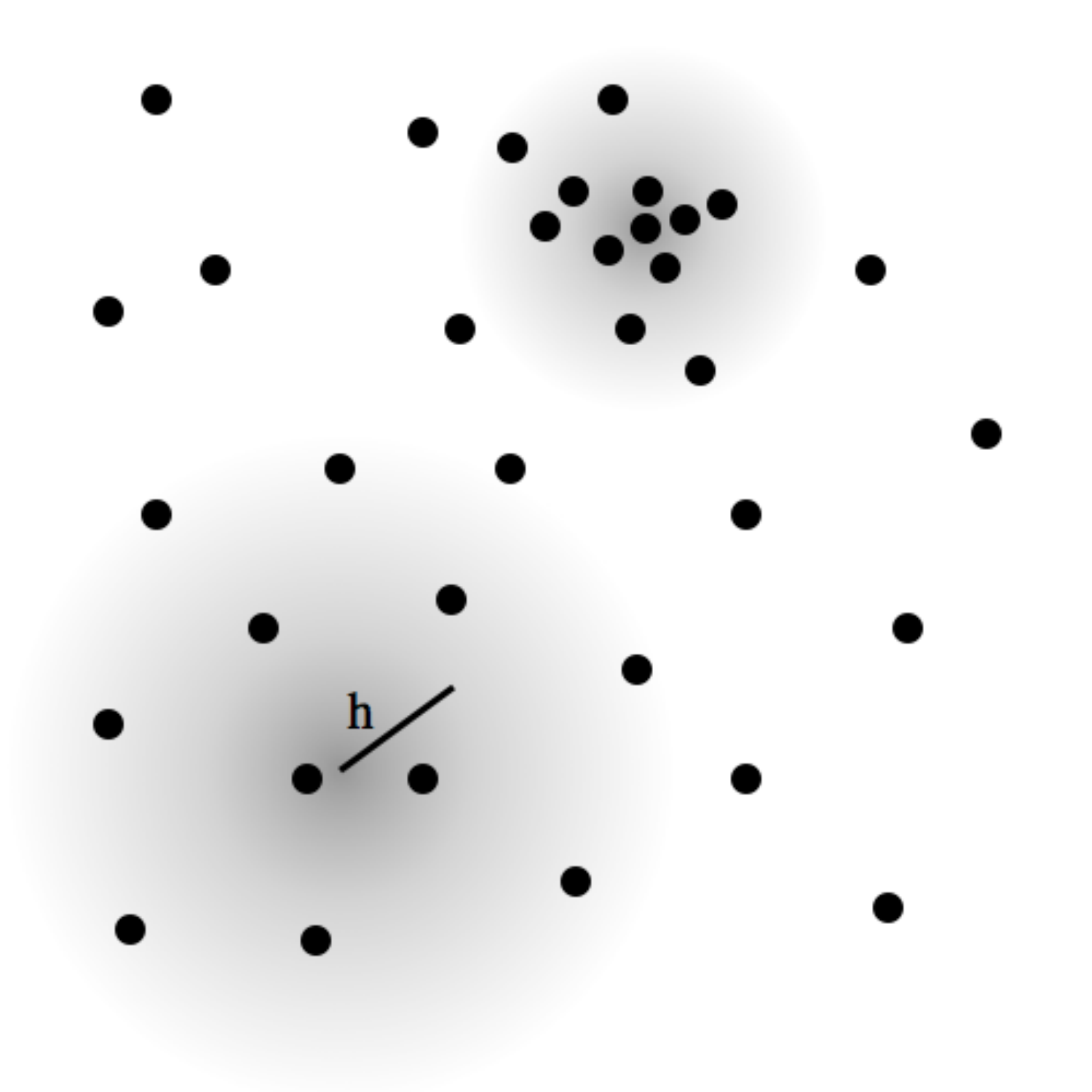} 
   \caption{Computing a continuous density field from a collection of point mass particles. a) In particle-mesh methods (left panel) the density is computed by interpolating the mass to a grid (or simply dividing the mass by the volume). However, this tends to over/under resolve clustered/sparse regions. b) An alternative not requiring a mesh is to construct a local volume around the sampling point, solving the clustering problem by scaling the sample volume according to the local number density of particles. c) This panel shows the approach adopted in SPH, where the density is computed via a weighted sum over neighbouring particles, with the weight decreasing with distance from the sample point according to a scale factor $h$. }
   \label{fig:density}
\end{center}
\end{figure}

\subsection{Approaches to computing the density}
\label{sec:approaches}
  Three common approaches are summarised in Fig.~\ref{fig:density}. Perhaps the most straightforward (Fig.~\ref{fig:density}a) is to construct a mesh of some sort and divide the mass in each cell by the volume. This basic approach forms the basis of hybrid particle-mesh methods such as Marker-In-Cell \citep[e.g.][]{hw65} and Particle-In-Cell \citep{he81} schemes, where one can further improve the density estimate using any of the standard particle-cell interpolation methods, such as Cloud-In-Cell (CIC), Triangular-Shaped-Cloud (TSC) etc. However there are clear limitations -- firstly that a fixed mesh will inevitably over/under-sample dense/sparse regions (respectively) when the mass distribution is highly clustered\footnote{More recently, this problem has been addressed by the use of adaptively refined meshes to calculate the density field \citep[e.g.][]{couchman91}.}; and secondly a loss of accuracy, speed and consistency because of the need to interpolate both to/from the particles, for example to compute forces.
 
  The second approach (Fig.~\ref{fig:density}b) is to remove the mesh entirely and instead calculate the density based on a local sampling of the mass distribution, for example in a sphere centred on the location of the sampling point (which may or may not be the location of a particle itself). The most basic scheme would be to divide the total mass by the sampling volume, i.e., 
\begin{equation}
\rho({\bf r}) = \frac{\sum_{b=1}^{N_{neigh}} m_{b}}{\frac43 \pi R^{3}}.
\end{equation}
 The problem of resolving clustered/sparse regions can be easily addressed in this method by adjusting the size of the sampling volume according to the local number density of sampling points, for example by computing with a fixed ``number of neighbours'' for each particle -- as shown in the Figure. However, this leads to a very noisy estimate, since the density estimate will be very sensitive to whether a distant particle on the edge of the volume is ``in'' or ``out'' of the estimate (with $\delta\rho \propto 1/N_{neigh}$ for equal mass particles). This leads naturally to the idea that one should progressively down-weight the contributions from neighbouring particles as their relative distance increases, in order that changes in distant particles have a progressively smaller influence on the local estimate (that is, the density estimate is \emph{smoothed}).

\subsection{The SPH density estimator}
\label{sec:density}
 This third approach forms the basis of SPH and is shown in Fig.~\ref{fig:density}c: Here the density is computed using a weighted summation over nearby particles, given by
\begin{equation}
\rho({\bf r}) = \sum_{b=1}^{N_{neigh}} m_{b} W({\bf r} - {\bf r}_{b}, h),
\end{equation}
where $W$ is an (as yet unspecified) weight function with dimensions of inverse volume and $h$ is a scale parameter determining the rate of fall-off of $W$ as a function of the particle spacing (also yet to be determined). Conservation of total mass $\int \rho {\rm dV} = \sum_{b=1}^{N_{part}} m_{b}$ implies a normalisation condition on $W$ given by
\begin{equation}
\int_{\rm V} W({\bf r}' - {\bf r}_{b}, h) {\rm dV'} = 1.
\label{eq:Wnorm}
\end{equation}

 The accuracy of the density estimate then rests on the choice of a sufficiently good weight function (hereafter referred to as the \emph{smoothing kernel}). Elementary considerations suggest that a good density kernel should have at least the following properties:
\begin{enumerate}
\item A weighting that is positive, decreases monotonically with relative distance and has smooth derivatives;
\item Symmetry with respect to (${\bf r} - {\bf r}'$) -- i.e., $W({\bf r}' - {\bf r}, h) \equiv W(\vert {\bf r}' - {\bf r} \vert, h)$; and
\item A flat central portion so the density estimate is not strongly affected by a small change in position of a near neighbour.
\end{enumerate}

 A natural choice that satisfies all of the above properties is the Gaussian:
\begin{equation}
W({\bf r} - {\bf r}',h) = \frac{\sigma}{h^{d}} \exp\left[-\frac{({\bf r} - {\bf r}')^{2}}{h^{2}}\right],
\end{equation}
where $d$ refers to the number of spatial dimensions and $\sigma$ is a normalisation factor given by $\sigma = [1/\sqrt{\pi},1/\pi,1/(\pi\sqrt{\pi})]$ in [1,2,3] dimensions. The Gaussian satisfies condition 1 particularly well since it is infinitely smooth (differentiable) -- and gives in practice an excellent density estimate. However it has the practical disadvantage of requiring interaction with all of the particles in the domain [with computational cost of $\mathcal{O}(N^2)$ if computing the density at the particle locations], despite the fact that the relative contribution from neighbouring particles quickly becomes negligible with increasing distance. Thus in practice it is better to use a kernel that is Gaussian-like in shape but truncated at a finite radius (e.g. a few times the scale length, $h$). Using kernels with such ``compact support'' means a much more efficient density evaluation, since the cost scales like $\mathcal{O}(N_{neigh} N)$, but inevitably leads to a more noisy density estimate since one is more sensitive to small changes in the local distribution.

\begin{figure}[t]
\begin{center}
   \includegraphics[width=0.7\columnwidth]{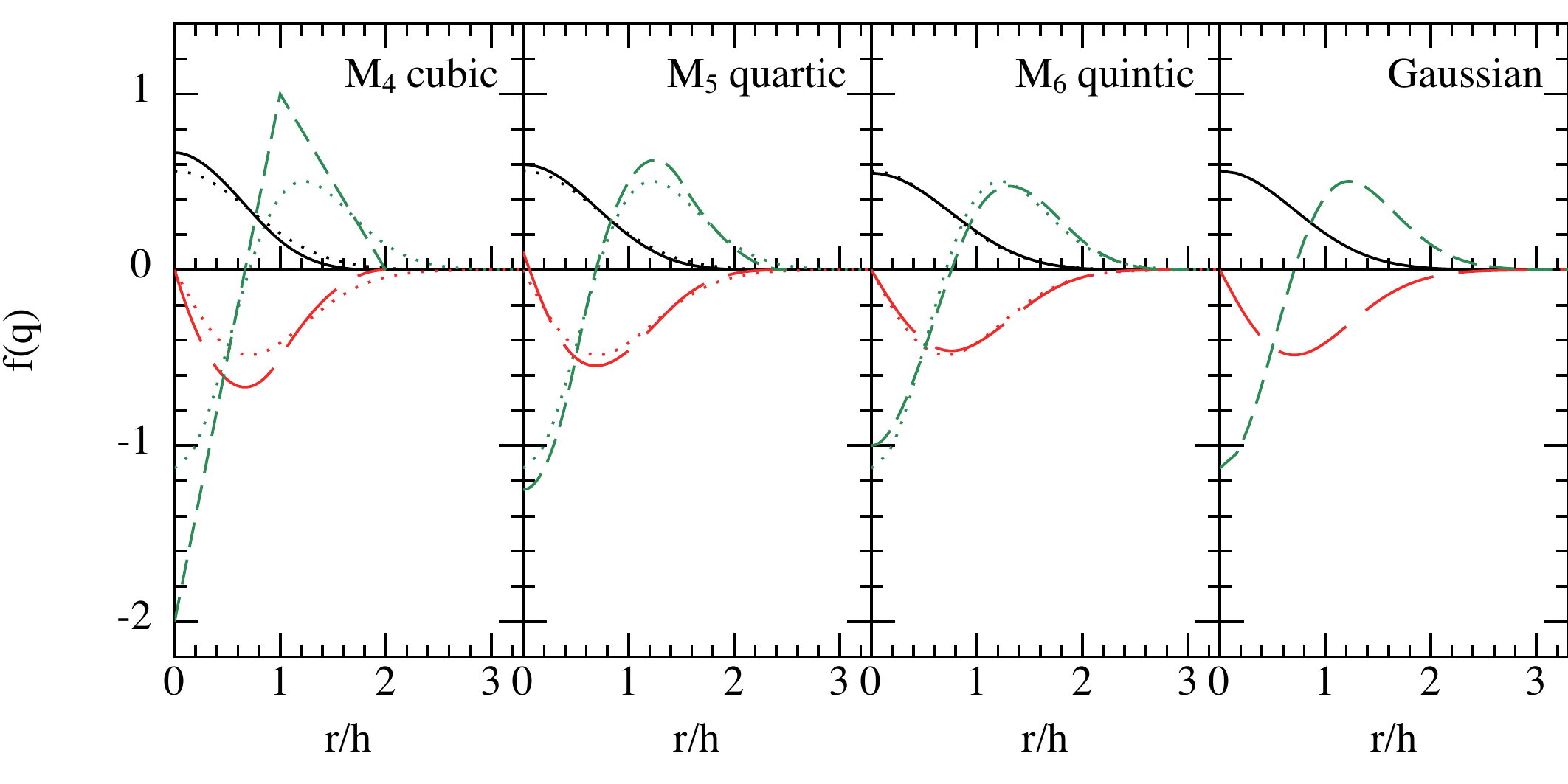}
   \caption{The $M_{4}$ (cubic, truncated at $2h$), $M_{5}$ (quartic, truncated at $2.5h$) and $M_{6}$ (quintic, truncated at $3h$) \citet{schoenberg46a} B-spline kernel functions (solid lines) and their first (long-dashed) and second (short-dashed) derivatives, compared to the Gaussian (right panel and dotted lines in other panels). Notice that although the ``number of neighbours'' increases in the $M_{5}$ and $M_{6}$ functions compared to the cubic spline, the smoothing scale $h$ retains the same meaning with respect to the Gaussian. Thus, using the higher order B-splines is a way to increase the smoothness of the kernel summations without altering the resolution length, and is very different to simply increasing the number of neighbours under the cubic spline.}
   \label{fig:Bsplines}
\end{center}
\end{figure}

\subsection{Kernel functions with compact support}
  There are many kernel functions which fit this bill. The most well-used (for SPH at least) are the \citet{schoenberg46a} B-spline functions \citep{ml85,monaghan85,monaghan05}, generated as the Fourier transform
\begin{equation}
M_{n} (x, h) = \frac{1}{2\pi} \int^{\infty}_{-\infty} \left[\frac{\sin{(kh/2)}}{k h/2} \right]^{n} \cos (k x) {\rm d}k.
\end{equation}
  
  These give progressively better approximations to the Gaussian at higher $n$, both by increasing the radius of compact support and by increasing smoothness, since each function $M_{n}$ is continuous up to the $\{n-2\}$th derivatives. Since we minimally require continuity in at least the first and second derivatives, the lowest order B-spline useful for SPH is the $M_{4}$ (cubic) spline truncated at $2h$:
\begin{equation}
w(q) = \sigma \left\{ \begin{array}{ll}
\frac{1}{4}(2-q)^3 - (1 - q)^{3}, & 0 \le q < 1; \\
\frac{1}{4}(2-q)^3, & 1 \le q < 2; \\
0. & q \ge 2, \end{array} \right. \label{eq:cubicspline}
\end{equation}
where for convenience we use $W(\vert {\bf r} - {\bf r}' \vert,h) \equiv \frac{1}{h^{d}} w(q)$, where $q=\vert{\bf r} -{\bf r}'\vert /h$ and $\sigma$ is a normalisation constant given by $\sigma = [2/3,10/(7\pi),1/\pi]$ in $[1, 2, 3]$ dimensions. Next are the $M_{5}$ quartic, truncated at $2.5h$:
\begin{equation}
w(q) = \sigma \left\{ \begin{array}{ll}
\left(\frac52 -q\right)^4 - 5\left(\frac32 -q\right)^4 + 10\left(\frac12-q\right)^4, & 0 \le q < \frac12; \\
\left(\frac52 -q\right)^4 - 5\left(\frac32 -q\right)^4, & \frac12 \le q < \frac32; \\
\left(\frac52 -q\right)^4, & \frac32 \le q < \frac52; \\
0. & q \ge \frac52, \end{array} \right. \label{eq:quarticspline} 
\end{equation}
 with normalisation $\sigma = [1/24,96/1199\pi,1/20\pi]$, and the $M_{6}$ quintic, truncated at $3h$:
\begin{equation}
w(q) = \sigma \left\{ \begin{array}{ll}
(3-q)^5 - 6(2-q)^5 + 15(1-q)^5, & 0 \le q < 1; \\
(3-q)^5 - 6(2-q)^5, & 1 \le q < 2; \\
(3-q)^5, & 2 \le q < 3; \\
0. & q \ge 3, \end{array} \right. \label{eq:quinticspline}
\end{equation}
with normalisation $\sigma = [1/120,7/478\pi,1/120\pi]$ \citep[e.g.][]{morrisphd}. These kernel functions and their first and second derivatives are shown for comparison with the Gaussian in Fig.~\ref{fig:Bsplines}.

  One important aspect to draw from our discussion so far is the clear meaning attached to the smoothing length $h$ as specifying the fall-off of the kernel weighting with respect to the particle separation. In particular, it is clear that referring to the ``number of neighbours'' does not have any meaning \emph{per se} for Gaussian and Gaussian-like kernels: For the Gaussian the number of neighbours is in principle infinite, but there nevertheless exists a well-defined smoothing scale, $h$. The higher order B-splines (Fig.~\ref{fig:Bsplines}) also demonstrate that it is possible to change the ``neighbour number''  -- by progressing to higher $n$ in the series -- \emph{without} changing the smoothing length. It is a widely-propagated myth that one can achieve formal convergence in SPH by ``increasing the number of neighbours'' (e.g. more slowly than the total number of particles). However, there are very important differences between simply ``stretching'' the cubic spline to accommodate a larger neighbour number -- which amounts to changing the ratio of $h$ to particle spacing -- and using a kernel that has a larger radius of compact support but retains the same $h$. That is, in no sense is the SPH density estimate (our ``approach 3'') the same as approach 2 shown in Fig.~\ref{fig:density}b. We will return to this point later.
 
 There are obviously other kernels, and other families of kernels, that satisfy the above properties \citep[e.g.][]{dehnen01}. However, more detailed investigations into kernels \citep[e.g.][]{fq96} tend to merely confirm the points made above -- namely that Bell-shaped, symmetric, monotonic kernels provide the best density estimates. We will examine the formal errors in the kernel density estimate shortly, but first we turn to the issue of setting the smoothing length, $h$.

\subsection{Setting the smoothing length}
\label{sec:hrho}
 Early SPH simulations \citep[e.g.][]{gm77} simply employed a spatially constant resolution length $h$, though one which was allowed to change as a function of time according to the densest part of a calculation\footnote{Similar to the spatially fixed but time-evolved gravitational softening lengths still employed in many cosmological simulations.}. However, as is evident from Fig.~\ref{fig:density}, it is clearly desirable to resolve both clustered and sparse regions evenly -- that is, with a roughly constant ratio of $h$ to the mean local particle separation. Thus, a natural choice for setting the smoothing length is to relate to the local number density of particles, i.e.,
\begin{equation}
h({\bf r}) \propto n({\bf r})^{-1/d}; \hspace{1cm} n({\bf r}) = \sum_{b} W\left[{\bf r} - {\bf r}_{b}, h({\bf r})\right].\label{eq:hn}
\end{equation}
For equal mass particles, this is equivalent to making $h$ proportional to the density itself (since $1/n \equiv m/\rho$). Since in turn density is itself a function of smoothing length, this leads to the idea of an iterative summation to simultaneously obtain the (mutually dependent) $\rho({\bf r})$ and $h({\bf r})$ \citep{sh02,monaghan02,pm07}. Computed at the location of particle $a$, we have a set of two simultaneous equations
\begin{equation}
\rho({\bf r}_{a}) = \sum_{b} m_{b} W({\bf r}_{a} - {\bf r}_{b}, h_{a}); \label{eq:rhosum} \hspace{1cm} h({\bf r}_{a}) = \eta \left( \frac{m_{a}}{\rho_{a}} \right)^{1/d}, \label{eq:hrho}
\end{equation}
where $\eta$ is a parameter specifying the smoothing length in units of the mean particle spacing $(m/\rho)^{1/d}$. These two equations can be solved simultaneously using standard root-finding methods such as Newton-Raphson or Bisection and most ``modern'' SPH codes employ such a procedure (for reasons that will become clear). Note that enforcing the relation given in (\ref{eq:hrho}) is approximately equivalent to keeping the ``mass inside the smoothing sphere'' constant \citep{sh02}, since for example in three dimensions
\begin{equation}
M_{tot}^{a} = \int_{V_{a}} \rho dV \approx \frac43 \pi R_{kern}^{3} \rho_{a}, \label{eq:Msphere}
\end{equation}
where $R_{kern}$ is the kernel radius ($2h$ for the cubic spline), so $M_{tot}=const$ implies $h^{3}\rho=const$. Since for equal mass particles $M_{tot} = m N_{neigh}$, this also means that the number of neighbours should be approximately constant if the relationship in (\ref{eq:hrho}) (or for unequal mass particles, Eq.~\ref{eq:hn}) is enforced. Indeed a ``number of neighbours'' parameter can be used in place of the parameter $\eta$, using
\begin{equation}
N_{neigh, 1D} = 2 \zeta \eta; \hspace{1cm} N_{neigh, 2D} = \pi (\zeta \eta)^{2};\hspace{1cm} N_{neigh, 3D} = \frac43 \pi (\zeta \eta)^{3},
\end{equation}
where $\zeta$ is the compact support radius in units of $h$ (i.e., $\zeta = 2$ for the cubic spline). However, this is problematic for several reasons. Firstly it gives the dangerous impression that $N_{neigh}$ is a free parameter unrelated to $h$, whereas changing $N_{neigh}$ explicitly changes $h$ -- more specifically, the ratio of $h$ to particle spacing (i.e., $\eta$) -- and corresponds to ``stretching'' the cubic spline as discussed above. Secondly, whereas $\eta$ carries the same meaning in $1$, $2$ and $3$ dimensions, the $N_{neigh}$ parameter changes, making it difficult to relate the results of one and two dimensional test problems to three dimensional simulations. Thirdly, $N_{neigh}$ is often used as an integer parameter, whereas it is clear from (\ref{eq:rhosum}) that the $h-\rho$ (or $h-n$) iterations can be performed to arbitrary accuracy (that is, to fractional neighbour numbers) -- which is also necessary if one is to assume that the relationship is differentiable. Finally, $N_{neigh}$ is only related to the \emph{true} number of neighbours so long as the (number) density of particles within the smoothing sphere is approximately constant (that is, so far as the integral in Eq.~\ref{eq:Msphere} can be approximated by $\frac43 \pi R^{3} \rho$). So at best a $N_{neigh}$ parameter only characterises the \emph{mean} neighbour number -- and there can be strong fluctuations about this mean, for example in strong density gradients.

 Earlier adaptive SPH implementations employed density estimates involving either an average smoothing length $\bar{h} = \frac12 \left( h_{a} + h_{b}\right)$ \citep[e.g.][]{benz90} or an average of the smoothing kernels $\overline{W}_{ab} = \frac12\left[W_{ab}(h_{a}) + W_{ab}(h_{b})\right]$ \citep[e.g.][]{hk89}. However, this inevitably leads to heuristic methods for setting the smoothing length itself -- for example by evolving the time derivative of Eq.~\ref{eq:hrho} with ``corrections'' to try to keep the neighbour number approximately constant \citep[e.g.][]{benz90} or simply enforcing a constant neighbour number either approximately or exactly \citep[e.g.][]{hk89}. It also considerably complicated attempts to incorporate derivatives of the smoothing length -- necessary for exact energy and entropy conservation -- into the equations of motion \citep{np94}. By contrast, the mathematical meaning of Eqs.~(\ref{eq:rhosum}) is clear and it is straightforward to take derivatives involving the smoothing length.

 Finally, the density estimate computed via (\ref{eq:rhosum}) is time-independent, depending only on particle positions and masses and thus explicitly answering our original question of how to compute a density field from point mass particles. This also means it has wide applicability to many other problems beyond SPH -- for example \citet{pf10} use it to construct a density field from Lagrangian tracer particles in grid-based simulations of supersonic turbulence; and it forms the basis of the adaptive gravitational force softening method introduced by \citet{pm07}.
 
\subsection{Errors in the density estimate}
\label{sec:densityerrors}
 The formal errors in the density estimate may be determined by writing the density summation as an integral -- that is, assuming $m\equiv \rho dV$ and that the summation is well sampled, giving
\begin{equation}
\langle \rho({\bf r}) \rangle = \int \rho({\bf r}') W({\bf r} - {\bf r}', h) dV'.
\end{equation}
where $\langle ..\rangle$ refers to a smoothed estimate. Expanding $\rho({\bf r}')$ in a Taylor series about ${\bf r}$, we have
\begin{equation}
\langle \rho({\bf r}) \rangle = \rho({\bf r}) \int W({\bf r} - {\bf r}', h) dV' + \nabla \rho({\bf r}) \cdot \int ({\bf r}' - {\bf r}) W({\bf r} - {\bf r}', h) dV' + \nabla^{\alpha}\nabla^{\beta}\rho({\bf r}) \int \delta {\bf r}^\alpha \delta {\bf r}^\beta W({\bf r} - {\bf r}', h) dV' + \mathcal{O}(h^{3})
\end{equation}
so that if the normalisation condition (\ref{eq:Wnorm}) is satisfied and a symmetric kernel $W({\bf r} - {\bf r}',h) = W({\bf r}' - {\bf r},h)$ is used such that the odd error terms vanish, the error in the density interpolant is $\mathcal{O}(h^{2})$. In principle it is also possible to construct kernels such that the second moment is also zero, resulting in errors of $\mathcal{O}(h^{4})$ \citep[see][]{monaghan85}. The disadvantage of such kernels is that the kernel function becomes negative in some part of the domain, resulting in a potentially negative density evaluation. Achieving such higher order in practice also requires that the kernel is extremely well sampled, leading to substantial additional cost requirements. One possibility given the iterations necessary to solve (\ref{eq:rhosum}) would be to automatically switch between high order and low order kernels during the iterations (e.g. if a negative density occurs), thus leading to high order interpolation in smooth regions but a low order interpolation where the density changes rapidly. The errors in the discrete version are discussed further in Sec.~\ref{sec:errors}.

\subsection{Alternatives to the SPH density estimate}
 Finally, it should be noted that the three general approaches described in Sec.~\ref{sec:density} are not the only methods that can be employed for estimating the density. A fourth alternative that has received recent attention involves the use of Delaunay or Voronoi tessellation -- the former proposed by \citet{pelupessyetal03} and the latter developed into a full hydrodynamics scheme by \citet{sez05} and \citet{hs10}. These are promising approaches that in principle can offer all the same advantages as SPH in terms of exact conservation -- since it can be derived similarly from a Hamiltonian formulation -- but with an improved density estimate and an exact partition of unity.

\section{From density to equations of motion}
\label{sec:densitytoequationsofmotion}
 The reader at this point may wonder why we have spent so long discussing nothing else except the density estimate. The reason is that this is the only real freedom one has if one wishes to obtain a fully conservative SPH algorithm, at least in the absence of dissipative terms. This is because the rest of the SPH algorithm can be derived entirely \emph{from} the density estimate.
 
 \subsection{The discrete Lagrangian}
\label{sec:L}
 The derivation starts with the discrete Lagrangian. As is usual, the Lagrangian is simply given by
\begin{equation}
L = T - V,
\end{equation}
where $T$ and $V$ are the kinetic and potential (in this case, thermal) energies respectively. For a system of point masses with velocity ${\bf v} \equiv d{\bf r}/dt$ and internal energy per unit mass $u$, we have
\begin{equation}
L = \sum_{b} m_{b} \left[\frac12 v_{b}^{2} - u_{b}(\rho_{b}, s_{b}) \right],
\label{eq:L}
\end{equation}
where in general the thermal energy $u$ can be specified as a function of the thermodynamic variables $\rho$ and $s$ (the density and entropy, respectively). Although (\ref{eq:L}) can be considered as a discrete version of the continuum Lagrangian for hydrodynamics \citep[e.g.][]{eckart60,salmon88,morrison98}
\begin{equation}
L = \int \left[ \rho v^{2} - \rho u(\rho, s) \right] {\rm dV},
\end{equation}
one is free to consider the discrete Hamiltonian system, it's associated symmetries and equations of motion directly -- that is, without explicit reference to the continuum system. In other words the Hamiltonian properties are directly present in the discrete system and the motions will be constrained to obey the symmetries and conservation properties of the discrete Lagrangian.

\subsection{Least action principle and the Euler-Lagrange equations}
\label{sec:S}
The equations of motion for such a system can be derived from the principle of least action, that is minimising the action
\begin{equation}
S = \int L {\rm dt},
\label{eq:S}
\end{equation}
such that $\delta S = \int \delta L {\rm dt} = 0$, where $\delta$ is a variation with respect to a small change in the particle coordinates $\delta {\bf r}$. Assuming that the Lagrangian can be written as a differentiable function of the particle positions ${\bf r}$ and velocities ${\bf v}$, we have
\begin{equation}
\delta S = \int \left( \frac{\partial L}{\partial {\bf v}} \cdot \delta{\bf v} + \frac{\partial L}{\partial {\bf r}} \cdot \delta{\bf r} \right) {\rm dt} = 0.
\end{equation}
Integrating by parts, using the fact that $\delta {\bf v} = d(\delta {\bf r})/dt$, where $d/dt \equiv \partial/\partial t + {\bf v}\cdot\nabla$ gives
\begin{equation}
\delta S = \int \left\{ \left[ - \frac{d}{dt} \left(\frac{\partial L}{\partial {\bf v}}\right) + \frac{\partial L}{\partial {\bf r}} \right] \cdot \delta{\bf r}\right\} {\rm dt} + \left[ \frac{\partial L}{\partial {\bf v}} \cdot \delta{\bf r} \right]^{t}_{t_{0}} = 0. \label{eq:deltaS}
\end{equation}
So if we assume that the variation vanishes at the start and end times, then since the variation $\delta {\bf r}$ is arbitrary, the equations of motion are given by the Euler-Lagrange equations, here taken with respect to particle $a$:
\begin{equation}
 \frac{d}{dt}\left(\frac{\partial L}{\partial {\bf v}_{a}}\right) - \frac{\partial L}{\partial {\bf r}_{a}} = 0.
 \label{eq:el}
\end{equation}
 We have somewhat laboured the point here because it is important to understand the assumptions we have made by employing the Euler-Lagrange equations to derive the equations of motion. The first is that in using Eq.~\ref{eq:el} we are not explicitly considering the discreteness of the time integral. So when we refer below to ``exact conservation'' (e.g. of energy and momentum) we mean ``solely governed by errors in the time integration scheme''\footnote{It should be noted that it is quite possible to also derive the time integration scheme from a Lagrangian -- for example, \citet{monaghan05} gives the appropriate Lagrangian for the symplectic and time-reversible leapfrog scheme.}.
 
  The other, more critical, assumption we have made in employing (\ref{eq:el}) is that the Lagrangian is differentiable. This means that we have explicitly excluded the possibility of discontinuous solutions to the equations of motion. What this means in practice is that any discontinuities present in the system (for example in the initial conditions) require careful treatment -- for example by adding dissipative terms that smooth discontinuities to a resolvable scale (i.e., a few $h$) such that they can be treated as no longer discontinuous. We give some practical examples of this in Sec.~\ref{sec:shocks}. A better way would be to account for the neglected surface integral terms directly in the Lagrangian \citep[e.g.][]{kats01}, though it is not clear how one would go about doing so in an SPH context.
 
\subsection{Equations of motion}
\label{sec:equationsofmotion}
 All that remains in order to derive the equations of motion is to compute the derivatives in (\ref{eq:el}) by writing the terms in the Lagrangian as a function of the particle coordinates and velocities. From (\ref{eq:L}) we have
\begin{equation}
\pder{L}{{\bf v}_a} = m_a {\bf v}_a; \hspace{1cm}
\pder{L}{{\bf r}_a} =  - \sum_{b} m_{b} \left.\frac{\partial u_{b}}{\partial \rho_{b}}\right\vert_{s} \frac{\partial \rho_{b}}{\partial {\bf r}_{a}}, \label{eq:dLdr}
\end{equation}
the latter since $u$ is a function of $\rho$ and $s$, and we assume that the entropy $s$ is constant (i.e., no dissipation).
 Note that the former gives the canonical momenta of the system (${\bf p} \equiv \partial L / \partial {\bf v}$). This step is straightforward for hydrodynamics, but it can also be used to derive the conservative momentum variable in the case of more complicated physics -- for example in relativistic SPH \citep[e.g.][]{mp01,rosswog09}.
 
\subsubsection{Thermodynamics of the fluid}
\label{sec:thermo}
 From the first law of thermodynamics we have
\begin{equation}
dU = TdS - PdV,
\label{eq:firstlaw}
\end{equation}
where $\delta Q \equiv TdS$ is the heat added to the system (per unit volume) and $\delta W \equiv P dV$ is the work done by expansion and compression of the fluid. We do not compute the volume directly in SPH, but instead we can use the volume estimate given by $V = m/\rho$ and thus the change in volume given by $dV = - m/\rho^{2} d\rho$. Using quantities per unit mass instead of per unit volume (i.e., $du$ instead of $dU$), we have
\begin{equation}
du = Tds + \frac{P}{\rho^{2}} d\rho,
\label{eq:du}
\end{equation}
such that at constant entropy, the change in thermal energy is given by
\begin{equation}
\left.\frac{\partial u_{b}}{\partial \rho_{b}}\right\vert_{s} =  \frac{P}{\rho^{2}}. \label{eq:dudrho}
\end{equation}

\subsubsection{The density gradient}
 So far we have not made any explicit reference to SPH or kernel interpolation. This arises because of the spatial derivative of the density (Eq.~\ref{eq:dLdr}), that we obtain by differentiating our density estimate (Eq.~\ref{eq:rhosum}). Noting that we are taking the gradient of the density estimate at particle $b$ with respect to the coordinates of particle $a$, this is given by
\begin{equation}
\pder{\rho_{b}}{{\bf r}_{a}} = \frac{1}{\Omega_{b}} \sum_{c} m_{c} \pder{W_{bc}(h_{b})}{{\bf r}_{a}} \left( \delta_{ba} - \delta_{ca}\right),
\label{eq:gradrho}
\end{equation}
where $W_{bc}(h_{b})\equiv W({\bf r}_{b} - {\bf r}_{c}, h_{b})$, $\delta_{ba}$ is a Dirac delta function referring to the particle indices and we have assumed that the smoothing length is itself a function of density [i.e., $h = h(\rho)$], giving a term accounting for the gradient of the smoothing length given by
\begin{equation}
\Omega_a \equiv \left[1 - \pder{h_a}{\rho_a}\sum_{b} m_{b}
\pder{W_{ab}(h_a)}{h_a}\right],
\end{equation}
where for the standard $h-\rho$ relationship (Eq.~\ref{eq:hrho}) the derivative is given by
\begin{equation}
\frac{\partial h}{\partial \rho} = - \frac{h}{\rho d},
\end{equation}
where $d$ is the number of spatial dimensions.

\subsubsection{Equations of motion}
 Using (\ref{eq:gradrho}) and (\ref{eq:dudrho}) in (\ref{eq:dLdr}) we have
\begin{equation}
\pder{L}{{\bf r}_a} =  - \sum_{b} m_{b} \frac{P_{b}}{\Omega_{b} \rho_{b}^{2}} \sum_{c} m_{c} \pder{W_{bc}(h_{b})}{{\bf r}_{a}} \left( \delta_{ba} - \delta_{ca}\right), \label{eq:dLdrfirst}
\end{equation}
which, upon simplification, gives the equations of motion from the Euler-Lagrange equations in the form
\begin{equation}
\frac{d{\bf v}_{a}}{dt} = -\sum_{b} m_{b} \left[ \frac{P_{a}}{\Omega_{a} \rho_{a}^{2}} \pder{W_{ab}(h_{a})}{{\bf r}_{a}} + \frac{P_{b}}{\Omega_{b} \rho_{b}^{2}} \pder{W_{ab}(h_{b})}{{\bf r}_{a}} \right]. \label{eq:mom}
\end{equation}
For a constant smoothing length these equations simplify to the standard SPH expression \citep{monaghan92}
\begin{equation}
\frac{d{\bf v}_{a}}{dt} = -\sum_{b} m_{b} \left( \frac{P_{a}}{ \rho_{a}^{2}} + \frac{P_{b}}{ \rho_{b}^{2}} \right) \nabla_{a} W_{ab}. \label{eq:momconsth}
\end{equation}

\subsubsection{Conservation properties}
 Although we cannot yet provide interpretation to the equations of motion so derived, we can note a number of interesting properties. The first is that the total linear momentum is conserved exactly, since
\begin{equation}
\frac{d}{dt} \sum_{a} m_{a} {\bf v}_{a} = \sum_{a} m_{a} \frac{d{\bf v}_{a}}{dt} = -\sum_{a} \sum_{b} m_{a} m_{b} \left( \frac{P_{a}}{ \rho_{a}^{2}} + \frac{P_{b}}{ \rho_{b}^{2}} \right) \nabla_{a} W_{ab} = 0,
\end{equation}
where the double summation is zero because of the antisymmetry in the kernel gradient\footnote{This can be easily seen by swapping the summation indices $a$ and $b$ in the double sum and adding half of the original term to half of the rearranged term, giving zero.} (the reader may verify that this is also true for Eq.~\ref{eq:mom}). Secondly, the total angular momentum is also exactly conserved, since
\begin{equation}
\frac{d}{dt} \sum_a {\bf r}_a \times m_a {\bf v}_a = \sum_{a} m_{a} \left( {\bf r}_a \times \frac{d{\bf v}_a}{dt}
\right) = -\sum_a \sum_b m_a m_b \left( \frac{P_a}{\rho_a^2} +
\frac{P_b}{\rho_b^2}\right) {\bf r}_a \times ({\bf r}_a - {\bf r}_b) \tilde{F}_{ab} = 0,
\end{equation}
where for convenience we have written the gradient of the kernel in the form $\nabla_{a} W_{ab} = {\bf r}_{ab} \tilde{F}_{ab}$, and the last term is zero again because of the antisymmetry in the double summation [since (${\bf r}_{a} \times {\bf r}_{b}) = -({\bf r}_{b} \times {\bf r}_{a}$)].

 The above conservation properties follow directly from the symmetries in the original Lagrangian and (by extension) the SPH density estimate (Eq.~\ref{eq:rhosum}) -- linear momentum conservation because the Lagrangian and density estimate are invariant to translations, and angular momentum conservation because they are invariant to rotations of the particle coordinates. This is important in thinking about possible modifications to the SPH scheme (for example using non-spherical kernels would result in the non-conservation of angular momentum because the density estimate would no longer be invariant to rotations).
 
  Finally, we note that although the equations of motion depend only on the relative positions of the particles, they \emph{do} depend on the absolute value of the pressure. That is, Eqs. (\ref{eq:mom}) and (\ref{eq:momconsth}) contain a force between the particles that is non-zero even when the pressure is constant. We discuss the importance of -- and problems associated with -- this `spurious' force in Sec.~\ref{sec:localconservation}.

\subsection{Energy equation}
 The remaining part of the (dissipationless) SPH algorithm -- the energy equation -- can also be derived from the Hamiltonian dynamics. Here we have the choice of evolving either the thermal energy $u$, the total specific energy $e = \frac12 v^{2} + u$ or an entropy variable $K = P/\rho^{\gamma}$. Equations for each of these can be derived, as given below. It is important to note, however, that -- provided the equations are derived from the Lagrangian formulation -- there is no difference in SPH between evolving any of these variables, \emph{except} due to the timestepping algorithm. This is rather different to the situation in an Eulerian code where the finite differencing of the advection terms mean that writing the equations in conservative form (i.e., using $e$) differs more substantially from evolving $u$ or $K$.

\subsubsection{Internal energy}
The evolution equation for the internal energy (in the absence of dissipation), from Eq.~\ref{eq:dudrho}, is given by
\begin{equation}
\frac{du_a}{dt} = \frac{P_a}{\rho_a^2}\frac{d\rho_a}{dt}. \label{eq:dudtdrhodt}
\end{equation}
Taking the time derivative of the density sum (Eq.~\ref{eq:rhosum}), we obtain an evolution equation for $u$ of the form
\begin{equation}
\frac{du_a}{dt} = \frac{P_a}{\Omega_{a} \rho_a^2} \sum_{b} m_{b} ({\bf v}_{a} - {\bf v}_{b})\cdot\nabla_{a} W_{ab} (h_{a}).
\label{eq:sphutherm}
\end{equation}

\subsubsection{Total energy}
\label{sec:totalE}
The conserved (total) energy is found from the Lagrangian via the Hamiltonian
\begin{equation}
H = \sum_a {\bf v}_a \cdot \pder{L}{{\bf v}_a} - L = \sum_{a} m_{a} \left( \frac{1}{2} v_a^2 + u_a \right),
\label{eq:H}
\end{equation}
which is simply the total energy of the SPH particles, $E$, since the Lagrangian does not explicitly depend on the time. Taking the (Lagrangian) time derivative of (\ref{eq:H}), we have
\begin{equation}
\frac{dE}{dt} = \sum_{a} m_{a} \left({\bf v}_a \cdot \frac{d{\bf v}_a}{dt} + \frac{du_a}{dt} \right).
\label{eq:totalE}
\end{equation}
Substituting (\ref{eq:mom}) and (\ref{eq:sphutherm}) and rearranging we find 
\begin{equation}
\frac{dE}{dt} = \sum_{a} m_{a} \frac{de_a}{dt} = -\sum_a \sum_b m_a m_b \left[ \frac{P_a}{\Omega_{a} \rho_a^2}{\bf v}_b\cdot\nabla_{a} W_{ab}(h_{a})
+ \frac{P_b}{\Omega_{b} \rho_b^2}{\bf v}_a\cdot\nabla_{a} W_{ab}(h_{b}) \right] = 0. \label{eq:dEdt}
\end{equation}
This equation shows that the total energy is also exactly conserved by the SPH scheme (where the double sum is zero again because of the antisymmetry with respect to the particle index, similar to the conservation of linear momentum discussed above). The conservation of total energy is a consequence of the symmetry of the Lagrangian (\ref{eq:L}) with respect to time as well as invariance under time translations. Eq.~\ref{eq:dEdt} also shows that the dissipationless evolution equation for the specific energy $e$ is given by
\begin{equation}
\frac{de_a}{dt} = -\sum_b m_b \left[ \frac{P_a}{\Omega_{a} \rho_a^2}{\bf v}_b\cdot\nabla_{a} W_{ab}(h_{a})
+ \frac{P_b}{\Omega_{b}\rho_b^2}{\bf v}_a\cdot\nabla_{a} W_{ab}(h_{b}) \right].
\label{eq:sphenergy}
\end{equation}

\subsubsection{Entropy}
 For the specific case of an ideal gas equation of state, where
\begin{equation}
P = K(s) \rho^\gamma,
\end{equation}
it is possible to use the function $K(s)$ as the evolved variable \citep{sh02}, where the evolution of $K$ is given by
\begin{equation}
\frac{dK}{dt} = \frac{\gamma-1}{\rho^{\gamma-1}}\left( \frac{du}{dt} - \frac{P}{\rho^2}\frac{d\rho}{dt} \right) = \frac{\gamma-1}{\rho^{\gamma-1}}\left( \frac{du}{dt} \right)_{diss}.
\label{eq:sphentropy}
\end{equation}
The thermal energy is then evaluated using
\begin{equation}
u = \frac{K}{\gamma - 1} \rho^{\gamma-1}.
\end{equation}
Since $dK/dt = 0$ in the absence of dissipation, using $K$ has the advantage that the evolution is independent of the time-integration algorithm. The disadvantage is that it is more difficult to apply to non-ideal equations of state. This is sometimes referred to as the `entropy-conserving' form of SPH \citep[after][]{sh02} -- which is somewhat misleading since the entropy per particle is also exactly conserved if (\ref{eq:sphutherm}) or (\ref{eq:sphenergy}) are used provided the smoothing length gradient terms are correctly accounted for (i.e., $du/dt - P/\rho^{2} d\rho/dt = 0$), apart from minor differences arising from the timestepping scheme. So the term `entropy-conserving' more correctly refers to the correct accounting of smoothing length gradient terms and a consistent formulation of the energy equation than whether or not an entropy variable is evolved.

\subsection{Summary}
 In summary, our full system of equations for $\rho$, ${\bf v}$ and $u$ is given by
\begin{eqnarray}
\rho_{a} & = & \sum_{b} m_{b} W({\bf r}_{a} - {\bf r}_{b}, h_{a});   \hspace{1cm} h = h(\rho),  \label{eq:rhosumfinal} \\
\frac{d{\bf v}_{a}}{dt} & = & -\sum_{b} m_{b} \left[ \frac{P_{a}}{\Omega_{a} \rho_{a}^{2}} \nabla_{a} W_{ab}(h_{a}) + \frac{P_{b}}{\Omega_{b} \rho_{b}^{2}} \nabla_{a} W_{ab}(h_{b}) \right], \label{eq:momfinal} \\
\frac{du_a}{dt} & = & \frac{P_a}{\Omega_{a} \rho_a^2} \sum_{b} m_{b} ({\bf v}_{a} - {\bf v}_{b})\cdot\nabla_{a} W_{ab} (h_{a}), \label{eq:dudtfinal}
\end{eqnarray}
where in place of (\ref{eq:dudtfinal}) we could equivalently use either (\ref{eq:sphenergy}) or (\ref{eq:sphentropy}). The reader will note that so far we have not even mentioned the continuum equations of hydrodynamics -- we have merely specified the physics that goes into the Lagrangian (Eq.~\ref{eq:L}), the thermodynamics of the fluid (Eq.~\ref{eq:du}) and the manner in which the density is calculated (Eq.~\ref{eq:rhosum}) and with these have directly derived the discrete equations of the Hamiltonian system. In order to interpret them we need to know how to translate our SPH equations (\ref{eq:rhosumfinal})--(\ref{eq:dudtfinal}) into their continuum equivalents.

 Before doing so, it is worth recapping briefly the assumptions we have made in arriving at (\ref{eq:rhosumfinal})--(\ref{eq:dudtfinal}) from the Lagrangian (Eq.~\ref{eq:L}). These are
\begin{enumerate}[i)]
\item That the time integration and thus time derivatives, $d/dt$, are computed exactly (though this assumption can in principle be relaxed);
\item That the Lagrangian, and by implication the density and thermal energies are differentiable;
\item That there is no change in entropy, such that the first law of thermodynamics $du = P dV$ is satisfied, and that the change in particle volume is given by $dV = -m/\rho^{2} d\rho$.
\end{enumerate}
 The second and third assumptions in particular come into play in dealing with shocks and other kinds of discontinuities, which will we discuss further in Sec~\ref{sec:shocks}.

\subsection{Alternative formulations}
 Within the constraints of a Hamiltonian SPH formulation, it is clear that there is only a very limited freedom to change the algorithm without breaking some of the conservation properties of the scheme. So there are only two basic ways to change the (dissipationless) algorithm that are consistent with the Hamiltonian approach (other than changing the first law of thermodynamics): i) change the way the density is calculated or ii) introduce additional physical terms, and associated constraints, into the Lagrangian. Examples of the former are the consistent formulation of variable smoothing length terms -- requiring the iterative solution of $h$ and $\rho$ in the density sum (Eq.~\ref{eq:rhosum}) -- by \citet{sh02} and \citet{monaghan02}, and an incorporation of boundary correction forces \citep{kulasegarametal04}. Examples of the latter include consistent derivations of relativistic SPH \citep{mp01,rosswog09}, adaptive gravitational force softening \citep{pm07}, sub-resolution turbulence models \citep{monaghan02,monaghan04} and MHD \citep{pm04b,price10}.

\section{Kernel interpolation theory and SPH derivatives}
\label{sec:interpolation}
 The usual way of introducing SPH is to start with a formal discussion of kernel interpolation theory. We have taken a rather different approach in this review and will introduce this theory primarily only to \emph{interpret} the equations that we have derived from the discrete Lagrangian, and also as a way of discussing how to go about introducing additional physics. However, we will \emph{not} use the linear error properties of the interpolation scheme to define the method -- apart from our construction of the density estimate discussed above. The reason is that focussing on linear errors over the Hamiltonian properties of SPH misses some of the subtle but important non-linear behaviour that makes SPH work in practice, which we will come to discuss. However, first let us proceed:
 
\subsection{Kernel interpolation: the basics}
\label{sec:kernelbasics}
   Kernel interpolation theory starts with the identity
\begin{equation}
A({\bf r}) = \int A({\bf r'}) \delta({\bf r}-{\bf r}') d{\bf r}',
\end{equation}
where $A$ is an arbitrary scalar variable and $\delta$ refers to the Dirac delta function. This integral is then
approximated by replacing the delta function with a smoothing kernel $W$ with finite width $h$, i.e.,
\begin{equation}
A({\bf r}) = \int A({\bf r}') W({\bf r} - {\bf r}', h) d{\bf r'} + O(h^2).
\label{eq:intint}
\end{equation}
where $W$ has the property
\begin{equation}
\lim_{h\to 0} W({\bf r} - {\bf r}',h) = \delta ({\bf r} - {\bf r}'),
\label{eq:wtodelta}
\end{equation}
and normalisation $\int_{V} W dV' = 1$, as we have already discussed in Sec.~\ref{sec:density}. Finally the integral interpolant (Eq.~\ref{eq:intint}) is discretised onto a finite set of interpolation points (the particles) by replacing the integral by a summation and the mass element $\rho dV$ with the particle mass $m$, i.e.
\begin{eqnarray}
\langle A({\bf r}) \rangle & = & \int \frac{A({\bf r'})}{\rho({\bf r'})}
W({\bf r}-{\bf r}', h) \rho({\bf r'})d{\bf r'}, \nonumber \\
& \approx & \sum^{N_{neigh}}_{b=1} m_{b} \frac{A_{b}}{\rho_b}
W({\bf r}-{\bf r}_{b}, h), \label{eq:sumint}
\end{eqnarray}
This `summation interpolant' is the basis of all SPH formalisms. The reader will note that choosing $A = \rho$ results in the SPH density estimate (Eq.~\ref{eq:rhosum}). In this paper we have argued that the density estimate (\ref{eq:rhosum}) is in some sense more fundamental than the summation interpolant (\ref{eq:sumint}), since the equations of motion can be derived without reference to (\ref{eq:sumint}). On the other hand, the summation interpolant gives a general way of a interpolating a quantity at any point in space $A({\bf r})$ from quantities defined solely on the particles themselves (i.e., $A_{b}$, $\rho_{b}$, $m_{b}$)\footnote{Eq.~\ref{eq:sumint} naturally also forms the basis for visualisation of SPH simulations, where one wishes to reconstruct the field in all of the spatial volume given quantities defined on particles: This is the approach implemented in \textsc{splash} \citep{splashpaper}.}, and in turn to a general way of formulating SPH equations. In particular, gradient terms may be straightforwardly calculated by taking the derivative of (\ref{eq:sumint}), giving
\begin{eqnarray}
\nabla A({\bf r}) & = & \frac{\partial}{\partial {\bf r}} \int
\frac{A({\bf r'})}{\rho({\bf r'})} W({\bf r}-{\bf r}', h)
\rho({\bf r'})d{\bf r'} + O(h^2), \label{eq:gradintint} \\
& \approx & \sum_{b} m_{b} \frac{A_{b}}{\rho_b} \nabla W({\bf r}-{\bf r}_{b}, h).\label{eq:gradsumint}
\end{eqnarray}
For vector quantities the expressions are similar, simply replacing $A$ with ${\bf A}$ in (\ref{eq:sumint}), giving
\begin{eqnarray}
{\bf A}({\bf r}) & \approx & \sum_{b} m_{b} \frac{{\bf A}_{b}}{\rho_b} W({\bf r}-{\bf r}_{b}, h), \\
\nabla\cdot {\bf A}({\bf r}) & \approx & \sum_{b} m_{b} \frac{{\bf A}_{b}}{\rho_b} \cdot \nabla W({\bf r}-{\bf r}_{b}, h), \label{eq:basicdiv} \\
\nabla\times {\bf A}({\bf r}) & \approx & - \sum_{b} m_{b} \frac{{\bf A}_{b}}{\rho_b} \times \nabla W({\bf r}-{\bf r}_{b}, h), \label{eq:basiccurl} \\
\nabla^{j} A^{i}({\bf r}) & \approx & \sum_{b} m_{b} \frac{A^{i}_{b}}{\rho_b} \nabla^{j} W({\bf r}-{\bf r}_{b}, h). \label{eq:basicgradvec} 
\end{eqnarray}
The problem is that using these expressions `as is' in general leads to quite poor gradient estimates, and we can do better by considering the errors in the above approximations. However, the basic interpolants given above give us a general way of interpreting SPH expressions such as those derived in Sec.~\ref{sec:densitytoequationsofmotion}. 

\subsection{Interpretation of the Hamiltonian SPH equations}
 We are now able to provide interpretation to our Hamiltonian-SPH equations (\ref{eq:rhosumfinal})--(\ref{eq:dudtfinal}) derived in Sec.~\ref{sec:densitytoequationsofmotion} using the basic identities (\ref{eq:gradsumint})--(\ref{eq:basiccurl}). We begin with the density summation (\ref{eq:rhosumfinal}). Taking the time derivative, we have
\begin{equation}
\frac{d\rho_{a}}{dt} = \frac{1}{\Omega_{a}} \sum_{b} m_{b} \left( {\bf v}_{a} - {\bf v}_{b} \right) \cdot \nabla_{a} W_{ab} (h_{a}),
\label{eq:drhodt}
\end{equation}
which for a constant smoothing length simplifies to
\begin{equation}
\frac{d\rho_{a}}{dt} = \sum_{b} m_{b} \left( {\bf v}_{a} - {\bf v}_{b} \right) \cdot \nabla_{a} W_{ab}(h).
\label{eq:drhodtconsth}
\end{equation}
Using (\ref{eq:basicdiv}) we can translate each of the terms according to
\begin{equation}
\frac{d\rho_{a}}{dt} = {\bf v}_{a}\cdot \sum_{b} \frac{m_{b}}{\rho_{b}} \rho_{b} \nabla_{a} W_{ab} - \sum_{b}  \frac{m_{b}}{\rho_{b}}  (\rho_{b} {\bf v}_{b}) \cdot \nabla_{a} W_{ab}  \approx {\bf v}_{a} \cdot \nabla \rho - \nabla\cdot(\rho {\bf v}) \approx - \rho_{a} (\nabla\cdot{\bf v})_{a}. \label{eq:drhodttranslate}
\end{equation}
So remarkably (\ref{eq:drhodtconsth}) (and by inference, \ref{eq:drhodt}) -- which we obtained simply by taking the time derivative of the density sum -- represents a (particular) SPH discretisation of the continuity equation. Indeed, the density summation is therefore an exact, time-independent, solution to the SPH continuity equation. This should not be particularly surprising, since the continuity equation derives from the conservation of mass, which is self-evidently enforced on the particles since $m$ is fixed.

 Our force term (\ref{eq:momfinal}), assuming a constant $h$ (Eq.~\ref{eq:momconsth}) can be translated according to
\begin{equation}
-\sum_{b} m_{b} \left(\frac{P_{a}}{\rho_{a}^{2}} + \frac{P_{b}}{\rho_{b}^{2}} \right) \nabla_{a} W_{ab} \approx -\frac{P}{\rho^{2}} \nabla \rho - \nabla \left( \frac{P}{\rho} \right) \approx -\frac{\nabla P}{\rho},
\end{equation}
where we have used the basic identity (\ref{eq:gradsumint}) to translate the two terms into their continuum equivalents. 

 Finally, the thermal energy equation (\ref{eq:dudtfinal}) can be translated, from (\ref{eq:dudtdrhodt}) and (\ref{eq:drhodttranslate}), giving
\begin{equation}
\frac{du}{dt} = - \frac{P}{\rho} \nabla\cdot{\bf v}.
\end{equation}

 In other words, it is evident that the Lagrangian has indeed given us valid discretisations for the equations of hydrodynamics, and therefore that the equations of our Hamiltonian system (\ref{eq:rhosumfinal})--(\ref{eq:dudtfinal}) do indeed solve these equations in discrete form -- a remarkable achievement given the relatively few assumptions that were made. Yet the discretisations we derived are clearly not the basic ones arising from kernel interpolation theory (\ref{eq:gradsumint})--(\ref{eq:basiccurl}). In order to examine the errors in these discretisations we need to understand the errors arising from the basic gradient operators, and how more general derivative operators can be constructed.

\subsection{Errors}
\label{sec:errors}
\begin{figure}[t]
\begin{center}
   \includegraphics[width=0.7\columnwidth]{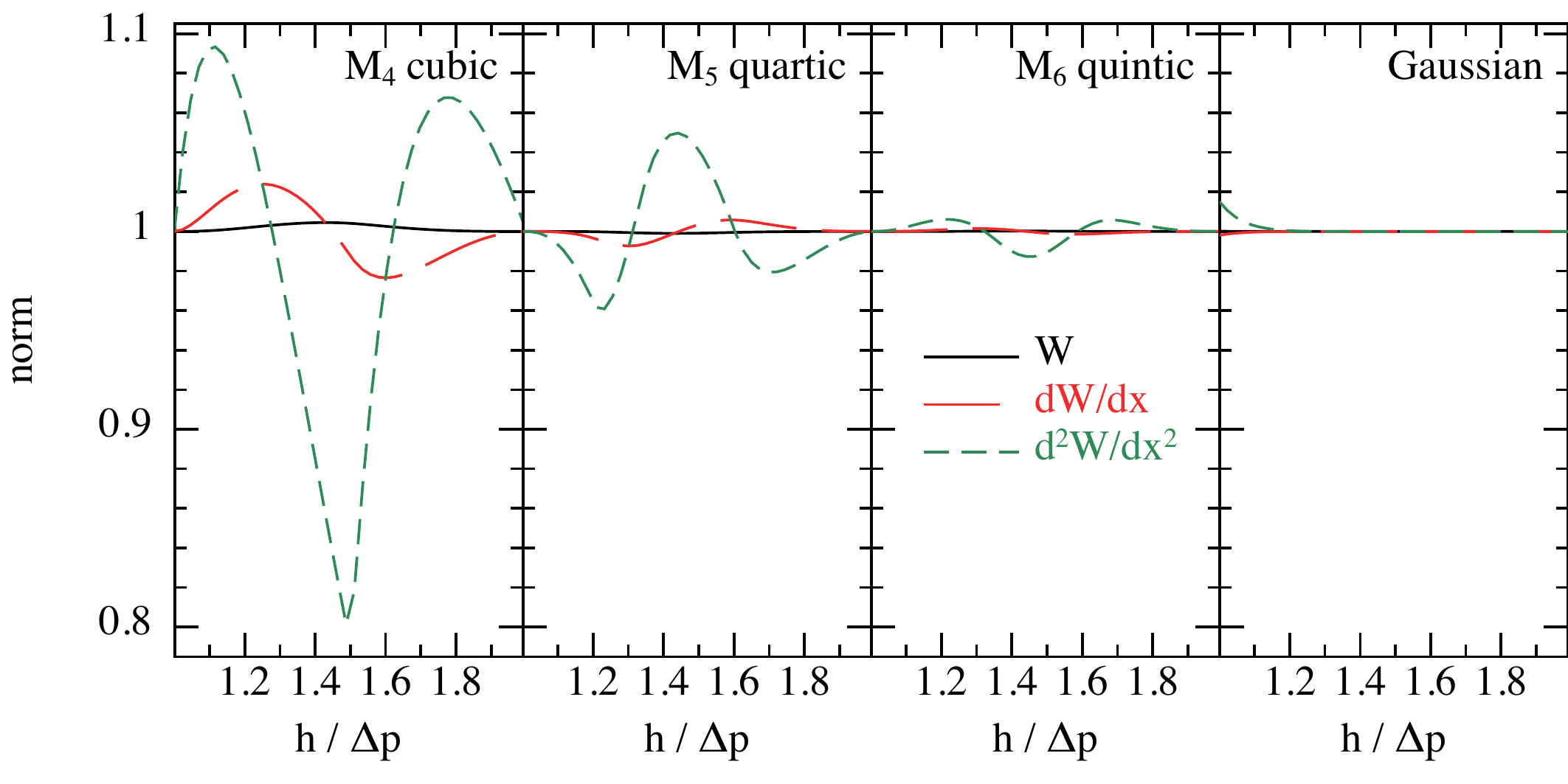}
   \caption{Accuracy with which the normalisation conditions on the kernel, the kernel gradient and the kernel second derivative are computed with fixed $h$ on a one-dimensional line of particles with the $M_{4}$--$M_{6}$ kernels (from left) compared to the Gaussian (right), as a function of the ratio of smoothing length to particle spacing $h/\Delta p$. Bell-shaped kernels with compact support lead to excellent density estimates (solid/black lines), reasonable gradient estimates (long dashed/red lines) but poor second derivative estimates (short dashed/green lines) when a finite number of neighbours are employed.}
   \label{fig:Bsplinesnorm}
\end{center}
\end{figure}

 The errors introduced by the approximation (\ref{eq:intint}) are similar to those in the density estimate (Sec.~\ref{sec:densityerrors}). That is, if we expand $A({\bf r}')$ in a Taylor series about ${\bf r}$ \citep{benz90,monaghan92}, we find
\begin{equation}
\langle A({\bf r}) \rangle = \int \left[A({\bf r}) + ({\bf r}'-{\bf r})^\alpha \pder{A}{{\bf r}^\alpha} + \frac12
({\bf r}'-{\bf r})^\alpha ({\bf r}'-{\bf r})^\beta \pder{^2 A}{{\bf r}^\alpha \partial{\bf r}^\beta} +
\mathcal{O}(({\bf r}-{\bf r}')^3)\right] W({\bf r}-{\bf r}', h) \mathrm{d{\bf r}'},
\end{equation}
such that for symmetric $W\equiv W(\vert {\bf r} - {\bf r}'\vert, h)$ and normalised ($\int W dV' = 1$) kernels we have
\begin{equation}
\langle A({\bf r}) \rangle = A({\bf r})  + \frac12
\pder{^2 A}{{\bf r}^\alpha \partial{\bf r}^\beta} \int ({\bf r}'-{\bf r})^\alpha ({\bf r}'-{\bf r})^\beta W(\vert {\bf r} - {\bf r}'\vert, h)
\mathrm{d{\bf r}'} + \mathcal{O}[({\bf r}'-{\bf r})^4],
\label{eq:kernelerrors}
\end{equation}
giving an interpolation that is second order accurate [$\mathcal{O}(h^{2})$] unless higher order kernels are used (see Sec.~\ref{sec:densityerrors}). However, the errors in the discrete version (Eq.~\ref{eq:sumint}) are not identical, since they depend on the degree to which the discrete summations approximate the integrals -- specifically on the degree to which the discrete normalisation conditions are satisfied. Following \citet{price04} we can perform a similar analysis on the summation interpolant (\ref{eq:sumint}, here assumed to be computed on particle $a$) by expanding $A_{b}$ in a Taylor series around ${\bf r}_{a}$, giving
\begin{equation}
\langle A_{a} \rangle = \sum_{b} m_{b} \frac{A_{b}}{\rho_{b}} W_{ab} = A_{a} \sum_{b} \frac{m_{b}}{\rho_{b}} W_{ab} + \nabla A_{a} \cdot \sum_{b} \frac{m_{b}}{\rho_{b}} ({\bf r}_{b} - {\bf r}_{a}) W_{ab} + \mathcal{O}(h^{2}).
\end{equation}
This shows that in practice, the interpolation will only truly be second order accurate if the conditions
\begin{equation}
\sum_{b} \frac{m_{b}}{\rho_{b}} W_{ab} \approx 1; \hspace{1cm}\textrm{and}\hspace{1cm} \sum_{b} \frac{m_{b}}{\rho_{b}} ({\bf r}_{b} - {\bf r}_{a}) W_{ab} \approx 0,
\label{eq:discretenorms}
\end{equation}
hold. The degree to which this is true depends strongly on the particle distribution within the kernel radius, and the properties of the kernel when a finite number of neighbours are employed -- in particular, the ratio of smoothing length to particle spacing $\Delta p$. Fig.~\ref{fig:Bsplinesnorm} shows the first condition computed for the B-spline kernels and the Gaussian as a function of $h/\Delta p$ in 1D (solid/black lines), showing that in general the above conditions are maintained very well, provided the particles are regular. Thus, maintaining a regular particle arrangement, together with an appropriate choice of $h/\Delta p$, can be very important in obtaining accurate results with SPH. Used another way, the conditions (\ref{eq:discretenorms}) are essentially the criteria for a `good density kernel' -- that is, a good density kernel is one which satisfies these conditions well given the typical particle distributions encountered in SPH. Note that the second of these is much easier to satisfy than the first, since it requires only a reasonably symmetric particle distribution to be satisfied. In general reasonable fulfilment of the first condition amounts to the conditions 1)--3) on the density kernel described in Sec.~\ref{sec:density}.

The errors resulting from the gradient interpolation (\ref{eq:gradintint}) may be estimated in a similar manner by again expanding $A({\bf r}')$ in a Taylor series about ${\bf r}$, giving
\begin{eqnarray}
\nabla A({\bf r}) & = & \int \left[A({\bf r}) + ({\bf r}' - {\bf r})^\alpha
\pder{A}{{\bf r}^\alpha} + \frac12 ({\bf r}'-{\bf r})^\beta ({\bf r}'-{\bf r})^\gamma
\pder{^2 A}{{\bf r}^\beta \partial {\bf r}^\gamma} + \mathcal{O}[({\bf r}-{\bf r}')^3] \right] \nabla
W(|\mathbf{r-r'}|, h) d{\bf r}', \nonumber \\ 
& = & A({\bf r}) \int \nabla W d{\bf r}' + \pder{A}{{\bf r}^\alpha}\int
({\bf r}'-{\bf r})^\alpha \nabla
W \mathrm{d{\bf r}'} + \frac12 \pder{^2 A}{{\bf r}^\beta \partial{\bf r}^\gamma} \int
 ({\bf r}'-{\bf r})^\beta ({\bf r}'-{\bf r})^\gamma \nabla W \mathrm{d{\bf r}'} +
\mathcal{O}[({\bf r}' - {\bf r})^3], \nonumber \\
& = & \nabla A({\bf r}) + \frac12 \pder{^2 A}{{\bf r}^\beta \partial {\bf r}^\gamma} \int
({\bf r}'-{\bf r})^\beta ({\bf r}'-{\bf r})^\gamma \nabla W(r) \mathrm{d{\bf r}'} +
\mathcal{O}[({\bf r}'-{\bf r})^3], \label{eq:gradinterrors}
\end{eqnarray}
where we have used the fact that $\int \nabla W d{\bf r}' = 0$ for even
kernels, whilst the second term
integrates to unity for even kernels satisfying the normalisation condition (\ref{eq:Wnorm}). The resulting errors in the integral interpolant for the gradient are therefore also of $\mathcal{O}(h^2)$. As previously, the errors in the discrete version (\ref{eq:gradsumint}) can be found by expanding $A_b$ in a Taylor series around ${\bf r}_a$, giving
\begin{eqnarray}
\langle \nabla A_a \rangle & = & \sum_b m_b \frac{A_{b}}{\rho_b} \nabla_{a} W_{ab}, \nonumber \\
& = & A_a \sum_b \frac{m_b}{\rho_b} \nabla_{a} W_{ab} + \pder{A_a}{{\bf r}^\alpha}  \sum_b \frac{m_b}{\rho_b} ({\bf r}_b - {\bf r}_a)^\alpha \nabla_{a} W_{ab} + \mathcal{O}(h^{2}) \label{eq:sumgraderrors}
\end{eqnarray}
where the summations represent SPH approximations to the integrals in the second line of (\ref{eq:gradinterrors}). So the gradient errors resulting from (\ref{eq:gradsumint}) would in principle be similarly governed by the extent to which these discrete summations approximate the integrals, i.e., how well the conditions
\begin{equation}
\sum_b \frac{m_{b}}{\rho_b} \nabla_{a} W_{ab} \approx {\bf 0}; \hspace{0.5cm}\textrm{and}\hspace{0.5cm} \sum_b \frac{m_b}{\rho_b} ({\bf r}_b - {\bf r}_a)^\alpha \nabla^{\beta}_{a} W_{ab} \approx \delta^{\alpha\beta};
\label{eq:gradWnorm}
\end{equation}
hold. The latter term is shown for the B-spline kernels by the long-dashed/red lines in Fig.~\ref{fig:Bsplinesnorm}. The difference is that for gradients we can explicitly use the error terms to construct more accurate gradient operators, and we do this below.

\subsection{First derivatives}
\label{sec:gradients}
 From (\ref{eq:sumgraderrors}) we immediately see that a straightforward improvement to the gradient estimate (\ref{eq:gradsumint}) can be obtained by a simple subtraction of the first error term (i.e., the term in (\ref{eq:sumgraderrors}) that is present even in the case of a constant function), giving \citep[e.g.][]{monaghan92}
\begin{equation}
\langle \nabla A_a\rangle = \sum_{b} m_{b} \frac{(A_b - A_a)}{\rho_b} \nabla_{a} W_{ab} \label{eq:bettergrad}, 
\end{equation}
which, interpreting each term according to (\ref{eq:gradsumint}), is an SPH estimate of
\begin{equation}
\nabla A \approx \langle \nabla A \rangle - A \langle \nabla 1\rangle.
\end{equation}
Since the first error term in (\ref{eq:sumgraderrors}) is removed, the interpolation is exact for constant functions and indeed this is obvious from the form of (\ref{eq:bettergrad}). The interpolation can be made exact for linear functions via a matrix inversion of the second error term in (\ref{eq:sumgraderrors}), i.e., solving
\begin{equation}
\chi^{\alpha\beta}\pder{A_a}{{\bf r}^\alpha} = \sum_b \frac{m_b}{\rho_b}(A_b - A_a)
\nabla^\beta W_{ab}, \hspace{1cm} \chi^{\alpha\beta} \equiv \sum_b\frac{m_b}{\rho_b} ({\bf r}_b - {\bf r}_a)^\alpha \nabla^\beta W_{ab}.
\label{eq:exactlinear}
\end{equation}
where $\nabla^\beta \equiv \partial / \partial {\bf r}^\beta$. This normalisation is somewhat cumbersome in practice, since $\chi$ is a matrix quantity, requiring considerable
extra storage (in three dimensions this means storing $3\times 3 = 9$ extra quantities for each particle) and also since calculation of this term requires prior knowledge of the density.

 A similar interpolant for the gradient follows by using
\begin{equation}
 \nabla A \approx \frac{1}{\rho} [\langle \nabla(\rho A)\rangle - A\langle \nabla \rho \rangle]  = \frac{1}{\rho_a} \sum_{b} m_{b} (A_b - A_a) \nabla_{a} W_{ab}, \label{eq:bettergradusual}
\end{equation}
which again is exact for a constant $A$. Expanding $A_b$ in a Taylor series, we see that in this case the interpolation of a linear function can be made exact by solving
\begin{equation}
 \chi^{\alpha\beta} \pder{A_a}{{\bf r}^\alpha} =\sum_{b} m_{b} (A_b - A_a) \nabla^\beta W_{ab}, \hspace{1cm} \chi^{\alpha\beta} \equiv \sum_{b} m_{b} ({\bf r}_b - {\bf r}_a)^\alpha \nabla^\beta W_{ab}. \label{eq:exactlinearrho}
\end{equation}
which has some advantages over (\ref{eq:exactlinear}) in that it can be computed without prior knowledge of the density.

 However, the gradient operator we derived in the equations of motion (\ref{eq:momconsth}) does not correspond to any of the above possibilities. This operator is given by
\begin{equation}
\nabla A \approx \rho \left[\frac{A}{\rho^2}\langle \nabla\rho \rangle + \langle \nabla\left(\frac{A}{\rho}\right) \rangle \right] = \rho_a \sum_{b} m_{b} \left( \frac{A_a}{\rho_a^2} +
\frac{A_b}{\rho_b^2}\right)\nabla_{a} W_{ab}. \label{eq:symmetrisedgrad}
\end{equation}
Expanding $A_{b}$ in a Taylor series about ${\bf r}_{a}$, we have
\begin{equation}
\rho_{a} A_a \sum_{b} m_{b} \left( \frac{1}{\rho_a^2} + \frac{1}{\rho_b^2}\right)\nabla_{a} W_{ab} + \pder{A_a}{{\bf r}^\alpha} \rho_{a} \sum_b \frac{m_b}{\rho_b^2} ({\bf r}_b - {\bf r}_a)^\alpha \nabla_{a} W_{ab} + \mathcal{O}(h^2), \label{eq:aboringline}
\end{equation}
from which we see that for a constant function the error in (\ref{eq:symmetrisedgrad}) is governed by the extent to which
\begin{equation}
\sum_{b} m_{b} \left( \frac{1}{\rho_a^2} + \frac{1}{\rho_b^2}\right)\nabla_{a} W_{ab} \approx 0.
\label{eq:errorterm}
\end{equation}
Although a simple subtraction of the first term in (\ref{eq:aboringline}) from the original expression (\ref{eq:symmetrisedgrad}) eliminates this error, this would not give the form we derived in Sec.~\ref{sec:equationsofmotion}. Indeed, retaining the exact conservation of momentum requires that such error terms are not eliminated, the consequences of which we will discuss in Sec.~\ref{sec:localconservation}.

\subsection{Generalised first derivative operators}
\label{sec:generalgradients}
 Finally, an infinite variety of gradient operators -- of two basic types -- can be constructed by noting that
\begin{equation}
\nabla A =\frac{1}{\phi} \left[\nabla(\phi A) - A \nabla \phi\right] \approx \sum_{b} \frac{m_{b}}{\rho_{b}}\frac{\phi_{b}}{\phi_{a}} \left(A_{b} - A_{a}\right) \nabla_{a} W_{ab}, 
\label{eq:phiantisym}
\end{equation}
and
\begin{equation}
\nabla A = \phi \left[\frac{A}{\phi^2} \nabla\phi  + \nabla\left(\frac{A}{\phi}\right) \right] \approx \sum_{b} \frac{m_{b}}{\rho_{b}} \left( \frac{\phi_{b}}{\phi_{a}} A_{a} +  \frac{\phi_{a}}{\phi_{b}} A_{b}  \right) \nabla_{a} W_{ab}, \label{eq:phisym}
\end{equation}
where $\phi$ is \emph{any} arbitrary, differentiable scalar quantity defined on the particles. Indeed, for a given $\phi$, the pair of operators defined by (\ref{eq:phiantisym}) and (\ref{eq:phisym}) can be shown to form a conjugate pair\footnote{The conjugate nature of the symmetric and antisymmetric SPH gradient operators was first noted by \citet{cr99} in the context of projection schemes for SPH.} -- and choosing one (e.g. for the density/thermal energy evolution) tends to lead to the other (e.g. in the equations of motion).  For example, (\ref{eq:bettergrad}) and (\ref{eq:bettergradusual}) correspond to using $\phi = 1$ and $\phi = \rho$ respectively in (\ref{eq:phiantisym}), whilst (\ref{eq:symmetrisedgrad}) correponds to using $\phi = \rho$ in (\ref{eq:phisym}) and arises in the equations of motion because it is the conjugate operator to (\ref{eq:bettergradusual}) that arises in the density gradient (\ref{eq:drhodtconsth}).

 Various `alternative' formulations of the SPH equations have been proposed that correspond to a particular choice of $\phi$ \citep[e.g.][]{hk89,rt01,mw03}. For example \citet{hk89} suggested using an acceleration equation of the form
\begin{equation}
\frac{d{\bf v}}{dt} = -\sum_{b} m_{b} \left( 2\frac{\sqrt{P_{a} P_{b}}}{\rho_{a} \rho_{b}} \right) \nabla_{a} W_{ab},
\label{eq:Pab}
\end{equation}
corresponding to the symmetric operator (Eq.~\ref{eq:phisym}) with $\phi = \sqrt{P}/\rho$. It can be readily shown that ensuring exact conservation of energy with such a formulation would require using the conjugate operator (i.e., Eq.~\ref{eq:phiantisym} with $\phi = \sqrt{P}/\rho$) in the thermal energy equation (although it should be noted that simultaneous exact conservation of energy and entropy is not possible with \emph{any} alternative formulation).


\subsection{First derivatives of vector quantities}
\label{sec:vectorderivs}
All of the above discussion applies also to vector derivatives, simply using (\ref{eq:basicdiv}) or (\ref{eq:basiccurl}) in place of (\ref{eq:gradsumint}) and likewise resulting in two basic operators for each type of derivative, given by
\begin{equation}
\langle \nabla\cdot{\bf A} \rangle_{a} \approx \sum_{b} \frac{m_{b}}{\rho_{b}}\frac{\phi_{b}}{\phi_{a}} \left({\bf A}_{b} - {\bf A}_{a}\right)\cdot \nabla_{a} W_{ab};
\hspace{1cm}
\langle\nabla\times{\bf A}\rangle_{a} \approx -\sum_{b} \frac{m_{b}}{\rho_{b}}\frac{\phi_{b}}{\phi_{a}} \left({\bf A}_{b} - {\bf A}_{a}\right)\times \nabla_{a} W_{ab},
\label{eq:antisymdivcurl}
\end{equation}
and
\begin{equation}
\langle \nabla\cdot {\bf A}\rangle_{a} \approx \sum_{b} \frac{m_{b}}{\rho_{b}} \left( \frac{\phi_{b}}{\phi_{a}} {\bf A}_{a} +  \frac{\phi_{a}}{\phi_{b}}{\bf A}_{b}  \right) \cdot \nabla_{a} W_{ab};
\hspace{1cm}
\langle \nabla\times {\bf A}\rangle_{a} \approx -\sum_{b} \frac{m_{b}}{\rho_{b}} \left( \frac{\phi_{b}}{\phi_{a}} {\bf A}_{a} + \frac{\phi_{a}}{\phi_{b}}{\bf A}_{b} \right)\times \nabla_{a} W_{ab},
\label{eq:symdivcurl}
\end{equation}
where as previously $\phi$ is an arbitrary (differentiable) scalar quantity. For general vector derivatives written in tensor notation the corresponding expressions are given by
\begin{equation}
\langle \nabla^{j} A^{i}\rangle_{a} \approx \sum_{b} \frac{m_{b}}{\rho_{b}}\frac{\phi_{b}}{\phi_{a}} \left(A^{i}_{b} - A^{i}_{a}\right) \nabla^{j}_{a} W_{ab};
\hspace{0.5cm}\textrm{or}\hspace{0.5cm}\langle \nabla^{j} A^{i}\rangle_{a} \approx \sum_{b} \frac{m_{b}}{\rho_{b}} \left( \frac{\phi_{b}}{\phi_{a}} A^{i}_{a} +  \frac{\phi_{a}}{\phi_{b}}A^{i}_{b}  \right) \nabla^{j}_{a} W_{ab};
\label{eq:gradA}
\end{equation}
These operators form the basic building blocks for formulating quite general SPH equations. Higher order operators can also be constructed for vector derivatives using matrix inversions, similar to (\ref{eq:exactlinear}) and (\ref{eq:exactlinearrho}). 

\subsection{Particle methods ``the wrong way'': an SPH formulation based on linear errors and exact derivatives}
\label{sec:relpressure}
 Based on the discussion given in Sec.~\ref{sec:errors}--\ref{sec:vectorderivs} a straightforward approach would be to simply go ahead and discretise the continuum equations of hydrodynamics (or magnetohydrodynamics) using the most accurate gradient estimates possible. For example employing (\ref{eq:bettergrad}) the hydrodynamic equations of motion $d{\bf v}/dt = -\nabla P / \rho$ could be written in the form
\begin{equation}
\frac{d{\bf v}}{dt} = \sum_{b} m_{b} \left( \frac{P_{a} - P_{b}}{\rho_{a} \rho_{b}} \right) \nabla_{a} W_{ab},
\label{eq:Pab}
\end{equation}
or any of the alternatives offered by choosing $\phi$ appropriately in (\ref{eq:phiantisym}). Indeed such a formulation was originally examined (and discarded) by \citet{morris96,morrisphd} but has been recently (re-)proposed by \citet{abel10} (the latter author employing $\phi = 1/\rho$). We could even proceed to more accurate derivatives using Eqs.~(\ref{eq:exactlinear}) or (\ref{eq:exactlinearrho}) that are exact to linear order. Yet, these operators are clearly different from the equation of motion we derived from the Lagrangian (\ref{eq:mom}), though on the basis of the linear error properties alone (Sec.~\ref{sec:gradients}) would seem to be a much better choice. On the other hand, it is clear that (\ref{eq:Pab}) does not exactly conserve linear (or angular) momentum, nor total energy.

 The key difference in approach is that the above analysis tells us about the \emph{linear} errors, whereas the Hamiltonian formulation tells us about the \emph{non-linear} properties of the system (i.e., symmetries, conservation and constraints on the behaviour of the global system). These turn out to be crucial for long term stability and accuracy, and we will look at what this means in practice in Sec.~\ref{sec:localconservation}. That said, ``linear errors do not lie'', so it will always be true that -- provided the particle distribution is regular -- formulations such as (\ref{eq:Pab}) or similar will be more accurate for linear or weakly non-linear problems (i.e., those run for a short time and/or not involving strong shocks) and with sufficient resolution can always be made to give accurate results.
 
  Ideally of course, one would want \emph{both} exact derivatives \emph{and} exact conservation. In SPH at least, it seems one cannot have both -- and to my knowledge this is yet to be convincingly demonstrated by any particle method, though it is perhaps possible using tesselation schemes.

\section{Why a bad derivative leads to good derivatives: The importance of local conservation}
\label{sec:localconservation}
 The paradox we face is that, whilst the Lagrangian derivation gave us valid discretisations of the equations of hydrodynamics, these are not the most accurate discretisations that are possible based on a linear error analysis. Indeed, on the basis of the linear error properties, the gradient estimate derived for the acceleration equation would be a very \emph{poor} choice of gradient operator, yet it is the only operator which respects all of the symmetry and conservation properties in the Lagrangian.
 
  To understand what these errors mean in practice we can perform a simple thought experiment (which we will also run as a numerical experiment): Consider a distribution of particles in a closed (e.g. periodic) box at constant pressure. In particular, we will consider a uniform random particle distribution. This is the worst-case error scenario, since errors in the interpolation will essentially be Monte-Carlo ($\propto 1/\sqrt{N}$). Now consider what will happen if use one of the more accurate gradient estimators, for example (\ref{eq:Pab}). Clearly, since the pressure is constant, there will be no force and hence no particle motion. That is, the force vanishes when the pressure is constant \emph{regardless of the particle distribution}. If instead we use the Hamiltonian formulation, then the pairwise force between particles is given by
\begin{equation}
\underbrace{-m_{a} m_{b}}_{-ve} \underbrace{\left[\frac{P}{\rho_{a}^{2}} + \frac{P}{\rho_{b}^{2}} \right]}_{+ve}  \underbrace{F_{ab}}_{-ve} (\hat{\bf r}_{a} - \hat{\bf r}_{b}).
\end{equation}
where we have written the kernel gradient using $\nabla_{a} W_{ab}\equiv \hat{\bf r}_{ab} F_{ab}$. Since for positive definite kernels (Fig.~\ref{fig:Bsplines}) the kernel derivative function is always negative, the overall result -- assuming positive pressure -- is a \emph{positive} (that is, repulsive) force between the particles directed along their line of sight. This force will only vanish when the error term (\ref{eq:errorterm}) becomes zero -- that is, when the particles are \emph{regular}. What this means is that \emph{the particles  care about their (bad) arrangement} and will rearrange themselves until the condition (\ref{eq:errorterm}) holds, corresponding to a particle distribution that is locally isotropic and regular. This state corresponds to the minimum energy state of the Hamiltonian system -- i.e., that which minimises the action (\ref{eq:S}), and in this state the symmetric gradient estimate (\ref{eq:symmetrisedgrad}) is computed with good accuracy (since by definition the motions act to minimise the errors in this estimate). \citet{monaghan05} gives some specific examples of this.

 In other words, the Hamiltonian SPH formulation -- or more generally any formulation which respects local conservation of momentum between particle pairs -- contains an \emph{intrinsic} ``re-meshing'' procedure and the particles are constrained to remain locally ordered at all times. With an `accurate' but non-conservative pressure estimate the particles are insensitive to their arrangement and can become arbitrarily randomised in the course of a simulation (according to the streamlines of the flow). Thus, methods based on an `exact derivatives' approach \citep[e.g.][]{dilts99,mh03} inevitably require the addition of explicit (and ad-hoc) re-meshing procedures, as the interpolation errors on a randomised particle distribution will be terrible regardless of the order of the interpolation scheme. So in practise the `accurate' gradient estimate will give much \emph{less} accurate results. This is the paradox of SPH: One deliberately chooses a bad gradient estimate (in the linear errors) in order to obtain a good gradient estimate (because the particles stay regular).
 
 One of the implications of this ``intrinsic remeshing'' is that not all initial conditions represent stable configurations for the particles. In particular, the cubic lattice -- though simple -- does not represent a minimum energy state and is only quasi-stable for certain ratios of the smoothing length to the particle separation \citep{morrisphd,bot04}. In general, given a small perturbation or sufficient time, the cubic lattice will transition to the more isotropic hexagonal-close-packed lattice arrangement and will do so almost immediately if $h/\Delta p$ is small (e.g. $\eta \lesssim 1.1$).

\begin{figure}[t]
\begin{center}
   \includegraphics[width=0.8\columnwidth]{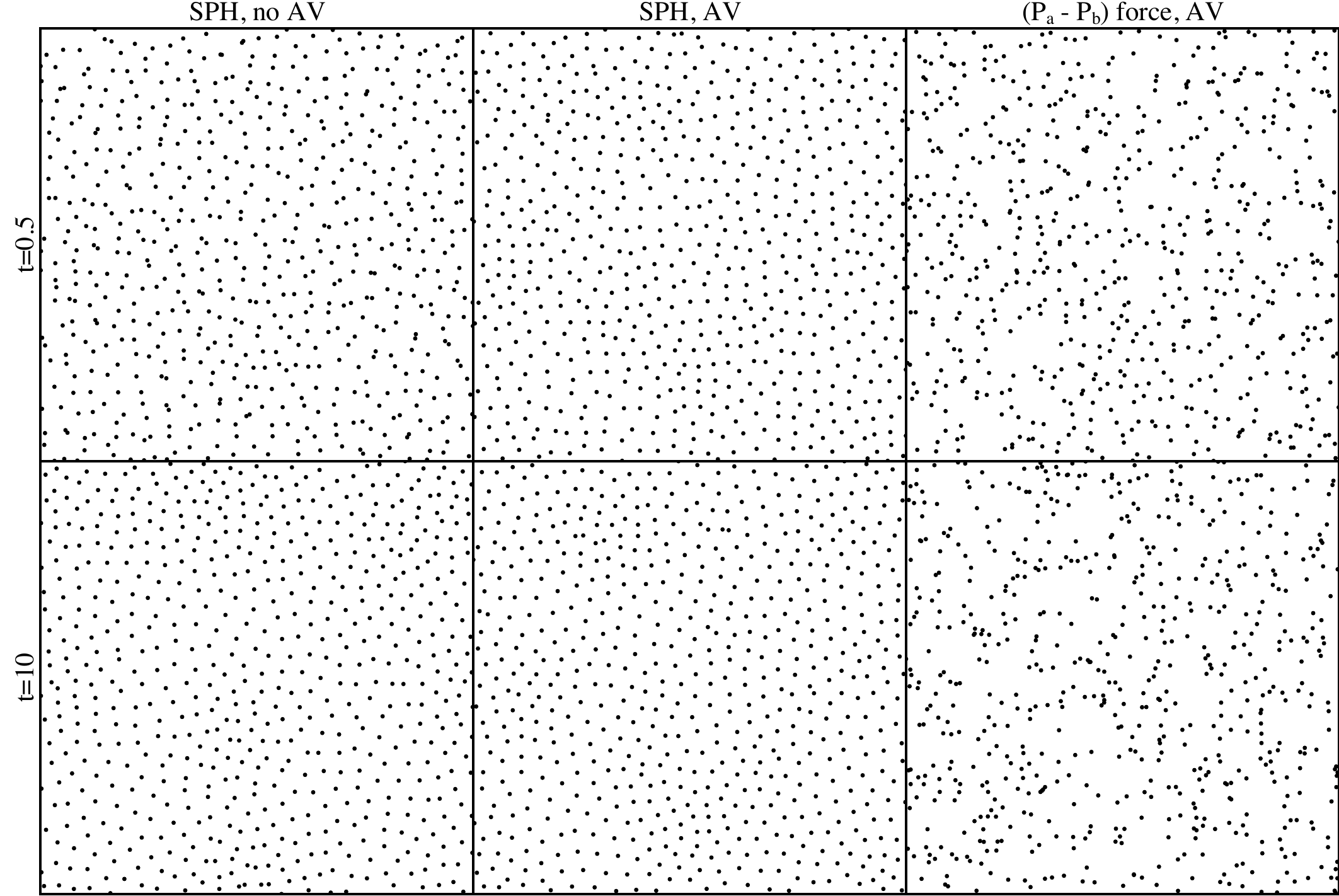}
   \caption{Settling of an initially random particle distribution due to pairwise conservation of momentum in the pressure gradient. The left and centre columns show the results with the standard (Hamiltonian) SPH method, both without (left) and with (centre) artificial viscosity, after $0.5$ (top) and 10 (bottom) sound crossing times. The right column shows the results when a ``relative pressure'' formulation is adopted. With a momentum-conserving force the particles are sensitive to their arrangement and will regularise accordingly (left and centre columns), whereas ``more accurate'' but non-conserving formulations (right) compute gradients that are insensitive to the particle arrangement and thus require explicit re-meshing procedures. Note that although the application of artificial viscosity helps the settling to proceed faster (centre), it is primarily a pressure-driven effect and occurs even if no viscosity is applied (left).}
   \label{fig:settling}
\end{center}
\end{figure}

\subsection{Example 1: Settling of a random particle distribution}
 Our first numerical example (using \textsc{ndspmhd}) demonstrates this `settling' in practice. The setup is a two dimensional domain $x,y \in [0,1]$ with periodic boundary conditions. The particles are given an initially uniform thermal energy, density is calculated according to the sum and the pressure is determined using $P = (\gamma - 1)\rho u$ with $\gamma = 5/3$. The thermal energy is set to give a sound speed $c_{s} = 1$, giving $u = 0.9$. The end result should therefore be a uniform pressure equilibrium. The particle distributions after 0.5 (top row) and 10 (bottom row) sound crossing times are shown in Fig.~\ref{fig:settling} using the Hamiltonian SPH formulation (\ref{eq:rhosumfinal})--(\ref{eq:dudtfinal}) (left and centre columns, without and with artificial viscosity respectively) and the equations of motion computed using a `relative pressure'  formulation (specifically, Eq.~\ref{eq:phiantisym} with $\phi = 1/\rho$ as employed by \citealt{abel10}) (right column). With a locally conservative formulation the random initial configuration settles rapidly into a regular particle distribution (left and centre columns), leading in turn to good gradient estimates. This settling occurs even in the absence of an artificial viscosity term (left panels), though adding viscosity does help speed the settling process (centre panels). However, using a ``more accurate'' but non-conservative gradient estimate there is no regularisation of the particle distribution, leading to poor gradient estimates due to the random nature of the particle distribution. The total energy is also conserved exactly by the SPH formulations, whilst in the relative pressure formulation the total energy grows exponentially.

\subsection{Example 2: A 2D shock tube}
\begin{figure}
\begin{center}
   \includegraphics[width=\columnwidth]{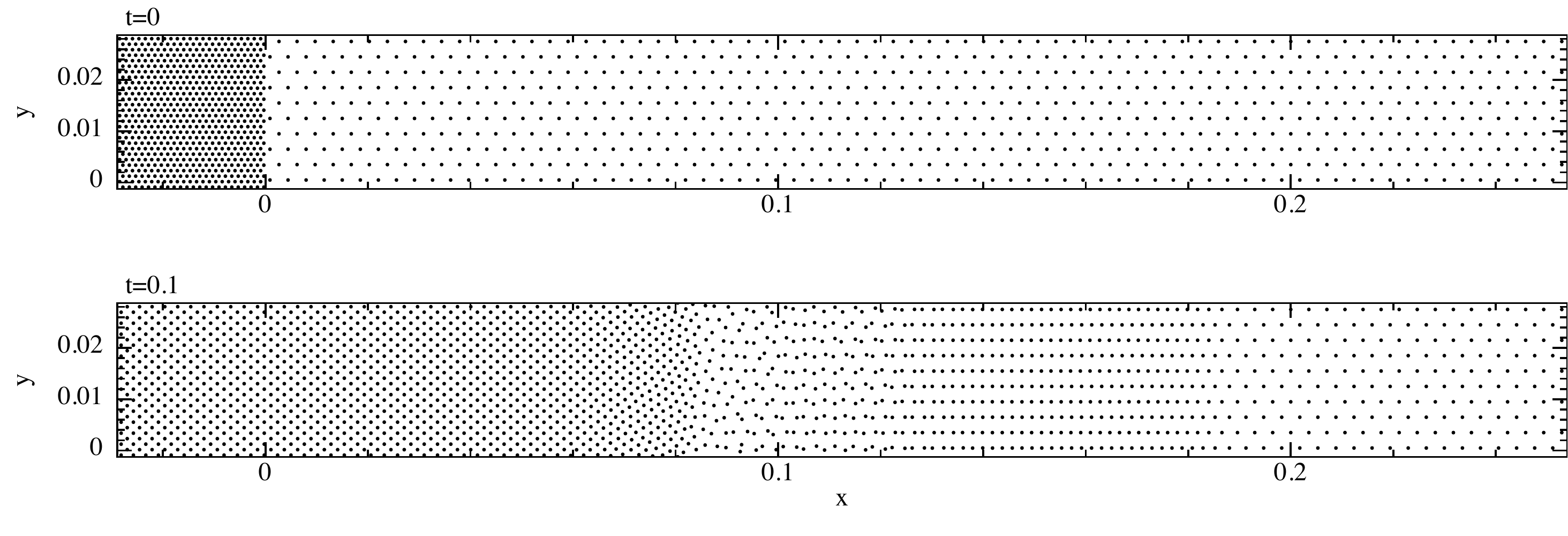}
   \caption{Particle settling in a two dimensional shock tube problem. The particles are initially arranged on hexagonal close packed lattices either side of the shock (top panel). As the shock propagates (bottom panel, showing $t=0.1$) it induces a one dimensional compression in the particles and thus a highly anisotropic particle arrangement, which ``remeshes'' to a more isotropic arrangement downstream from the shock, involving small motions of particles in the $y-$direction. }
   \label{fig:shockparts2D}
\end{center}
\end{figure}

 The other ``classic'' example of particle settling is the behaviour of SPH particles in a multidimensional shock tube problem, where there is a 1D compression of the particle distribution (e.g. along the x axis). Since the shock induces a highly anisotropic compression -- and thus a highly non-preferred particle arrangement -- the mutual repulsion of SPH particles will eventually produce a post-shock ``remeshing'' of the particle distribution, involving transverse motions of the particles. An example is given in Fig.~\ref{fig:shockparts2D}, showing the particle distribution at $t=0.1$ in a two dimensional Sod shock tube problem (described further in Sec.~\ref{sec:sodshockkh}) in which the particles were initially placed on two hexagonal close packed lattices upstream and downstream of the shock (initially placed at the origin). The particles can be seen to ``break'' (at $x\approx 0.14$) from the highly anisotropic compression-induced `lines' at $0.14 \lesssim x \lesssim 0.19$, leading to a more isotropic particle distribution further downstream ($x \lesssim 0.12$). This example also illustrates the fact that one inevitably has some motions at the resolution scale that are not related to the physical problem, but related to the implicit ``regularisation'' of the particle distribution present in locally conservative SPH formulations.

\subsection{Corollary: Negative pressures and the tensile instability}
\label{sec:tensile}
 The corollary of the above is that the particles require a \emph{positive} pressure in order to remain ordered. If the net pressure (or stress) becomes negative, the net force between a particle pair will become \emph{attractive}, causing a catastrophic numerical instability. For example, with a pressure gradient of the form
 \begin{equation}
\frac{d{\bf v}_{a}}{dt} = -\sum_{b} m_{b} \left( \frac{P_{a} - P_{0}}{ \rho_{a}^{2}} + \frac{P_{b} - P_{0}}{ \rho_{b}^{2}} \right) \nabla_{a} W_{ab},
\end{equation}
the pairwise force will become negative when $P_{0} > P$, and in this situation the particles clump together unphysically. This is known as the `tensile instability' \citep{monaghan00} and occurs in SPH when a stress tensor is employed that can result in (physically) negative stresses. In particular, this is the case for MHD \citep{pm85} and in elastic dynamics \citep*{gms01}. The occurrence of the tensile instability was one of the main initial difficulties with the development of MHD in SPH and is discussed in detail in Sec.~\ref{sec:mhdtensile}.

\subsection{The pairing instability: Why one cannot simply use `more neighbours'.}
 Another, more benign, instability in the particle distribution occurs with the cubic spline and other bell-shaped kernels depending on the ratio of smoothing length to particle spacing. This is due to the shape of the kernel gradient term for these kernels (see Fig.~\ref{fig:Bsplines}), and is a consequence of the fact that these kernels are designed to give good density estimates (Sec.~\ref{sec:density}), rather than necessarily being the best choice for calculating gradients. In particular, the kernel gradient in these kernels contains a maximum (negative) value at $r/h \sim 2/3$ and tends to zero at the origin (Fig.~\ref{fig:Bsplines}). This characteristic is desirable for a good density estimate -- as it means one is insensitive to a small change in the position of a near neighbour -- but means that the mutual repulsive force tends to zero for neighbouring particles placed ``within the hump'' of the kernel gradient. The net result is that two particles spaced closer than the location of the ``hump'' in the gradient form a ``pair'', eventually falling on top of each other.
\begin{figure}[t]
\begin{center}
   \includegraphics[width=0.8\columnwidth]{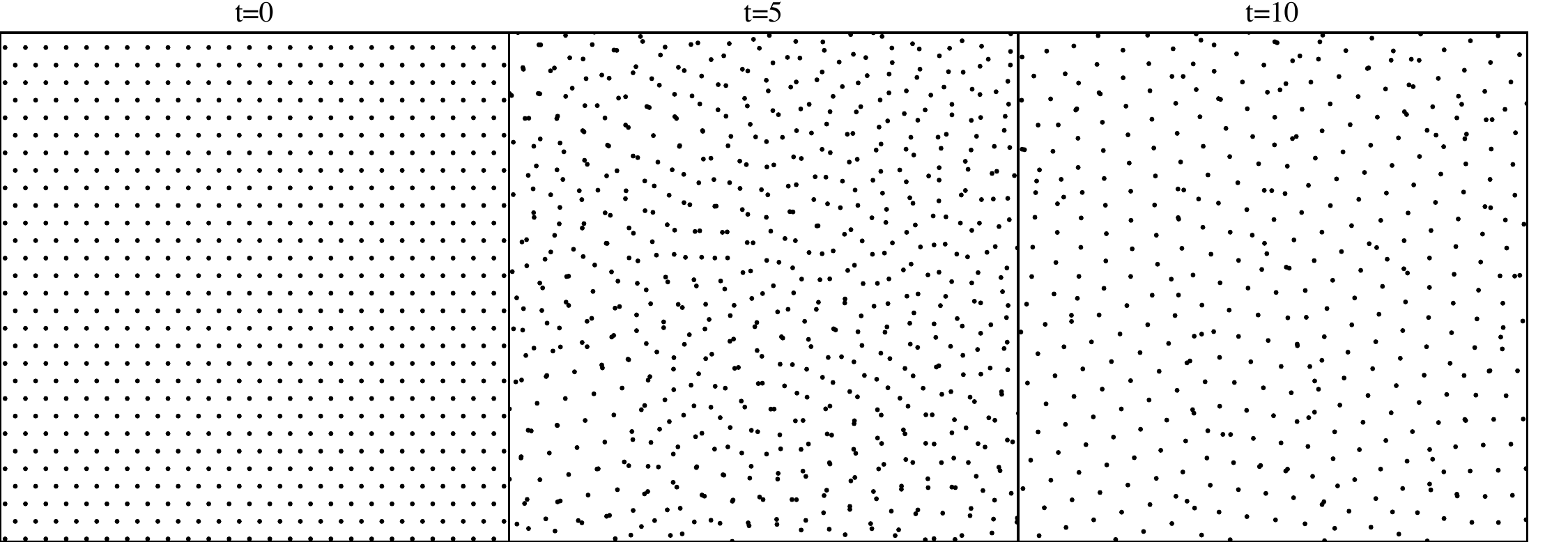}
   \caption{The pairing instability in action: The (2D) setup is similar to that shown in Fig.~\ref{fig:settling} except that the particles are initially placed on a close packed lattice and we use the $M_{4}$ cubic spline kernel with a large ratio of smoothing length to particle spacing (here $\eta = 1.5$, corresponding to $\sim 28$ and $100$ neighbours in 2 and 3D respectively). After a few sound crossing times ($t=5$, centre panel) particles form `pairs' which proceed to merge into a locally hexagonal ``glass-like'' lattice arrangement with almost exactly half the resolution of the initial conditions (right panel, shown after 10 sound crossing times). Although fairly benign -- and easily avoided by a sensible choice of $\eta$ -- the pairing instability is the main reason one cannot simply ``stretch'' the cubic spline to large neighbour numbers to achieve convergence. Instead, one should use a kernel with a larger radius of compact support but the same ratio of smoothing length to particle spacing, such as the $M_{5}$ or $M_{6}$ splines.}
   \label{fig:pairing}
\end{center}
\end{figure}

  For the cubic and other B-spline kernels complete merging occurs when $h \gtrsim 1.5 \Delta p$ (i.e., $\eta \gtrsim 1.5$ or $\gtrsim 100$ Neighbours in 3D for the cubic spline), corresponding to the placement of the first neighbour ``inside the hump''. There is also an intermediate regime $1.225 \lesssim \eta \lesssim 1.5$ ($62$--$100$ Neighbours in 3D) where a close-packed or cubic lattice is unstable to pair formation, but where the pairs do not completely merge. These empirical regimes are confirmed by detailed stability analysis of the SPH equations in 2D \citep{morrisphd,bot04} that explicitly show that instability occurs -- though with small energies -- for large $h/\Delta p$.
  
   Fig.~\ref{fig:pairing} (and our example 3) shows the pairing instability in action: The setup is as for example 1 but with $\eta = 1.5$ in the cubic spline kernel instead of $\eta = 1.2$ and with particles placed initially on a hexagonal close-packed lattice (an otherwise very stable configuration: left panel). After a few sound crossing times (centre panel) particles begin to form pairs, with these pairs eventually merging completely (right panel) to give a locally hexagonal ``glass-like'' configuration, but with exactly half the resolution of the initial conditions! 
   
   Though fixes have been proposed\footnote{\citet{tc92} suggested modifying the gradient of the cubic spline kernel, using
\begin{equation}
w'(q)= -\sigma \left\{ \begin{array}{ll}
-1 & 0 \le q < 2/3; \\
 - 3 q + \frac94 q^{2}, & 2/3 \le q < 1; \\
\frac{3}{4}(2-q)^2, & 1 \le q < 2; \\
0. & q \ge 2. \end{array} \right. \label{eq:tc92}
\end{equation}
with $W$ itself unchanged and $\sigma$ equal to the usual normalisation factor for the cubic spline (i.e., $1/\pi$ in 3D). That is, the ``hump'' is removed by simply making the kernel gradient constant within $r/h < 2/3$. Whilst it cures the pairing instability, one should be careful about employing such a gradient in practice since the kernel gradient (\ref{eq:tc92}) is no longer correctly normalised (i.e., Eq.~\ref{eq:gradWnorm}b no longer holds, even in the continuum limit) meaning that as the region within $r/h < 2/3$ is increasingly well sampled the numerical sound speed and other quantities will be systematically wrong. Though one could attempt to re-normalise the new gradient kernel, this results in a low weighting in the outer regions that in turn leads to poor gradient estimates.

 Similarly, whilst perhaps a satisfactory `gradient kernel' could be derived without a pairing instability, in the derivation from a Lagrangian there is no freedom over the kernel gradient since it derives directly from the gradient of density -- that is, if one separates the gradient kernel from the density kernel then either the total energy (from Eq.~\ref{eq:totalE}) or the entropy will no longer exactly be conserved (the latter if $du/dt \neq P/\rho^{2} d\rho/dt$ and thus $dK/dt \ne 0$ in Eq.~\ref{eq:sphentropy}).}, none are entirely satisfactory. However, the pairing instability, unlike the tensile instability, is quite benign. For example the density change associated with the transition in Fig.~\ref{fig:pairing} is of order $1\%$ for the cubic spline and $0.1\%$ for the $M_{6}$ quintic -- but entails a factor-of-two loss in spatial resolution and is therefore a waste of computational resources. Furthermore, it can be easily avoided by a sensible choice of $\eta$ (we recommend $\eta = 1.2$ for the B-spline kernels, corresponding to $N_{neigh}=57.9$ for the cubic spline in 3D). The pairing instability is the main reason one cannot simply ``stretch'' the cubic spline to large neighbour numbers in order to obtain convergence and demonstrates at least one good reason why $\eta$ (or $N_{neigh}$) should not be regarded as a free parameter in SPH simulations.

\section{Second derivatives and dissipation terms in SPH and SPMHD}
\label{sec:2ndderivs}
\subsection{The SPH Laplacian}
 Our remaining ``basic'' issue regarding both SPH and SPMHD regards the formulation of second derivative terms. As for first derivative terms we can start with the basic summation interpolant (\ref{eq:sumint}) and simply take derivatives analytically, e.g.
\begin{equation}
\langle \nabla^{2} A\rangle_{a} \approx \sum_{b} m_{b} \frac{A_{b}}{\rho_b} \nabla^{2}_{a} W_{ab}.\label{eq:gradgradsumint}
\end{equation}
 Expanding $A_{b}$ in a Taylor series about ${\bf r}_{a}$, we have
\begin{equation}
\sum_{b} m_{b} \frac{A_{b}}{\rho_b} \nabla^{2}_{a} W_{ab}  = A_{a} \sum_{b} \frac{m_{b}}{\rho_b} \nabla^{2}_{a} W_{ab} + \nabla^{\alpha} A_{a} \sum_{b} \frac{m_{b}}{\rho_b} {\delta \bf r}^{\alpha} \nabla^{2}_{a} W_{ab} +  \frac12 \nabla^{\alpha} \nabla^{\beta} A_{a} \sum_{b} \frac{m_{b}}{\rho_b} {\delta \bf r}^{\alpha} {\delta \bf r}^{\beta} \nabla^{2}_{a} W_{ab} + \ldots.
\end{equation}
As previously, we can immediately improve the estimate by subtracting the first error term from both sides, giving an interpolant of the form
\begin{equation}
\sum_{b} \frac{m_{b}}{\rho_b} (A_{b} - A_{a}) \nabla^{2}_{a} W_{ab}  = \nabla^{\alpha} A_{a} \sum_{b} \frac{m_{b}}{\rho_b} {\delta \bf r}^{\alpha} \nabla^{2}_{a} W_{ab} +  \frac12\nabla^{\alpha} \nabla^{\beta} A_{a} \sum_{b} \frac{m_{b}}{\rho_b} {\delta \bf r}^{\alpha} {\delta \bf r}^{\beta} \nabla^{2}_{a} W_{ab} + \ldots,
\label{eq:2ndderiverrors}
\end{equation}
which vanishes when $A$ is constant. However, the accuracy of our second derivative estimate will depend on the remaining error terms in (\ref{eq:2ndderiverrors}), corresponding to the normalisation conditions:
\begin{equation}
\sum_{b} \frac{m_{b}}{\rho_{b}} \delta {\bf r} \nabla^{2} W_{ab} = {\bf 0};  \hspace{0.5cm}\textrm{and}\hspace{0.5cm} \frac12 \sum_{b} \frac{m_{b}}{\rho_{b}} \delta {\bf r}^{\alpha} \delta {\bf r}^{\beta} \nabla^{2} W_{ab} = \delta^{\alpha\beta}, \label{eq:del2norm}
\end{equation}
such that
\begin{equation}
\nabla^{2} A_{a} \approx \sum_{b} \frac{m_{b}}{\rho_b} (A_{b} - A_{a}) \nabla^{2}_{a} W_{ab}.
\label{eq:2ndderivsph}
\end{equation}

 The problem is that the conditions (\ref{eq:del2norm}) are very poorly satisfied using the second derivative of a compact bell-shaped kernel function (short-dashed/green lines in Fig.~\ref{fig:Bsplines}, showing the second term in \ref{eq:del2norm}), since the second derivative changes sign inside the kernel domain (for the cubic spline it is also discontinuous) and must therefore be extremely well sampled in each direction to give accurate results. Thus in practice we require a better functional form -- a ``second derivative kernel''.

\begin{figure}[t]
\begin{center}
   \includegraphics[width=0.35\columnwidth]{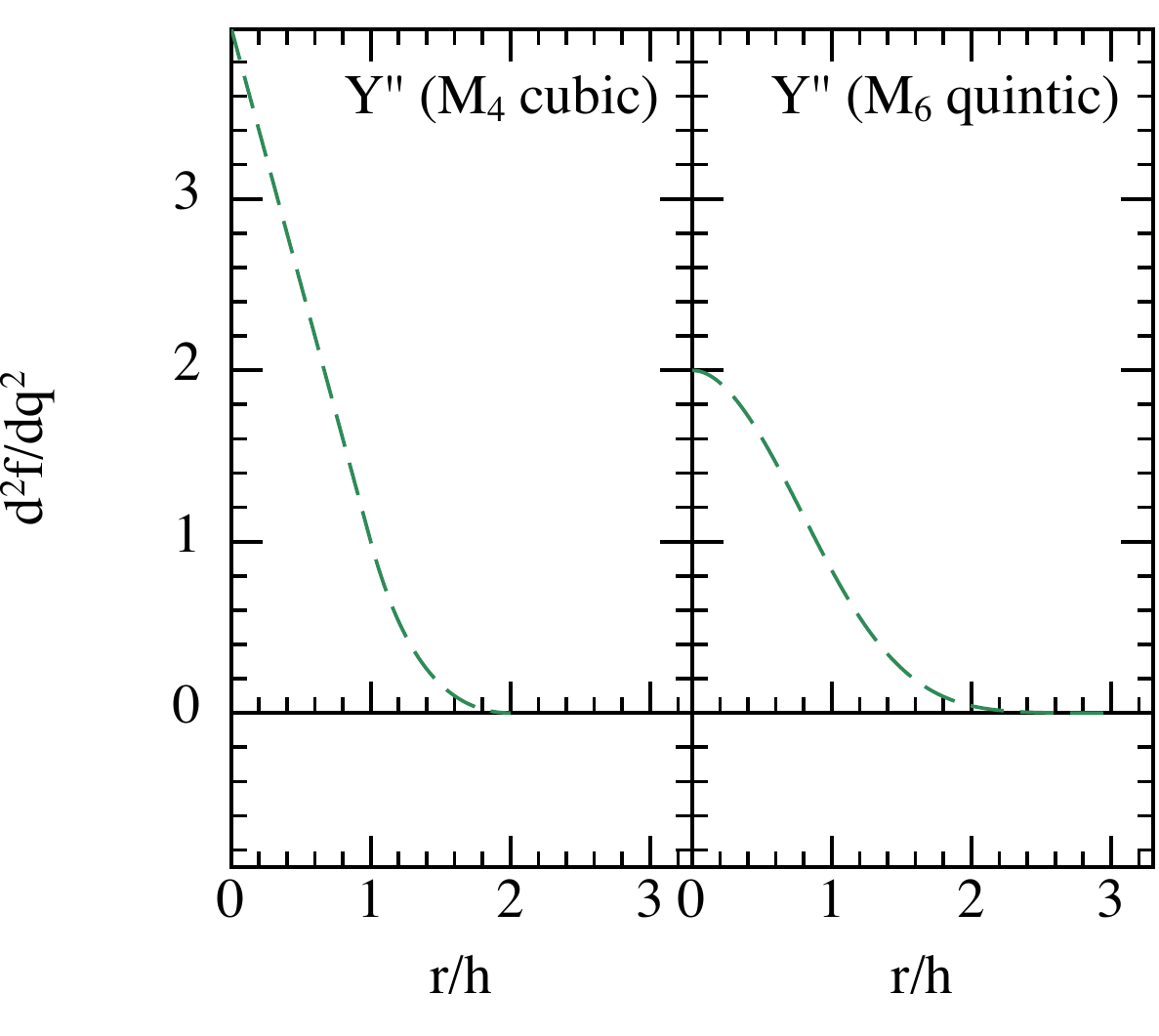}
   \includegraphics[width=0.35\columnwidth]{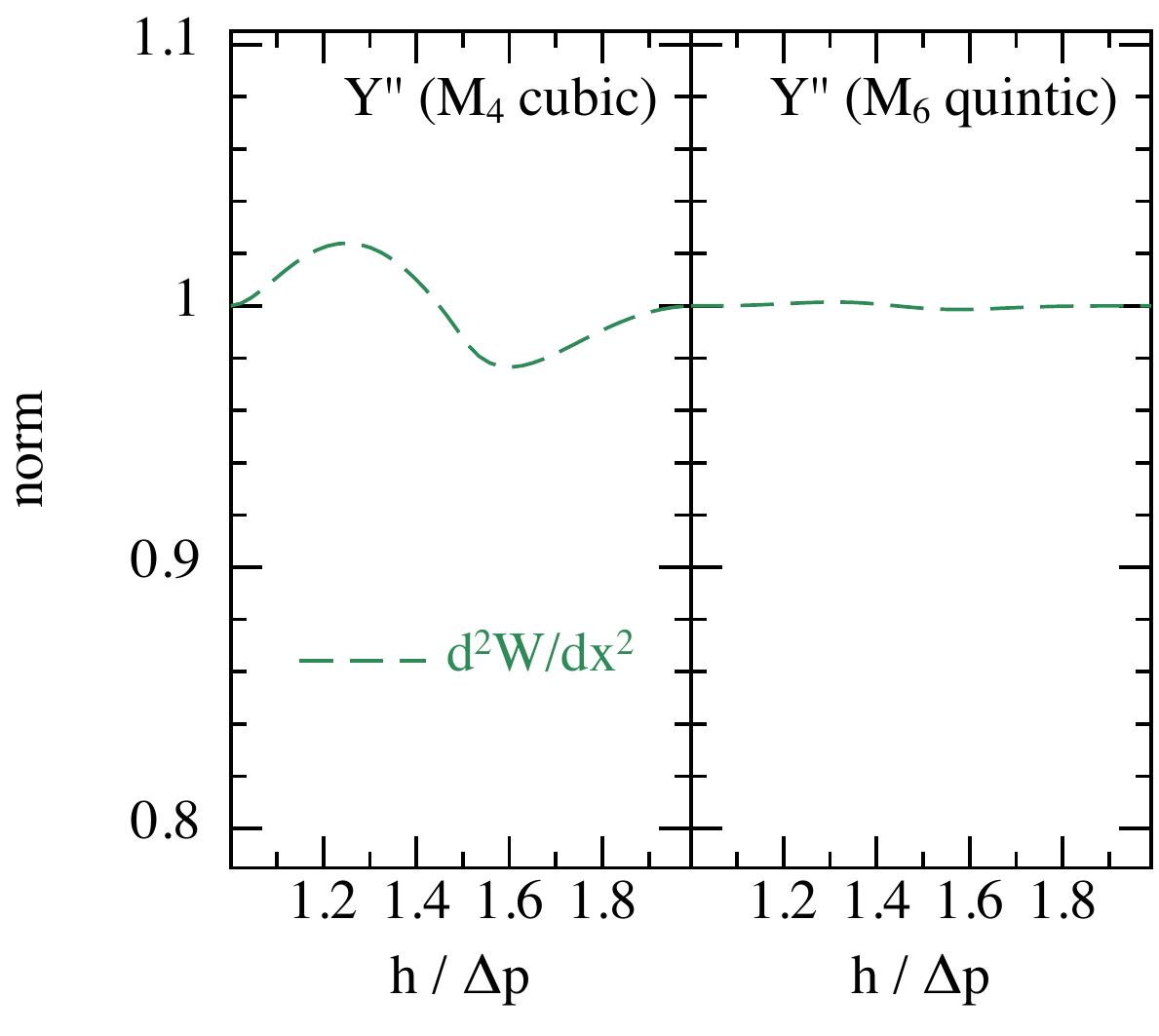}
   \caption{Second derivative kernel functions $Y''(q) \equiv -2 w'(q)/q$ constructed from the first derivatives of the $M_{4}$ cubic and $M_{6}$ quintic kernel functions (left figure, compare to the standard second derivatives shown in Fig.~\ref{fig:Bsplines}), together with the accuracy with which the second derivative normalisation conditions (\ref{eq:del2norm}) are satisfied for a fixed ratio of $h/\Delta p$ (compare with the standard second derivative functions shown in Fig.~\ref{fig:Bsplinesnorm}).}
   \label{fig:Ykernels}
\end{center}
\end{figure}

 The actual formulation commonly employed derives from early work by Monaghan and first published by \citet{brookshaw85}, where the SPH Laplacian is written in the form
\begin{equation}
\nabla^{2} A_{a} \approx  2 \sum_{b} \frac{m_{b}}{\rho_{b}} (A_{a} - A_{b}) \frac{F_{ab}}{\vert r_{ab} \vert},
\label{eq:sphlaplacian}
\end{equation}
where we use the definition $\nabla_{a} W_{ab} \equiv \hat{\bf r}_{ab} F_{ab}$ such that $F_{ab}$ is the scalar part of the kernel gradient term. Thus in effect we use the \emph{first derivative} kernel function divided by the particle spacing to give a second derivative. Whilst the interpretation of (\ref{eq:sphlaplacian}) from the integral representation is in general rather complicated \citep[see, e.g.][]{er03,jsd04,price04,monaghan05}, it can be more easily understood as if we had simply used (\ref{eq:2ndderivsph}) with a new kernel $Y_{ab}$ instead of $W_{ab}$, defined according to
\begin{equation}
\nabla^{2} Y_{ab} \equiv - \frac{2 F_{ab}}{\vert r_{ab} \vert}.
\label{eq:del2Y}
\end{equation}
The left figure in Fig.~\ref{fig:Ykernels} shows $Y''$ constructed in the above manner from the $M_{4}$ and $M_{6}$ kernel gradients, and may be compared with the standard kernel second derivatives shown in Fig.~\ref{fig:Bsplines}. The right figure shows the normalisation condition (\ref{eq:del2norm}) for these `kernels'. Comparison with Fig.~\ref{fig:Bsplinesnorm} shows that, indeed, the second derivative is much better estimated with kernel second derivative functions that are monotonically decreasing and positive, such as those constructed via (\ref{eq:del2Y}) from the bell-shaped kernels.

 Understanding the \citet{brookshaw85} Laplacian as equivalent to (\ref{eq:2ndderivsph}) with an alternative kernel also helps to interpret more complicated expressions. For example, it is quite straightforward to show that
\begin{equation}
\sum_{b} \frac{m_{b}}{\rho_{b}} (\kappa_{a} + \kappa_{b}) (A_{a} - A_{b}) \frac{F_{ab}}{\vert r_{ab} \vert} \approx \nabla\cdot(\kappa \nabla A),
\end{equation}
by writing $- 2 F_{ab}/\vert r_{ab}\vert \equiv \nabla^{2} Y_{ab}$ and interpreting each term via (\ref{eq:gradgradsumint}). \citet{cm99} proposed an alternative average of the $\kappa$ terms given by
\begin{equation}
\sum_{b} \frac{m_{b}}{\rho_{b}} \frac{4 \kappa_{a} \kappa_{b}}{(\kappa_{a} + \kappa_{b})} (A_{a} - A_{b}) \frac{F_{ab}}{\vert r_{ab} \vert} \approx \nabla\cdot(\kappa \nabla A),
\end{equation}
which was formulated to give smooth derivatives when $\kappa$ is discontinuous. The above expression forms the basis of formulations of thermal conductivity in SPH \citep{cm99}, though has been similarly used to model a wide range of dissipative terms including viscosity \citep[e.g.][]{clearyha02,huadams06}, salt diffusion \citep*{monaghanetal05} and in an astrophysical context for the treatment of radiation in the Flux-Limited Diffusion approximation \citep{wb04,wb05}.

\subsection{Vector second derivatives}
 Second derivatives of vector quantities do not quite follow the same analogy as the Laplacian, because in general $\nabla^{i} \nabla^{j} W_{ab}$ involves a mix of the first and second derivatives of the dimensionless kernel function (see appendix A in \citealt{price10}). Thus, the proof is more involved \citep[see][for details]{er03,monaghan05}, but the basic expressions for vector second derivatives are given by
\begin{eqnarray}
\langle \nabla^{2} {\bf A} \rangle & = & - 2 \sum_{b} \frac{m_{b}}{\rho_{b}} ({\bf A}_{a} - {\bf A}_{b}) \frac{F_{ab}}{\vert r_{ab} \vert}, \label{eq:del2A} \\
\langle \nabla (\nabla\cdot {\bf A}) \rangle & = & - \sum_{b} \frac{m_{b}}{\rho_{b}} \left[(\delta^{k}_{k} + 2)({\bf A}_{ab}\cdot\hat{\bf r}_{ab})\hat{\bf r
}_{ab} - {\bf A}_{ab} \right]  \frac{F_{ab}}{\vert r_{ab} \vert}, \label{eq:graddivA}
\end{eqnarray}
where $\delta^{k}_{k} \equiv d$ i.e., the number of spatial dimensions. A corollary of the above is that a second derivative computed purely along the line of sight between the particles (e.g. constructed so as to conserve angular momentum) corresponds to
\begin{equation}
-\sum_{b} \frac{m_{b}}{\rho_{b}} ({\bf A}_{ab}\cdot \hat{\bf r}_{ab}) \frac{F_{ab}}{\vert r_{ab} \vert} = \frac{1}{d + 2}\nabla (\nabla\cdot {\bf A}) +\frac{1}{2(d + 2)} \nabla^{2} {\bf A}.
\label{eq:avderiv}
\end{equation}

 An alternative approach is to calculate vector second derivatives by taking two first derivatives. Although more expensive, this has been the approach adopted in several formulations of physical viscosity, e.g. for SPH modelling of accretion discs \citep{flebbeetal94,watkins96}.

\subsection{Artificial dissipation terms in SPH and SPMHD}
\label{sec:shocks}

\subsubsection{Interpretation of SPH artificial viscosity terms}
 This brings us to the formulation and interpretation of artificial dissipation terms in SPH. The `standard' formulation of artificial viscosity is given by \citep{monaghan92}
\begin{equation}
\left(\frac{d{\bf v}}{dt}\right)_{diss} = -\sum_{b} m_{b} \frac{-\alpha \bar{c}_{s,ab} \mu_{ab} + \beta \mu_{ab}^{2}}{\bar{\rho}_{ab}} \hat{\bf r}_{ab} F_{ab}; \hspace{1cm} \mu_{ab} = \frac{h {\bf v}_{ab}\cdot{\bf r}_{ab}}{{\bf r}_{ab} + \epsilon \bar{h}_{ab}^{2}}, \label{eq:avm92}
\end{equation}
where barred quantities correspond to an average, i.e., $\bar{\rho}_{ab} = (\rho_{a} + \rho_{b})/2$, $c_{s}$ is the sound speed, $\alpha$ and $\beta$ are dimensionless parameters (typically $\alpha = 1$ and $\beta = 2$) and $\epsilon \sim 0.01$ is a small parameter to prevent divergences. This form was chosen because it is Galilean invariant, vanishes for rigid body rotation and conserves total linear and angular momentum \citep{monaghan92}. If we neglect the non-linear ($\beta$) term, set $\epsilon = 0$ and assume that $c_{s}$, $h$ and $\rho$ are approximately constant over the kernel radius, then using (\ref{eq:avderiv}) we can directly translate this expression into the continuum form
\begin{equation}
\frac{1}{d + 2} \alpha c_{s} h \nabla(\nabla\cdot{\bf v}) + \frac{1}{2(d + 2)} \alpha c_{s} h \nabla^{2} {\bf v}.
\end{equation}
Comparison with the compressible Navier-Stokes equations shows that the artificial viscosity is therefore equivalent to a physical viscosity with shear and bulk coefficients proportional to the resolution length, i.e., \citep[e.g.][]{al94,murray96,monaghan05,lp10}
\begin{equation}
\nu \approx \frac{1}{2(d+2)} \alpha c_{s} h;  \hspace{1cm} \zeta = \frac{5}{3} \nu \approx \frac{5}{6(d+2)} \alpha c_{s} h.
\end{equation}
 Since for shock-capturing the viscosity is only applied when particles are approaching (${\bf v}_{ab}\cdot{\bf r}_{ab} < 0$), in a uniform shear flow -- or accretion disc -- the viscosity coefficients will be approximately half of these values.

\begin{figure}[t]
\begin{center}
\begin{minipage}{0.45\columnwidth}
   \includegraphics[width=\columnwidth]{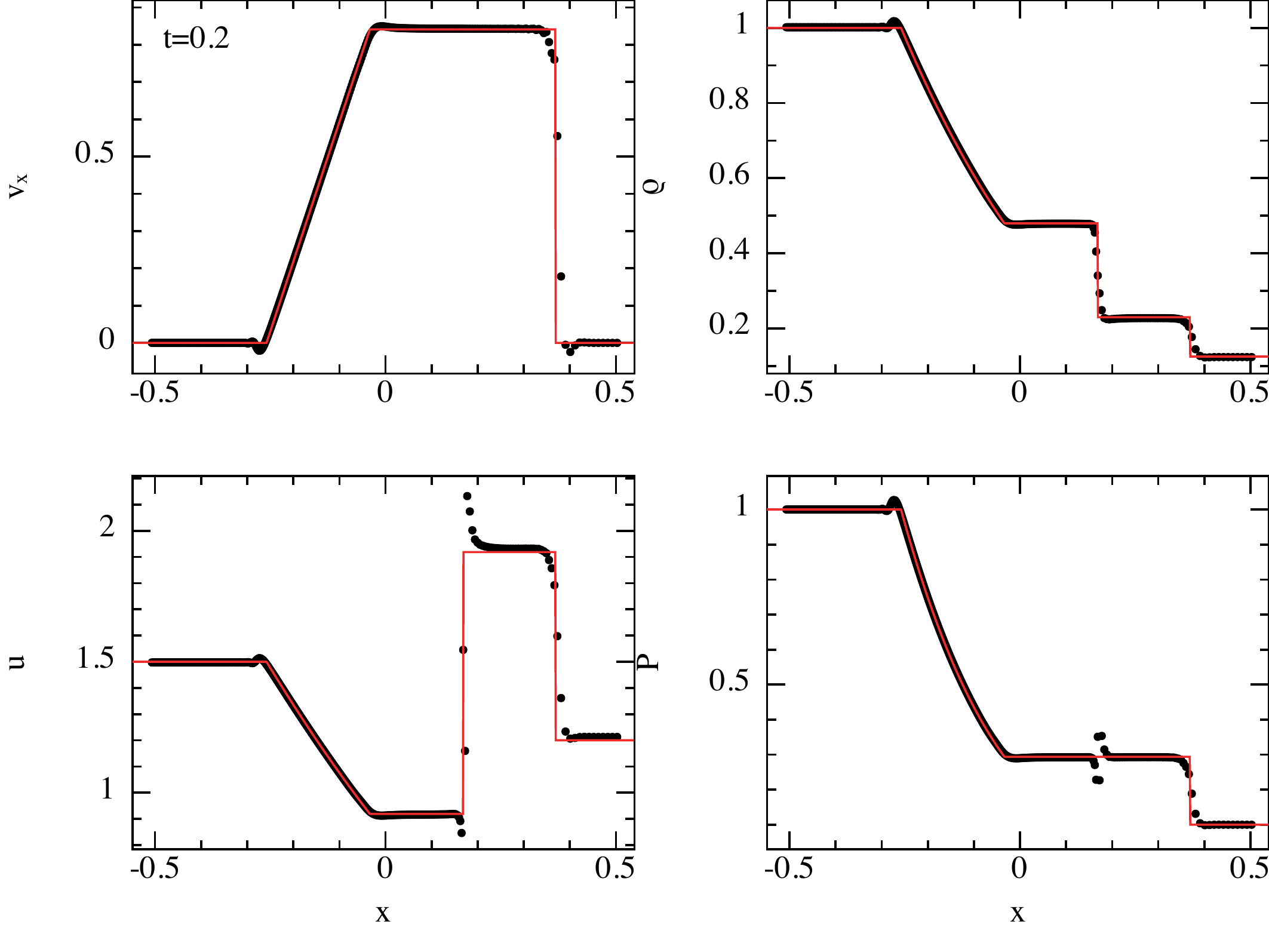}
\end{minipage}
\hspace{1cm}
\begin{minipage}{0.45\columnwidth}
   \includegraphics[height=0.66\columnwidth]{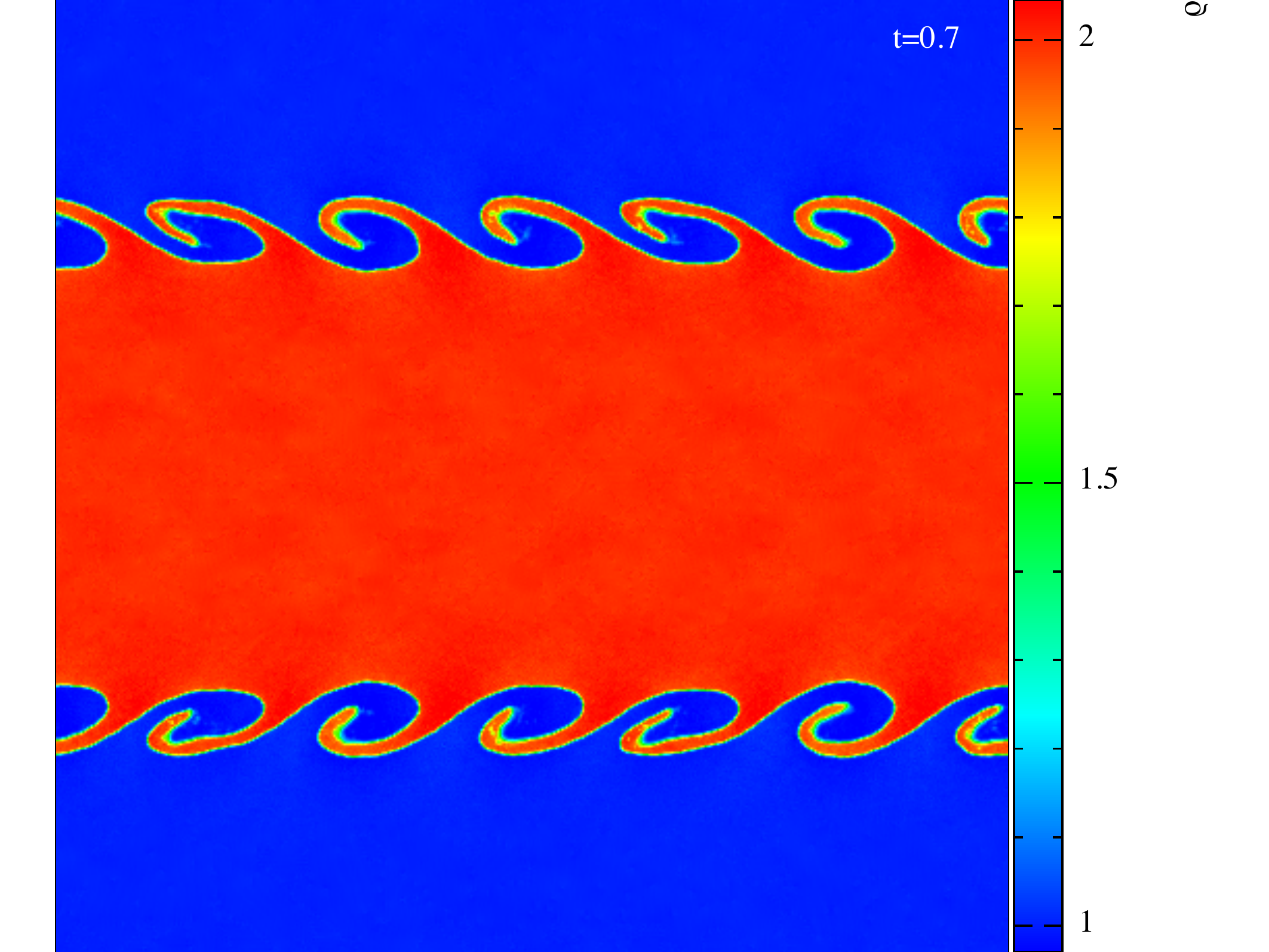}
   \vspace{0.09\columnwidth}
\end{minipage}
   \caption{Treatment of discontinuities in SPH: A 1D Sod shock tube problem (left) and 2D Kelvin-Helmholtz instability (right), applying only artificial viscosity terms and with unsmoothed initial conditions. Whilst in the 1D shock tube problem (left panel) the shock is smoothed over several particle spacings by the viscosity term, there are problems at the contact discontinuity causing a `blip' in the pressure. This leads to a suppression of mixing across contact discontinuities in 2D (right panel)}
   \label{fig:sodshockkh}
\end{center}
\end{figure}

\begin{figure}
\begin{center}
\begin{minipage}{0.45\columnwidth}
   \includegraphics[width=\columnwidth]{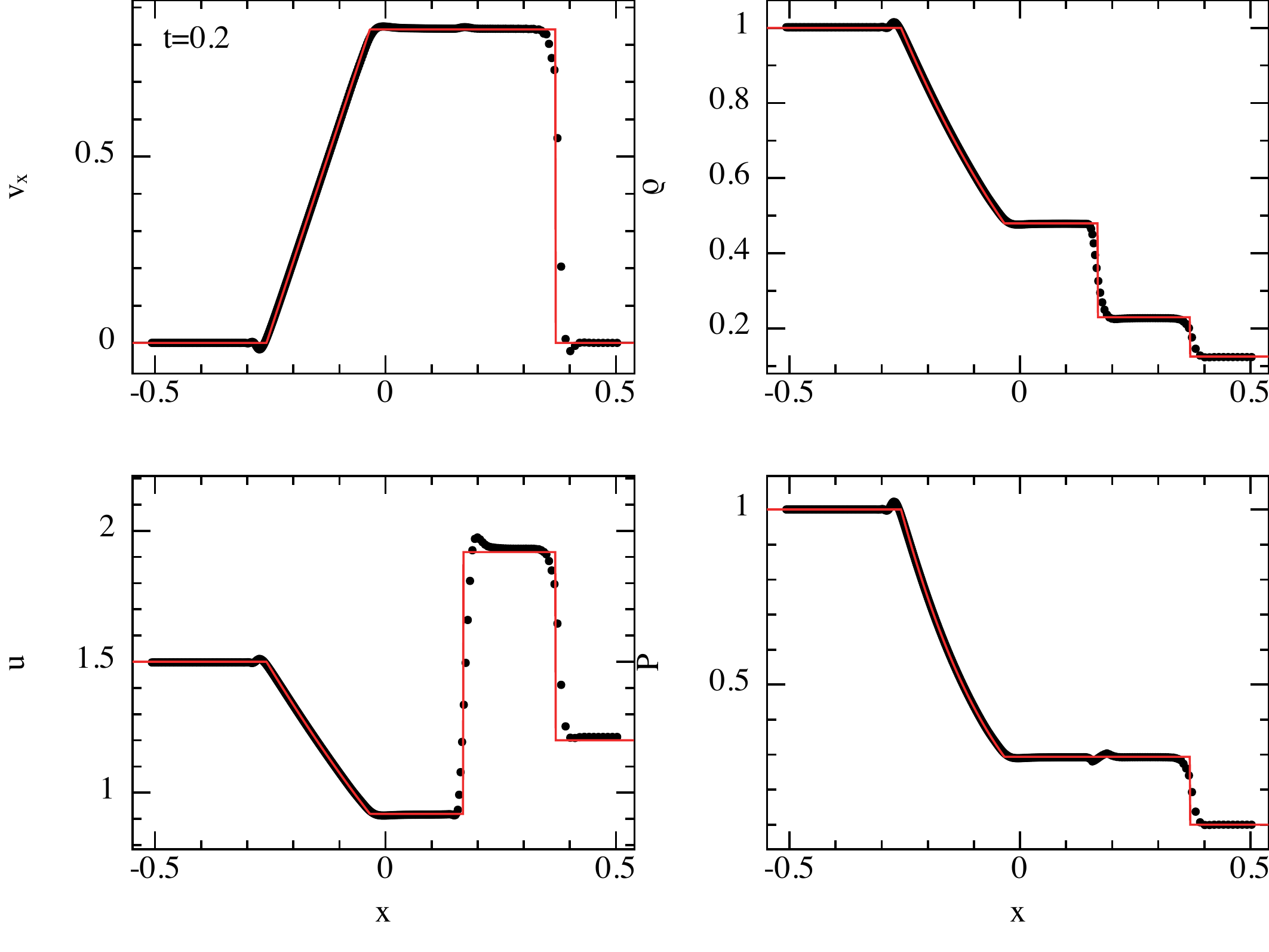}
\end{minipage}
\hspace{1cm}
\begin{minipage}{0.45\columnwidth}
   \includegraphics[height=0.66\columnwidth]{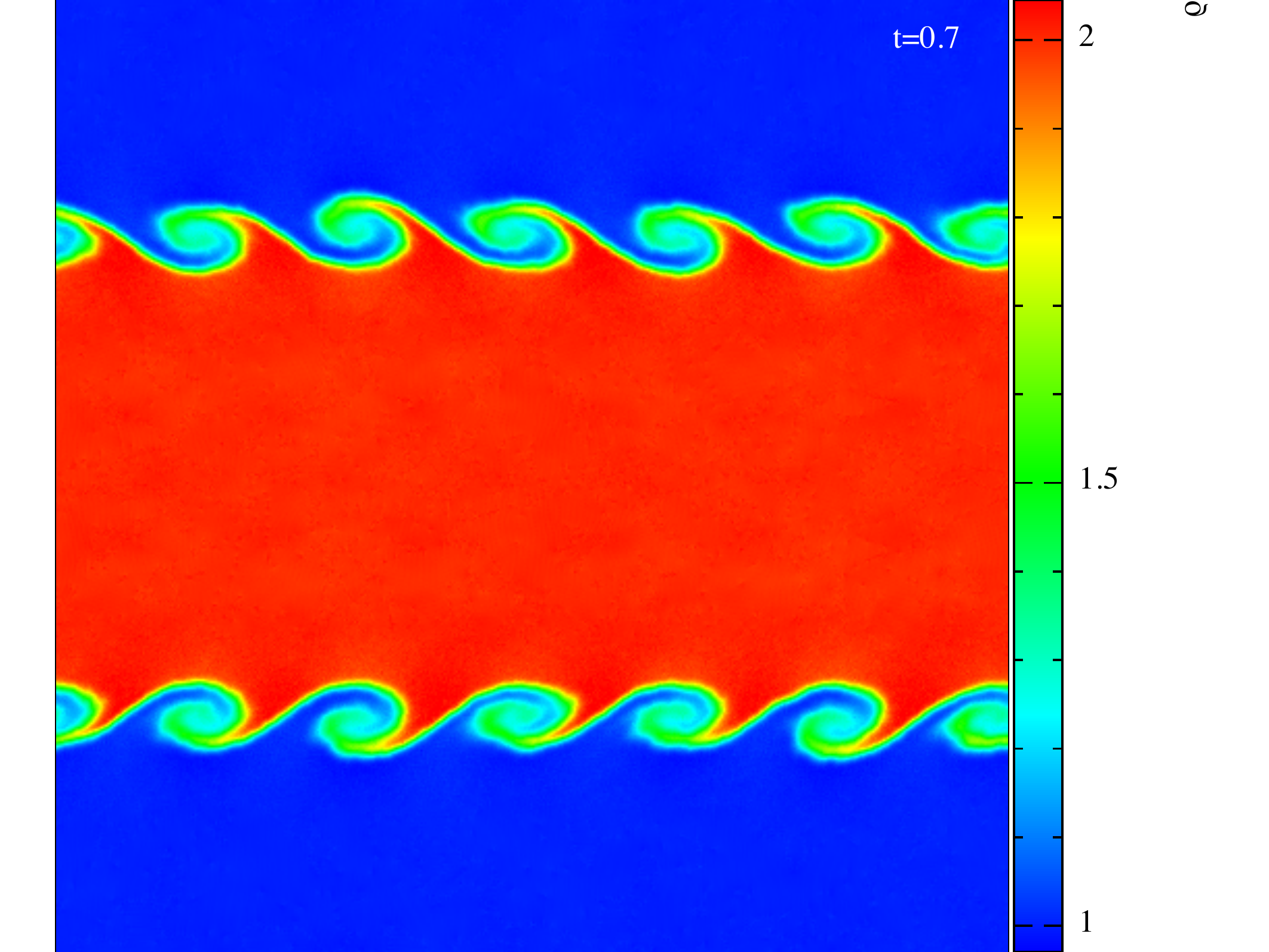}
   \vspace{0.09\columnwidth}
\end{minipage}

   \caption{As in Fig.~\ref{fig:sodshockkh} but with artificial conductivity applied as well as artificial viscosity. This smoothes the contact discontinuity, removing the pressure `blip' (left panel) and restoring the mixing in the 2D K-H problem (right panel).}
   \label{fig:conductivity}
\end{center}
\end{figure}
 
\begin{figure}
\begin{flushleft}
\hspace{0.1cm}
   \includegraphics[width=0.45\columnwidth]{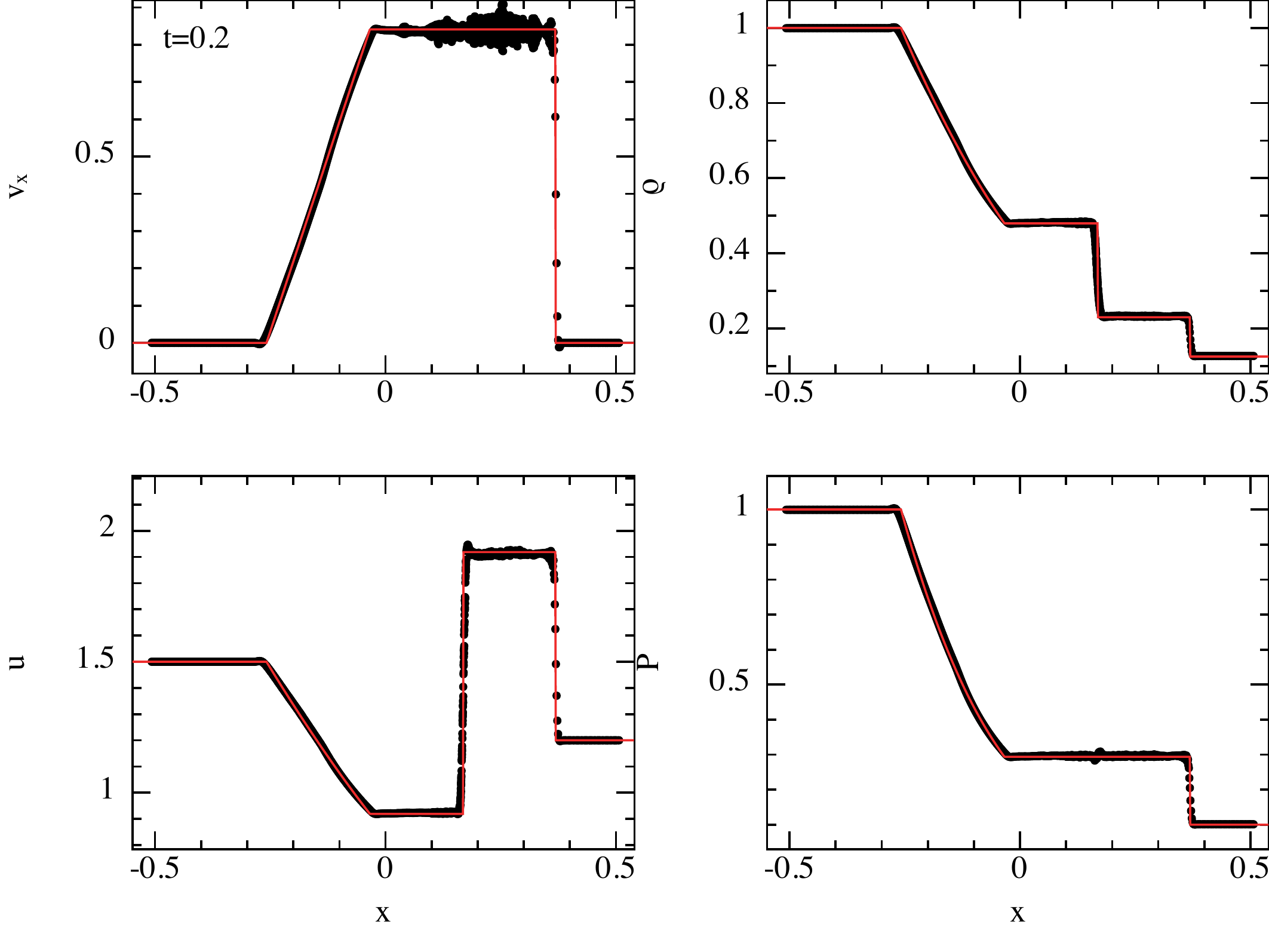}
\hspace{0.25cm}
   \includegraphics[width=0.45\columnwidth]{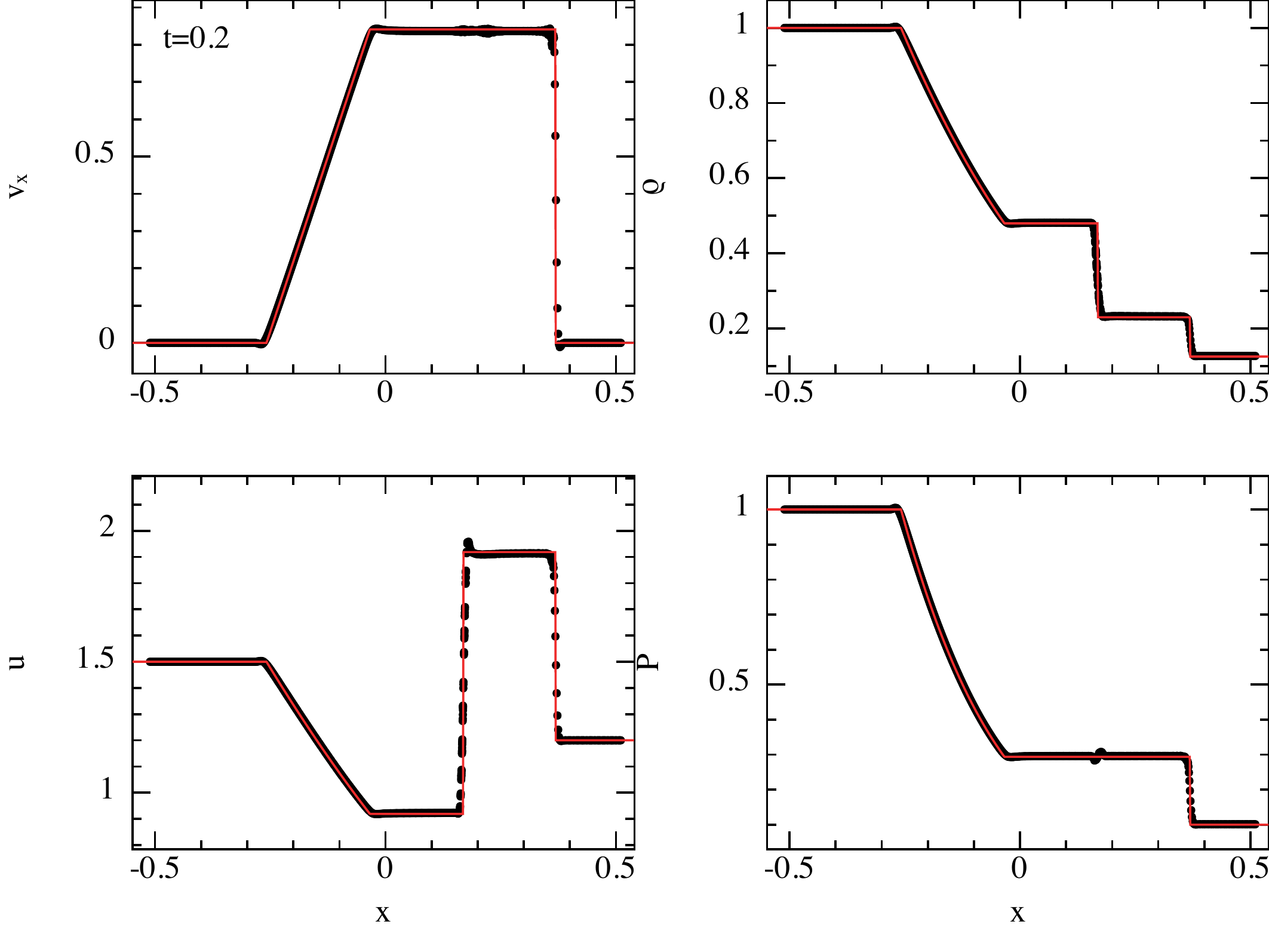}
\vspace{-0.4cm}
   \caption{The Sod shock tube in 2D. Using the cubic spline kernel (left panel), there is additional `noise' in the 2D velocity field (compared to 1D, Fig.~\ref{fig:conductivity}) due to the transverse ``remeshing'' motions of particles behind the shock front in multidimensions (see Fig.~\ref{fig:shockparts2D}). However, this can be quite effectively minimised by using a smoother kernel (right panel, using the $M_{6}$ quintic).}
   \label{fig:sodshock2D}
\end{flushleft}
\end{figure}

\subsubsection{General formulation of dissipative terms in SPH and SPMHD}
\label{sec:dissipation}
  A more general formulation of dissipative terms was proposed by \citet{monaghan97}, based on an analogy with Riemann solvers and the need to formulate dissipative terms for ultra-relativistic  shocks \citep{cm97}.  The general principle is that dissipative terms in the conservative variables involves jumps in those variables (the ``left'' and ``right'' states of the Riemann problem), multiplied by eigenvalues that can be interpreted as signal velocities.
 
  For the equations of hydrodynamics the conservative variables are the density, specific momentum and energy ($\rho$, ${\bf v}$ and $e = \frac12 v^{2} + u$ respectively), and the terms take the form\footnote{Note that in SPH no dissipation is required in the density equation provided the density is computed from a sum. This is because (\ref{eq:rhosum}) represents an integral form of the continuity equation and can be shown to differ from (\ref{eq:drhodt}) by a surface integral term that is non-zero only at boundaries or equivalently, discontinuities in the flow. See \citet{price08} for more details.} \citep{monaghan97,price08}
\begin{eqnarray}
\left(\frac{d{\bf v}_{a}}{dt}\right)_{diss} & = & \sum_b m_b \frac { \alpha v_{sig} ({\bf v}_a -
{\bf v}_b ) \cdot \hat{\bf r}_{ab}}{\bar{\rho}_{ab} } \hat{\bf r}_{ab}\overline{F_{ab}}, \label{eq:av} \\
\left(\frac{de_a}{dt}\right)_{diss} & = & \sum_b m_b \frac{(e^*_a - e^*_b)}{\bar{\rho}_{ab}} \overline{F_{ab}},
\label{eq:ave}
\end{eqnarray}
where $e_{a}^{*} =  \frac12 \alpha v_{sig}({\bf v}_{a} \cdot \hat{\bf r}_{ab})^{2} + \alpha_{u}  v_{sig}^{u} u_{a} $ refers to an energy including only components along the line of sight joining the particles, $\overline{F_{ab}} \equiv [F_{ab}(h_{a}) + F_{ab}(h_{b})]/2$ (equivalent to writing $\bar{F}_{ab} = \hat{\bf r}_{ab}\cdot \overline{\nabla_a W_{ab}}$) and $\alpha$ and $\alpha_{u}$ are the dimensionless artificial viscosity and thermal conductivity parameters, respectively. The signal speed $v_{sig}$ refers to the maximum (averaged) signal speed between a particle pair. For hydrodynamics we use
\begin{equation}
v_{sig} = 
\left\{\begin{array}{ll}
\frac12[ c_{s,a} + c_{s,b} - \beta {\bf v}_{ab}\cdot\hat{\bf r}_{ab}]; & {\bf v}_{ab}\cdot\hat{\bf r}_{ab} \le 0; \\
0; & {\bf v}_{ab}\cdot\hat{\bf r}_{ab} > 0.
\end{array}\right.
\label{eq:vsig}
\end{equation}
 Eq.~(\ref{eq:av}), with the above $v_{sig}$, provides a standard artificial viscosity term similar to (\ref{eq:avm92}) (careful expansion shows they differ only by a factor of $h/\vert r_{ab} \vert$ and the no-longer-necessary $\epsilon$ term). The dissipation term in the total energy also contains a term involving $(u_{a} - u_{b})$ which acts to smooth jumps in the thermal energy (e.g. at a contact discontinuity). The interpretation of this term is clearer from the contribution to the thermal energy evolution, given by
\begin{equation}
\left(\frac{du}{dt}\right)_{diss}= -\sum_b \frac{m_{b}}{\bar{\rho}_{ab}}\left[ \frac{1}{2}\alpha v_{sig}
({\bf v}_{ab}\cdot\hat{\bf r}_{ab})^2 +  \alpha_u v_{sig}^{u} (u_a - u_b) \right] \bar{F}_{ab}, \label{eq:dudtdiss}
\end{equation}
where the second term, from (\ref{eq:sphlaplacian}), can be directly interpreted as a $\nabla^{2} u$ term. Note that the signal velocity $v_{sig}^{u}$ used in the artificial conductivity does not have to be the same as that used for the viscosity. In particular, \citet{price08} proposed using $v_{sig}^{u} = \sqrt{\vert P_{a} - P_{b} \vert /\bar{\rho}_{ab}}$ to equalise the pressure across contact discontinuities, whilst \citet{wadsleyetal08} proposed a conductivity term equivalent to using $v_{sig}^{u} = \vert {\bf v}_{ab}\cdot \hat{\bf r}_{ab}\vert$ (we adopt the former for the tests shown in this paper).
 
 \subsubsection{Switches for viscosity terms}
  One of the key issues in practice is to ensure that sufficient dissipation is applied to discontinuities, but that such dissipation is effectively turned off in smooth parts of the flow by designing appropriate switches. \citet{mm97} suggested allowing the parameter $\alpha$ to be individual to each particle, with an evolution equation of the form
\begin{equation}
\frac{d\alpha}{dt} = \mathcal{S} + \frac{\alpha - \alpha_{min}}{\tau}; \hspace{1cm} \tau = \frac{h}{\sigma c_{s}},
\label{eq:dalphadt}
\end{equation}
where $\mathcal{S}$ is a source term that grows large at the discontinuity [e.g. $\mathcal{S} = \max(0,-\nabla\cdot{\bf v})$ for shocks], $\tau$ is the decay time, set such that $\alpha$ decays to $\alpha_{min}$ over several smoothing lengths (typically $\sigma = 0.1$ and $\alpha_{min} = 0.1$). A similar switch can be employed for the thermal conductivity parameter $\alpha_{u}$, with \citet{pm05} adopting a source term given by $\mathcal{S}_{u} = 0.1h \nabla^{2} u$. More sophisticated switches for shock detection are also possible, with a promising recent alternative suggested by \citet{cd10}. Directly employing the Riemann solution is another possibility \citep[e.g.][]{inutsuka02,cw03}.

\subsubsection{Examples 4 and 5: One and two dimensional shock tubes, and Kelvin-Helmholtz instabilities}
\label{sec:sodshockkh}
 Two specific examples of the dissipative terms in practice are shown in Figs.~\ref{fig:sodshockkh} (applying only viscosity) and \ref{fig:conductivity} (applying artificial viscosity \emph{and} conductivity), showing the results of a one dimensional Sod shock tube problem (left subfigure) and a two dimensional Kelvin-Helmholtz (K-H) instability problem (right subfigure). The 1D shock tube is setup using a total of 450 particles in $-0.5 < x < 0.5$ with conditions to the left of the origin given by $[\rho_{L},P_{L},v_{L}] = [1,1,0]$ and to the right by $[\rho_{R},P_{R},v_{R}] = [0.125,0.1,0]$ with $\gamma = 5/3$. Importantly, purely discontinuous initial conditions are employed so that the contact discontinuity is not already smoothed. The K-H instability problem is setup with a 2:1 density ratio and equal mass particles, identical to the setup described in \citet{price08} and using $512 \times 512$ particles in the low density fluid. Whilst shocks are smoothed by the artificial viscosity term (Figs.~\ref{fig:sodshockkh} and \ref{fig:conductivity}), with only viscosity the jump in thermal energy at the contact discontinuity is not treated, resulting in a `blip' in the pressure profile (Fig.~\ref{fig:sodshockkh}). This manifests in the 2D K-H problem as an `artificial surface tension' effect, caused by the very same kind of pressure blip across the boundary (contact discontinuity) between the dense and the light fluids, suppressing mixing of the two. With conductivity applied (Fig.~\ref{fig:conductivity}) the pressure is smooth across the contact discontinuity in both problems, which for the K-H problem means that the two fluids mix correctly. The lack of treatment of contact discontinuities in standard SPH codes (i.e., with no artificial conductivity term) explains the discrepancy between grid and SPH results somewhat infamously highlighted by \citet{agertzetal}.
 
  Fig.~\ref{fig:sodshock2D} shows the same shock tube in 2D (as already discussed briefly in Fig.~\ref{fig:shockparts2D}). The main difference to 1D is that the ``noise'' due to the particle resettling behind the shock front is visible (left subfigure). It is often asserted that SPH performs poorly on 2D shocks for this reason, however the noise can be very effectively minimised (at some additional cost) by employing the $M_{6}$ quintic kernel instead of the cubic spline (right subfigure), giving results comparable to the 1D version and illustrating in practice how the higher $M_{n}$ kernels can be used to obtain convergence in SPH.


\section{Smoothed Particle Magnetohydrodynamics from a Lagrangian}
\label{sec:spmhdfromL}
 We can follow the same general approach to constructing an SPMHD algorithm as for hydrodynamics (Sec.~\ref{sec:densitytoequationsofmotion}): Write down the Lagrangian, use appropriate physical constraints and use this to consistently derive the resultant equations of motion.

\subsection{MHD Lagrangian}
  For MHD, the Lagrangian is given by \citep{pm04b}
\begin{equation}
L_{\rm MHD} = \sum_{b} m_{b} \left[\frac{1}{2} v_b^2 - u_b(\rho_b,s_b) -\frac{1}{2\mu_0}
\frac{B_b^2}{\rho_b}\right],
\label{eq:Lmhd}
\end{equation}
corresponding simply to the subtraction of a magnetic energy term from the hydrodynamic verison (Eq.~\ref{eq:L}). In the continuum limit this corresponds to the standard MHD Lagrangian used by many authors \citep[e.g.][]{newcomb62,field86}
\begin{equation}
L_{\rm MHD} = \int \left(\frac{1}{2}\rho v^2 -\rho u-\frac{1}{2\mu_0}B^2\right) \mathrm{dV}.
\end{equation}

 The difference to the hydrodynamic case is that, unlike the thermal energy term, neither the magnetic field ${\bf B}$ nor the change in the magnetic field can be written directly as a function of the particle coordinates, so we cannot straightforwardly employ the Euler-Lagrange equations (\ref{eq:el}). Instead, we can use the more general form of the variational principle given by $\delta S = \int \delta L dt = 0$ (Sec.~\ref{sec:S}), where from (\ref{eq:Lmhd}) we have
\begin{equation}
\delta L = m_a {\bf v}_a\cdot\delta{\bf v}_a - \sum_{b} m_{b}\left[\left.\pder{u_b}{\rho_b}\right\vert_s\delta\rho_b +
\frac{1}{2\mu_0} \left(\frac{B_b}{\rho_b}\right)^2\delta\rho_b + \frac{1}{\mu_0} {\bf B}_b\cdot\delta\left(\frac{{\bf B}_b}{\rho_b}\right)\right],
\label{eq:deltaL}
\end{equation}
and the perturbation is with respect to a small change in the particle coordinates $\delta {\bf r}$. So we can derive the equations of motion provided that we are able to express the change in the magnetic field $\delta{\bf B}$ as a function of the \emph{change} in particle coordinates -- equivalent to being able to write down an expression for the Lagrangian time derivative $d{\bf B}/dt$ [or equivalently, $d({\bf B}/\rho)/dt$] since $d/dt \equiv \delta/\delta t$. In other words, in order to derive the SPMHD equations of motion it is necessary to specify not only the density estimate but also the manner in which the magnetic field is evolved.

\subsection{SPMHD formulation of the induction equation}
 Given the induction equation for (ideal) MHD, written in the Lagrangian form
\begin{equation}
\frac{d}{dt} \left(\frac{{\bf B}}{\rho} \right) = \left(\frac{{\bf B}}{\rho} \cdot\nabla \right) {\bf v},
\label{eq:mhddBrhodt}
\end{equation}
and using our general formulations of SPH derivatives given in Sec.~\ref{sec:vectorderivs}, it is straightforward to write down an SPH version of the form
\begin{equation}
\frac{d}{dt} \left(\frac{{\bf B}_{a}}{\rho_{a}} \right) = -\sum_{b} m_{b} ({\bf v}_{a} - {\bf v}_{b} ) \frac{{\bf B}_{a}}{\Omega_{a} \rho_{a}^{2}} \cdot \nabla W_{ab} (h_{a}), \label{eq:dBrhodt}
\end{equation}
which is equivalent to using the antisymmetric derivative (\ref{eq:gradA}a) with $\phi = \rho$.

\subsection{Equations of motion}
\label{sec:mhdequationsofmotion}
The perturbations required in (\ref{eq:deltaL}), from (\ref{eq:drhodt}) and (\ref{eq:dBrhodt}) are therefore given by
\begin{eqnarray}
\delta\rho_b & = & \frac{1}{\Omega_{b}}\sum_{c} m_{c}\left(\delta {\bf r}_b - \delta {\bf r}_c\right)\cdot \nabla_b W_{bc}(h_{b}) \label{eq:deltarho} \\
\delta\left(\frac{{\bf B}_b}{\rho_b}\right) & = & -\sum_{c} m_{c} (\delta {\bf r}_b - \delta {\bf r}_c) \frac{{\bf B}_b}{\Omega_{b}\rho_b^2} \cdot\nabla_b W_{bc}(h_{b})
\label{eq:deltaBrho}
\end{eqnarray}
giving, from (\ref{eq:deltaL}) and using (\ref{eq:dudrho}) 
\begin{eqnarray}
\delta S = \int \delta L {\rm dt} & = & \int \left\{ m_{a}{\bf v}_{a}\cdot\delta {\bf v}_{a}  -\sum_b m_b\left[\left( \frac{P_b + \frac12 B_{b}^{2}/\mu_{0}}{\Omega_{b}\rho_b^2}\right) \sum_c m_c \nabla_b W_{bc} (h_{b}) (\delta_{ba} -
         \delta_{ca}) \right]  \cdot {\delta {\bf r}_a}  \right. \nonumber \\
         & & \phantom{\int\{}+ \left. \sum_b m_b\left[ \frac{{\bf B}_{b}}{\mu_0} \sum_c m_c \frac{{\bf B}_b}{\Omega_{b} \rho_b^{2}}\cdot \nabla_b W_{bc}(h_{b})(\delta_{ba} - \delta_{ca})\right]\cdot  {\delta {\bf r}_a} \right\} {\rm dt} = 0,
\end{eqnarray}
where $\delta_{ba} \equiv \delta {\bf r}_{b} / \delta {\bf r}_{a}$. Integrating the velocity term by parts, simplifying the double summations using the Kronecker deltas and the antisymmetry of the kernel gradient, and assuming the perturbations $\delta {\bf r}_{a}$ are arbitrary, we find that the SPMHD equations of motion are given by
\begin{eqnarray}
\frac{d{\bf v}_a}{dt} & = & -\sum_{b} m_{b} \left[ \frac{P_a + \frac{1}{2\mu_{0}} B_{a}^{2}}{\Omega_{a}\rho_a^2} \nabla_{a} W_{ab}(h_{a}) + \frac{P_b + \frac{1}{2\mu_{0}} B_{b}^{2}}{\Omega_{b}\rho_b^2} \nabla_{a} W_{ab}(h_{b})\right] \nonumber \\
& & + \frac{1}{\mu_{0}}\sum_b m_b\left[ \frac{{\bf B}_{a}({\bf B}_{a}\cdot \nabla_{a} W_{ab}(h_{a}))}{\Omega_{a}\rho_{a}^{2}} + \frac{{\bf B}_{b}({\bf B}_{b}\cdot \nabla_{a} W_{ab}(h_{b}))}{\Omega_{b}\rho_{b}^{2}} \right]. \label{eq:spmhdmom}
\end{eqnarray}
 In tensor notation these can be written more compactly in the form
\begin{equation}
\frac{dv^{i}_a}{dt} = \sum_{b} m_{b} \left[\frac{S^{ij}_{a}}{\Omega_{a}\rho_{a}^{2}} \nabla^{j}_{a} W_{ab} (h_{a}) + \frac{S^{ij}_{b}}{\Omega_{b}\rho_{b}^{2}} \nabla^{j}_{a} W_{ab} (h_{b}) \right], \label{eq:spmhdmomtensor}
\end{equation}
where $S^{ij}$ is the MHD stress tensor, defined according to
\begin{equation}
S^{ij} \equiv -\left( P + \frac{1}{2\mu_{0}} B^{2} \right) \delta^{ij} + \frac{1}{\mu_{0}}B^{i}B^{j}. \label{eq:sij}
\end{equation}
 
As for the hydrodynamic case (Sec.~\ref{sec:densitytoequationsofmotion}) it is readily seen that the equations of motion conserve linear momentum exactly, due to the pairwise symmetry in the force. However, the MHD equations, unlike their hydrodynamic counterparts, do not exactly conserve angular momentum, since the perturbation to the magnetic field, (\ref{eq:deltaBrho}) -- and hence the anisotropic force term derived from it -- is not invariant to rotations. It is interesting to note that the anisotropic magnetic force term in (\ref{eq:spmhdmom}) derives entirely from the numerical representation of the induction equation (\ref{eq:dBrhodt}), whilst the isotropic term derives purely from the magnetic energy term in the Lagrangian and the density perturbation (\ref{eq:deltarho}).

\subsection{Energy equation}
 The evolution equations for thermal energy and entropy -- in the absence of dissipation -- are identical to their hydrodynamic counterparts. The total energy evolution can be deduced from the Hamiltonian as in Sec.~\ref{sec:totalE}. The corresponding expression for MHD is given by
\begin{equation}
\frac{dE}{dt} = \sum_{a} m_{a} \left[ {\bf v}_a \cdot \frac{d{\bf v}_a}{dt} + \frac{du_a}{dt} + 
\frac{1}{2}\frac{B_a^2}{\rho_a^2}\frac{d\rho_a}{dt} + {\bf B}_a\cdot\frac{d}{dt}\left(\frac{{\bf B}_a}{\rho_a}\right)\right].
\label{eq:mhdtotalE}
\end{equation}
 Using (\ref{eq:spmhdmomtensor}), (\ref{eq:sphutherm}), (\ref{eq:drhodt}) and (\ref{eq:dBrhodt}), it can be shown that the total energy evolves according to
 \begin{equation}
\frac{dE}{dt} = \sum_{a} m_{a} \sum_{b}m_{b} \left[ \left(\frac{S^{ij}}{\Omega\rho^2}\right)_a v^i_b \nabla^j_a W_{ab}(h_a) + \left(\frac{S^{ij}}{\Omega\rho^2}\right)_b v^i_a \nabla^j_a W_{ab}(h_b) \right] = 0,
\end{equation}
and is thus also conserved exactly. This further implies an evolution equation for the specific energy $e \equiv \frac12 v^{2} + u + \frac{1}{2\mu_{0}} B^{2}/\rho$ of the form
\begin{equation}
\frac{de_a}{dt} = \sum_{b} m_{b} \left[ \left(\frac{S^{ij}}{\Omega\rho^2}\right)_a v^i_b \nabla^j_a W_{ab}(h_a) + \left(\frac{S^{ij}}{\Omega\rho^2}\right)_b v^i_a \nabla^j_a W_{ab}(h_b) \right], \label{eq:mhddedt}
\end{equation}

\subsection{Interpretation of the Hamiltonian SPMHD equations}
 Having used the SPH forms of the continuity and induction equations in the form
\begin{eqnarray}
\frac{d\rho}{dt} & = & - \rho \frac{\partial v^{i}}{\partial x^{i}}, \\
\frac{d}{dt} \left( \frac{B^{i}}{\rho} \right) & = & - \frac{B^{j}}{\rho} \frac{\partial v^{i}}{\partial x^{j}},
\end{eqnarray}
it is straightforward to show (using \ref{eq:basicgradvec}) that our expressions (\ref{eq:spmhdmomtensor}) and (\ref{eq:mhddedt}) derived above are SPH representations of the MHD acceleration and energy equations in the form
\begin{eqnarray}
\frac{dv^{i}}{dt} & = & \frac{1}{\rho}\frac{\partial S^{ij}}{\partial x^{j}}, \label{eq:accelmhd} \\
\frac{de}{dt} & = & \frac{\partial (v^{i} S^{ij})}{\partial x^{j}},
\end{eqnarray}
where $S^{ij}$ is the MHD stress tensor given by (\ref{eq:sij}). That we have explicitly \emph{used} the induction equation to derive the equations of motion and energy is useful to our later discussion regarding what is a consistent formulation of monopole terms in the MHD equations (Sec.~\ref{sec:sourceterms}).

\subsubsection{MHD Example 1: Advection of a current loop}

\begin{figure}[t]
\begin{center}
   \includegraphics[width=\columnwidth]{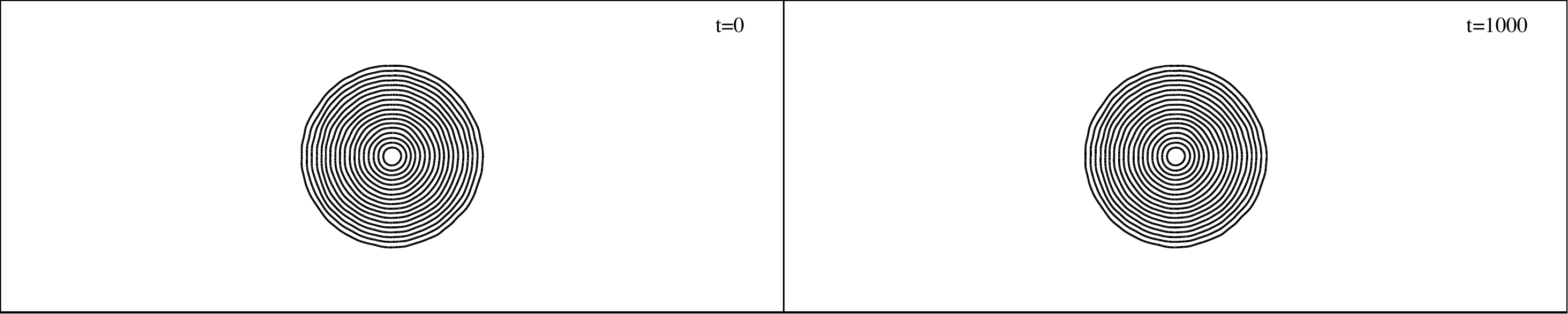}
   \caption{Advection of a current loop across a 2D domain. Magnetic field lines are shown at $t=0$ (left panel) and after 1000 crossings of the computational domain (right panel). In the absence of the explicit addition of resistivity terms, there is \emph{no} change in the magnetic field and the advection is computed exactly in SPMHD, irrespective of the direction of propagation, numerical resolution or advection velocity.}
   \label{fig:jadvect}
\end{center}
\end{figure}

Our first MHD example demonstrates that some problems that prove very difficult for Eulerian schemes are almost trivial in a Lagrangian scheme such as SPMHD. The problem involves the advection of a loop of current across the computational domain, was introduced by \citet{gs05} to test their \textsc{Athena} MHD code and presents a challenging problem for grid-based MHD schemes. The setup used here is identical to that in \citet{rp07} and we refer the reader to that paper (or the \textsc{ndspmhd} setup file) for full details. The current loop itself is given by the vector potential $A_{z} = A_{0} (R-\sqrt{x^{2} + y^{2}})$, with $A_{0}$ set to give a weak field (plasma $\beta$ of $2\times 10^{6}$) (here we compute and evolve the magnetic field, ${\bf B}$). All particles in the domain ($-1 < x < 1$, $-0.5 < y < 0.5$) are given a constant initial velocity along the box diagonal, with the magnitude set such that $t=1$ represents one crossing of the domain. The results are shown in Fig.~\ref{fig:jadvect}, at $t=0$ (left panel) and after $1000$ (this is not a misprint!) crossings of the computational domain (right panel). In the absence of the explicit addition of resistivity terms, there is \emph{no} change in the current or the magnetic energy and the advection is computed exactly.

\section{The tensile instability in MHD}
\label{sec:mhdtensile}
 The rather large caveat to using the equations of motion derived in Sec.~\ref{sec:mhdequationsofmotion} is that in MHD the total stress can become negative, meaning that the force between particles can become attractive rather than repulsive and the equations will be unstable to the tensile instability discussed in Sec.~\ref{sec:tensile}. The regime of instability is evident if we consider the force in just one spatial dimension (and at constant h), given by
\begin{equation}
\frac{dv^{x}_a}{dt} = -\sum_{b} m_{b} \left[ \frac{P_a - \frac{1}{2\mu_{0}} B_{x}^{2}}{\rho_a^2} + \frac{P_b - \frac{1}{2\mu_{0}} B_{x}^{2}}{\rho_b^2} \right]  \frac{dW_{ab}}{dx},
\label{eq:mom1D}
\end{equation}
so the force will become attractive (along the field lines) when $\frac12 B^{2}/\mu_{0} > P$, i.e., when magnetic pressure exceeds gas pressure. A more detailed stability analysis \citep[e.g.][]{pm85,morrisphd,morris96,bot04} confirms that this is also the criterion for instability in more than one spatial dimension. Thus, whilst for particular applications that remain in the regime where magnetic pressure is smaller than gas pressure it is possible to use the conservative formulation \citep*[e.g.][]{dbl99}, in most cases stabilising the tensile instability is the first and most basic requirement for a stable SPMHD algorithm.

 The physical reason for the MHD instability is that the momentum-conserving force corresponds to
\begin{equation}
\frac{d{\bf v}}{dt} = \frac{\nabla\cdot{\bf S}}{\rho} \equiv  -\frac{\nabla P}{\rho} + \frac{(\nabla\times {\bf B}) \times {\bf B}}{\mu_{0}\rho} + \frac{{\bf B} \nabla\cdot {\bf B}}{\mu_{0}\rho},
\label{eq:mhdconservative}
\end{equation}
which differs from equations of motion written in terms of the Lorentz force (where ${\bf J} = \nabla\times{\bf B}/\mu_{0}$),
\begin{equation}
\frac{d{\bf v}}{dt} = -\frac{\nabla P}{\rho} + \frac{{\bf J} \times {\bf B}}{\rho},
\label{eq:JcrossB}
\end{equation}
by the  ``monopole term'' in (\ref{eq:mhdconservative}). This term gives a force directed along the magnetic field and proportional to the (numerically non-zero) divergence of the magnetic field, and is the source of the instability in the conservative formulation if it cannot be counteracted by pressure. As a result, Eq.~\ref{eq:JcrossB} was used as the basis of a number of early SPMHD formulations in which the force was simply computed using standard curl operators such as (\ref{eq:basiccurl}) or (\ref{eq:antisymdivcurl}) \citep[e.g.][]{meglicki95,bp96,cg01,hw04a}. However, the poor conservation properties of such formulations means that MHD shocks are not well captured.

 Dealing with such `source terms' is not an issue unique to SPMHD and requires careful consideration in all numerical MHD formulations (see Sec.~\ref{sec:sourceterms}). Of course, $\nabla\cdot{\bf B} = 0$ should be zero physically due to the non-existence of magnetic monopoles in the Universe, so both (\ref{eq:JcrossB}) and (\ref{eq:mhdconservative}) are valid in the continuum limit. The problem is that the divergence term is not exactly zero numerically -- so one is forced to make a choice. As noted by \citet{toth00}, it is also not a simple matter of enforcing the $\nabla\cdot{\bf B} = 0$ constraint in some way, since what is necessary  to achieve a force that is both conservative \emph{and} exactly perpendicular to ${\bf B}$ is to constrain $\nabla\cdot{\bf B}$ to be zero \emph{in the discretisation that the term appears in the force equation}\footnote{This was initially thought impossible by \citealt{toth00}, though \citealt{toth02} later showed such a discretisation could be achieved for grid codes. However, there does not appear to be an equivalent formulation in SPMHD, since the tensile instability occurs even in one dimension where the divergence constraint can be trivially enforced using $B_{x} = const$, but $\partial B_{x}/\partial x \neq 0$ in the force equation (c.f. Eq.~\ref{eq:mom1D}).}. In SPMHD this is equivalent to requiring both exact derivatives and exact conservation which, as discussed in Sec.~\ref{sec:relpressure}, does not appear to be possible.

\subsection{Fix 1: subtract a constant from the stress}
 The original paper by \citet{pm85} proposed a simple fix involving a prior sweep over the particles to find the maximum (negative) stress, which would then simply be subtracted (as a constant) from the stress in the equations of motion, giving
 \begin{equation}
\frac{dv^{i}_a}{dt} = \sum_{b} m_{b} \left[\frac{S^{ij}_{a} - S^{ij}_{max}}{\Omega_{a}\rho_{a}^{2}} \nabla^{j}_{a} W_{ab} (h_{a}) + \frac{S^{ij}_{b}- S^{ij}_{max}}{\Omega_{b}\rho_{b}^{2}} \nabla^{j}_{a} W_{ab} (h_{b}) \right],
\end{equation}
which conserves momentum but not total energy. The caveats are that there is a computational cost involved to compute $S^{ij}_{max}$ and if this term is large it can lead to unphysical effects in the simulation. On the other hand, this is a simple technique that removes the instability and has relatively few side effects provided the correction is small. It is particularly useful if, for example, the simulation is dominated by large (constant) external stresses, whereby explicitly subtracting the external component of the stress (e.g. due to an externally imposed magnetic field) can serve to stabilise the formulation.

\begin{figure}[t]
\begin{center}
   \includegraphics[width=0.45\columnwidth]{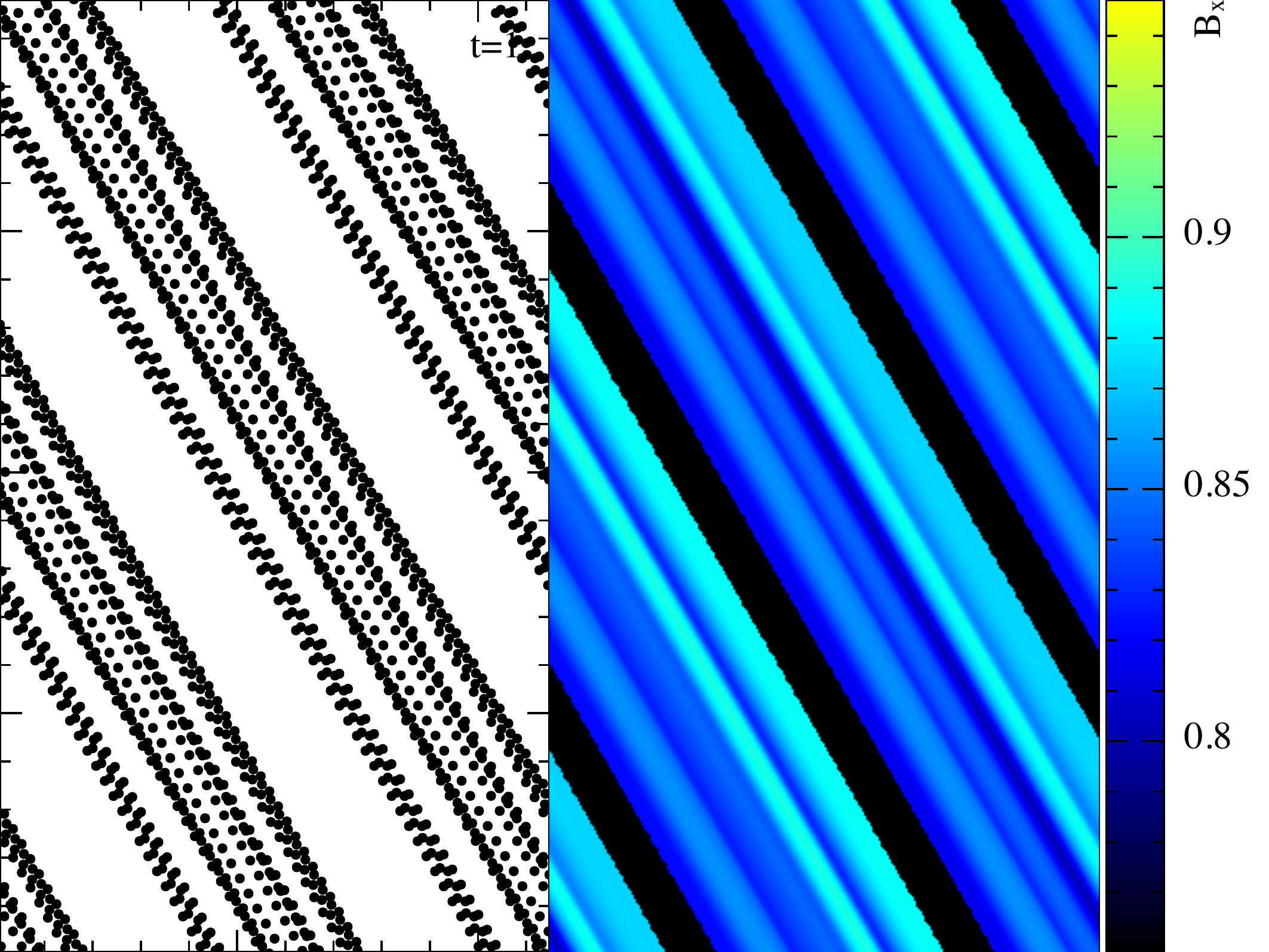}
   \hspace{0.05\columnwidth}
   \includegraphics[width=0.45\columnwidth]{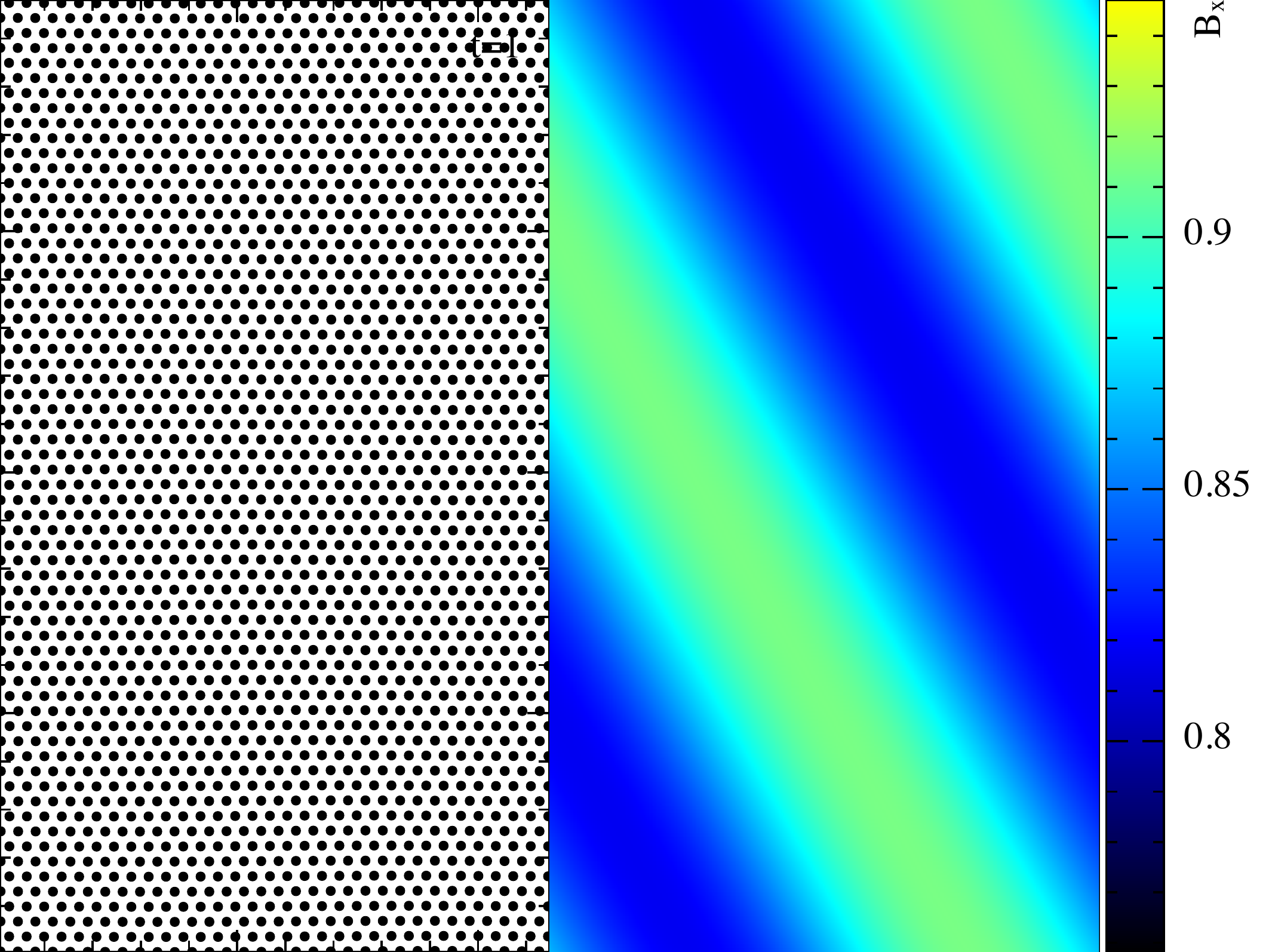}
   \caption{The tensile instability in SPMHD: The figures show the particle distribution (left panel in each figure) and $x$-component of the magnetic field (right panel in each figure) in the 2.5D circularly polarized Alfv\'en wave test using the (unstable) conservative SPMHD force (left figure) and with a stable formulation (right figure), shown after 1 wave crossing time. Untreated, the MHD tensile instability results in a catastrophic clumping of particles along the magnetic field lines (left figure) which proceeds to destroy the calculation. Physically, it can be attributed to non-zero divergence terms in the conservative form of the MHD force, and can be stabilised by explicitly subtracting the unphysical `source term' (right figure).}
   \label{fig:alfvenwave}
\end{center}
\end{figure}

\subsection{Fix 2: use a more accurate but non-conservative gradient estimate in the anisotropic force}
 \citet{morrisphd} proposed a compromise approach whereby one retains conservative form in the isotropic part of the force, but adopts a more accurate but non-conservative derivative estimate (that vanishes when the stress is constant) in the anisotropic term. Adapted for variable smoothing lengths, the resultant force is given by \citep{pm05,price10}
 \begin{equation}
\frac{dv_{a}^{i}}{dt} = - \sum_b m_b\left[\frac{P_a + \frac{1}{2\mu_{0}} B^2_a}{\Omega_{a}\rho_a^2}\nabla^{i}_{a} W_{ab}(h_{a}) + \frac{P_b + \frac{1}{2\mu_{0}} B^2_b}{\Omega_{b}\rho_b^2}\nabla_{a}^{i} W_{ab}(h_{b})\right]  + \frac{1}{\mu_0}\sum_b m_b
\frac{(B_iB_j)_b - (B_iB_j)_a}{\rho_a\rho_b}\overline{\nabla^{j} W}_{ab},
\label{eq:morrisforce}
\end{equation}
where $\overline{\nabla W_{ab}} \equiv [\nabla W_{ab}(h_{a}) + \nabla W_{ab}(h_{b})]/2$. The `Morris approach' does not exactly conserve momentum or total energy, but the errors due to non-conservation are in practice quite small even on strong shock tube problems \citep[see][]{price04}. Note, however, that retaining local conservation in the \emph{isotropic} term is important for the reasons discussed in Sec.~\ref{sec:localconservation} and also in order to obtain the correct jump conditions on MHD shocks. The caveat to this approach is that the remaining non-conservation of momentum can become problematic, especially in simulations run for a long time period, where secular effects can accumulate and cause a loss of symmetry, and that it cannot be turned ``off'' in the otherwise stable weak-field regime.

\subsection{Fix 3: subtract the unphysical source term}
\label{sec:borve}
 \citet{bot01} suggested an approach whereby conservation form is retained, but the monopole source term is explicitly subtracted. Extending their expression to incorporate variable smoothing length terms, the corrected equations of motion are given by
\begin{equation}
\frac{dv^{i}_a}{dt} = \sum_{b} m_{b} \left[\frac{S^{ij}_{a}}{\Omega_{a}\rho_{a}^{2}} \nabla^{j}_{a} W_{ab} (h_{a}) + \frac{S^{ij}_{b}}{\Omega_{b}\rho_{b}^{2}} \nabla^{j}_{a} W_{ab} (h_{b}) \right] - \hat{B}_{a}^{i} \sum_{b} m_{b} \left[\frac{B^{j}_{a}}{\Omega_{a}\rho_{a}^{2}} \nabla^{j}_{a} W_{ab} (h_{a}) + \frac{B^{j}_{b}}{\Omega_{b}\rho_{b}^{2}} \nabla^{j}_{a} W_{ab} (h_{b}) \right]. \label{eq:borvemethod}
\end{equation}
 The corrective term with $\hat{B}^{i}_{a} = {B}^{i}_{a}$ is equivalent to the source term added to the momentum equation in the eight-wave Riemann solver \citep{pea99} -- with the key point from an SPH standpoint being that the discretisation used for the correction term is identical to that in which the divergence term occurs in the force equation. \citet{bot04} also showed that it is not necessarily required to always choose $\hat{B}^i_{a} = B^{i}_{a}$ in order to achieve stability, showing in particular that $\hat{B}^{i}_{a} = \frac12 B^{i}_{a}$ is sufficient. \citet*{bot06} give a more general method for setting  $\hat{B}^{i}$ such that -- amongst other things -- no correction is applied when the gas pressure exceeds the magnetic pressure (we refer the reader to their paper for more details). Whilst adding the correction term violates exact conservation of momentum and energy, it does so only insofar as the divergence term in (\ref{eq:borvemethod}) is non-zero. Like the Morris approach, (\ref{eq:borvemethod}) is a good general method for removing the tensile instability in SPMHD -- but has the advantage of being able to be switched \emph{off} in the regime where the conservative formulation is stable.
 
 \subsection{Other methods}
 Finally, other methods have been proposed for dealing with the tensile instability in other physical problems, such as elastic dynamics. In particular \citet{monaghan00} suggested explicitly adding a short range repulsive force between particles to counteract the unphysical attractive force, similar to what would occur in atoms. \citet{pm04a} initially applied this approach to the MHD equations and it was shown by \citet{price04} to be equivalent to a modification of the kernel gradient in the anisotropic part of the MHD force (the second term in Eq.~\ref{eq:spmhdmom}). However it proved problematic in highly compressible calculations where the smoothing lengths could vary dramatically (leading to a difficulty in defining ``short range'') and at large negative stress was found to result in large errors in the numerical sound speed \citep[see][]{price04}. It was therefore abandoned by \citet{pm05} in favour of either the Morris or \citet{bot01} approaches.

\subsubsection{MHD Example 2: Circularly polarized Alfv\'en wave in 2.5D}
An example of the tensile instability in practice is shown in Fig.~\ref{fig:alfvenwave}, which shows the particle distribution (left panels in each figure) and $x$-component of the magnetic field (right panels in each figure) in the 2.5D circularly polarized Alfv\'en wave test from \citet{toth00}. The initial conditions \citep{toth00,pm05} are $\rho = 1$, $P = 0.1$, $v_\parallel = 0$, $B_\parallel = 1$, $v_\perp = B_\perp = 0.1\sin{(2\pi r_\parallel)}$ and $v_z = B_z = 0.1 \cos{(2\pi r_\parallel)}$ with $\gamma = 5/3$ (where $r_\parallel = x \cos{\theta} + y\sin{\theta}$), with the wave vector directed at an angle of $\theta = 30^{\circ}$ with respect to the $x$-axis in the computational domain $0 < x < 1/\cos{\theta}$, $0 < y < 1/\sin{\theta}$ with periodic boundary conditions.

 Using a conservative SPMHD formulation (left figure, shown at $t=1$) whole lines of particles can be seen to have attracted each other in the direction of $B_\parallel$, which proceeds to destroy the calculation at later times. With either the Morris or \citeauthor{bot01} approach (right figure, also shown at $t=1$) the particle arrangement -- and hence the wave -- is stable and propagates correctly.

\section{Dissipation terms and shocks in SPMHD}
\label{sec:mhddiss}
 The second important issue for SPMHD is the formulation of dissipation terms appropriate for MHD shocks. We can follow the same general principle as in Sec.~\ref{sec:dissipation}, starting with (\ref{eq:ave}) as the basic equation, except that in MHD the energy variable is given by
\begin{equation}
e^{*}_{a} = \frac12 \alpha v_{sig} ({\bf v}_{a} \cdot \hat{\bf r}_{ab})^{2} + \alpha_{u}  v_{sig}^{u} u_{a}  + \alpha_{B} v_{sig}^{B} \frac{B^{2}_{a}}{2 \mu_{0} \bar{\rho}_{ab}}, \label{eq:mhdestar}
\end{equation}
and the signal velocity in the kinetic energy term is generalised to
\begin{equation}
v_{sig} = 
\left\{\begin{array}{ll}
\frac12[ v_{a} + v_{b} - \beta {\bf v}_{ab}\cdot\hat{\bf r}_{ab}]; & {\bf v}_{ab}\cdot\hat{\bf r}_{ab} \le 0; \\
0; & {\bf v}_{ab}\cdot\hat{\bf r}_{ab} > 0,
\end{array}\right.
\label{eq:mhdvsig}
\end{equation}
where $v$ is the fastest wave speed for MHD (i.e., the fast MHD mode), given by
\begin{equation}
v_a = \frac{1}{\sqrt{2}}\left[\left(c_{s,a}^2 + \frac{B^2_a}{\mu_0\rho_a}\right)
+ \sqrt{\left(c_{s,a}^2 + \frac{B^2_a}{\mu_0\rho_a}\right)^2 - 4\frac{c_{s,a}^2 ({\bf B}_{a}\cdot\hat{\bf r}_{ab})^2}{\mu_0\rho_a}}
\right]^{1/2}.
\end{equation}
 The key constraint in deriving dissipative terms in the other equations is that the contribution to the entropy from the dissipative terms in the total energy equation must be positive definite in order to satisfy the second law of thermodynamics. For the kinetic energy term this is satisfied provided a term is added to the acceleration equation (i.e., Eq.~\ref{eq:av}), giving a positive definite contribution to the thermal energy (Eq.~\ref{eq:dudtdiss}). It may also be shown that the thermal energy term results in a positive definite contribution to the entropy (Appendix~B of \citealt{pm04a}). For the magnetic energy term, positivity means that the overall dissipative contribution to the thermal energy equation is given by
\begin{equation}
\left(\frac{du}{dt}\right)_{diss} = -\sum_b \frac{m_{b}}{\bar{\rho}_{ab}}\left[ \frac{1}{2}\alpha v_{sig}({\bf v}_a - {\bf v}_b)^2 + \alpha_u v_{sig}^{u} (u_a - u_b)  + v_{sig}^{B} \frac{\alpha_B}{2\mu_0\bar{\rho}_{ab}} ({\bf B}_a - {\bf B}_b)^2\right] \overline{F_{ab}}, \label{eq:mhddudtdiss}
\end{equation}
which is positive definite because $F_{ab}$ is negative definite (Fig.~\ref{fig:Bsplines}) and the terms inside the brackets are positive. Using (\ref{eq:mhddudtdiss}), we can deduce the term which must arise in the induction equation using
\begin{equation}
 \frac{d}{dt}\left(\frac{B^2}{2\mu_0\rho} \right)_{diss} =  \left(\frac{de}{dt}\right)_{diss} - {\bf v}\cdot\left(\frac{d{\bf v}}{dt}\right)_{diss} - \left(\frac{du}{dt}\right)_{diss},
\end{equation}
giving a term in the induction equation of the form
\begin{equation}
\left(\frac{d{\bf B}_{a}}{dt}\right)_{diss} = \rho_a \sum_b m_b \frac{\alpha_B v^{B}_{sig}}{\bar{\rho}_{ab}^2}
\left({\bf B}_a -{\bf B}_b \right) \bar{F}_{ab} \label{eq:dBdtdiss},
\end{equation}
or, evolving ${\bf B}/\rho$,
\begin{equation}
\frac{d}{dt} \left(\frac{{\bf B}_{a}}{\rho_{a}}\right)_{diss} = \sum_b m_b \frac{\alpha_B v^{B}_{sig}}{\bar{\rho}_{ab}^2}
\left({\bf B}_a -{\bf B}_b \right) \bar{F}_{ab} \label{eq:dBrhodtdiss}.
\end{equation}
 The interpretation of this term is straightforward using (\ref{eq:del2A}): We have derived an artificial resistivity term
\begin{equation}
 \left(\frac{d{\bf B}_{a}}{dt}\right)_{diss} = \eta_{M} \nabla^{2}{\bf B}_{a}; \hspace{1cm} \eta_{M} \approx \frac12 \alpha_{B} v_{sig}^{B} \vert r_{ab} \vert .
 \label{eq:artresistinterpret}
\end{equation}
For the artificial resistivity it is also better to use a signal velocity that differs from the one used in the viscosity. A simple choice (used in \textsc{ndspmhd}) is to use an averaged Alfv\'en speed, e.g.
\begin{equation}
 v_{sig}^{B} = \frac12 \sqrt{v_{A,a}^{2} + v_{A,b}^{2}};  \hspace{1cm} v_{A}^{2} = \frac{B^{2}}{\mu_{0}\rho}.
\end{equation}
\citet{pm04a,pm05} also show how an artificial resistivity can be derived that uses only components of the magnetic field perpendicular to the line of sight. However, \citet{pm05} found that the version using the jump in total energy (Eqs.~\ref{eq:mhdestar}, \ref{eq:dBdtdiss}) performed better in practice. The parameter $\alpha_{B}$ can also be evolved using a switch similar to (\ref{eq:dalphadt}), where \citet{pm05} suggest a source term of the form
\begin{equation}
\mathcal{S} = \mathrm{max}\left(\frac{\vert \nabla \times {\bf B}
\vert}{\sqrt{\mu_0\rho}}, \frac{\vert \nabla \cdot {\bf B}
\vert}{\sqrt{\mu_0\rho}} \right).
\label{eq:alphaBsource}
\end{equation}
Note that artificial thermal conductivity and resistivity should in general be applied regardless of whether or not the particle pair is approaching or receding.

\begin{figure}[t]
\begin{center}
   \includegraphics[width=0.75\columnwidth]{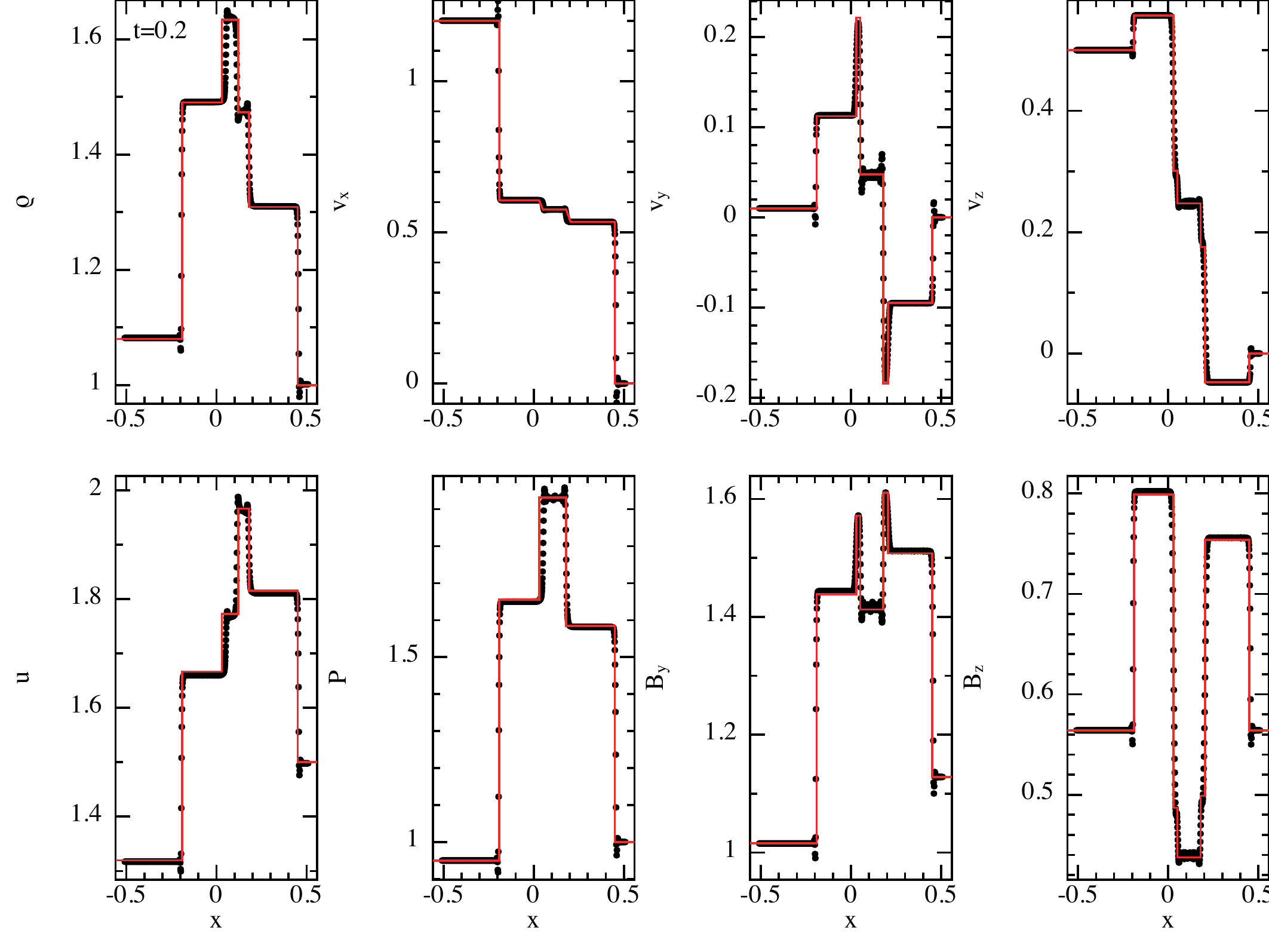}
   \caption{A ``1.75D'' MHD shock tube problem, showing the formation of 7 discontinuities: A slow, Alfv\'en and fast wave propagating in each direction, plus the contact discontinuity. Capture of all of these discontinuities requires the application of artificial viscosity, thermal conductivity and resistivity to treat jumps in $v$, $u$ and $B$ respectively. The $M_{4}$ cubic spline kernel has been used.}
   \label{fig:mhdshock}
\end{center}
\end{figure}

\begin{figure}[t]
\begin{center}
   \includegraphics[width=0.7\columnwidth]{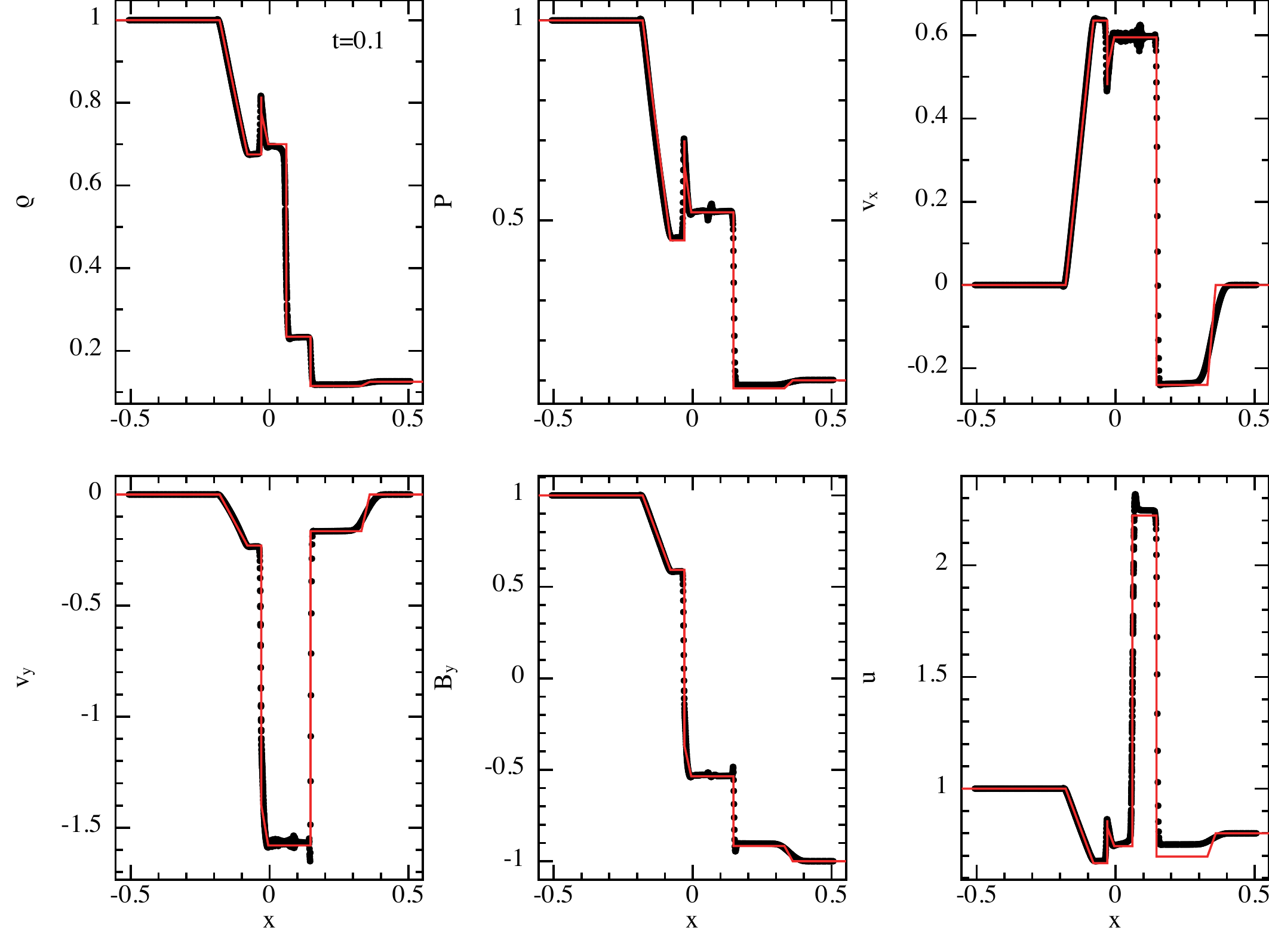}
   \caption{The Brio-Wu MHD shock tube problem in 2D, computed using $800\times 24$ particles and the $M_{6}$ quintic kernel. With the quintic, results comparable to the 1D solution can be obtained. The numerical solution from \citet{balsara98} is given by the solid (red) line. All particles are plotted.}
   \label{fig:briowu2D}
\end{center}
\end{figure}

\subsection{MHD Example 3: shock tube problems in MHD}
 The application of specific dissipative terms for shock-capturing is illustrated in Fig.~\ref{fig:mhdshock}, which shows a ``1.75D'' MHD shock tube problem (i.e., 1D but with 3D magnetic and velocity fields) from \citet{rj95}, using 800 particles and the standard $M_{4}$ cubic spline kernel. The setup is $(\rho,P,v_x,v_y,v_z,B_y,B_z) = [1.08,0.95,1.2,0.01,0.5,3.6/(4\pi)^{1/2},2/(4\pi)^{1/2}]$ for $x<0$, whilst for $x\ge0$ we have $(\rho,P,v_x,v_y,v_z,B_y,B_z)=[1,1,0,0,0,4/(4\pi)^{1/2},2/(4\pi)^{1/2}]$, with $B_x = 2/(4\pi)^{1/2}$ and $\gamma=5/3$. The main point is that this shows all 7 possible discontinuities in MHD: Three waves (slow, Alfv\'en and fast modes) propagating in each direction, plus the contact discontinuity. Since each requires treatment, we require artificial viscosity, thermal conductivity and resistivity to treat the jumps in velocity, thermal energy and magnetic field respectively.

Fig.~\ref{fig:briowu2D} shows a 2D version of the \citet{bw88} shock tube test, considered a standard test for MHD codes \citep[e.g.][]{sea92,dw94,rj95,balsara98}, using $800 \times 24$ particles and the $M_{6}$ quintic kernel. Initially $(\rho,P,v_x,v_y,B_y) = [1,1,0,0,1]$ for $x<0$, with $(\rho,P,v_x,v_y,B_y)=[0.125,0.1,0,0,-1]$ for $x\ge 0$ with $B_x = 0.75$ everywhere and $\gamma=2.0$. Profiles of all SPMHD particles are plotted at $t=0.1$ whilst the numerical solution from \citet{balsara98} is given by the solid (red) line. The solution is comparable to that achievable in 1D \citep[e.g.][]{rp07} -- at this resolution the only noticeable deviation from the Balsara solution is the slight mismatch in thermal energy at $x\approx 0.15$--$0.3$, a result of the low spatial resolution (10$\times$ lower than in the $x<0$ region) in this low density region.

\section{The divergence constraint in SPMHD}
\label{sec:divB}
 The third major issue, common to all numerical MHD codes, is how to enforce the $\nabla\cdot{\bf B} = 0$ constraint. The problem arises because the constraint occurs only as an \emph{initial condition} in the MHD equations, since from the induction equation we have
\begin{equation}
\frac{\partial {\bf B}}{\partial t} = -\nabla \times ({\bf v} \times {\bf B}); \hspace{1cm} \frac{\partial}{\partial t} \left(\nabla\cdot{\bf B}\right) = 0,
\label{eq:induction}
\end{equation}
 The problem is that the truncation error in the numerical solution means that such a condition cannot be maintained indefinitely. What to do about it falls into three general approaches: ``ignore'', ``clean'' or ``prevent'' the numerical growth of monopole terms. We will discuss each of these in the context of SPMHD, but first turn to the issue of formulating consistent equations given that $\nabla\cdot{\bf B}\neq 0$.

\subsection{$\nabla\cdot{\bf B} \neq 0$: Monopole terms in the evolution equations}
\label{sec:sourceterms}
  We have already touched on the `source terms' issue in the context of the tensile instability in SPMHD, which arises from the fact that the MHD force written as the divergence of a stress tensor differs from the exactly-perpendicular-to-B Lorentz force by a term proportional to $\nabla\cdot{\bf B}$ -- and that this term does not vanish even in the trivial case of 1D where $B_{x} = const$. A similar issue arises in the induction equation (\ref{eq:induction}), which in ``conservative form'' is given by
\begin{equation}
\frac{\partial {\bf B}}{\partial t} = \nabla\cdot ({\bf v}{\bf B} - {\bf B}{\bf v}) =  -({\bf v}\cdot\nabla){\bf B} + ({\bf B}\cdot\nabla){\bf v} - {\bf B}(\nabla\cdot{\bf v}) + {\bf v} (\nabla\cdot{\bf B}),
\end{equation}
identical to (\ref{eq:induction}), and conserving the \emph{volume} integral of the flux $\int {\bf B} {\rm dV}$. However, if the monopole term ${\bf v}(\nabla\cdot{\bf B})$ is neglected, then we have (using the Lagrangian time derivative)
\begin{equation}
\frac{d{\bf B}}{dt} = ({\bf B}\cdot\nabla){\bf v} - {\bf B}(\nabla\cdot{\bf v}),
\label{eq:dBdt}
\end{equation}
taking the divergence of which shows that the monopoles evolve according to an equation similar to the continuity equation for density
\begin{equation}
\frac{\partial}{\partial t} \left(\nabla\cdot{\bf B}\right) + \nabla\cdot ({\bf v}\nabla\cdot{\bf B}),
\end{equation}
meaning that the volume integral $\int_{\rm V} (\nabla\cdot{\bf B}) {\rm dV}$, and by Green's theorem the \emph{surface} integral of the flux, $\oint_{S} {\bf B}\cdot{\rm d{\bf S}}$ is conserved. So adopting (\ref{eq:dBdt}) represents a `monopole conserving' form of the induction equation, such that the surface flux integral will be conserved even if the divergence errors are non-zero. On the basis of this idea \citet{pea99} suggested the 8-wave formulation, with the induction equation given by (\ref{eq:dBdt}) and the momentum and energy equations given by
\begin{eqnarray}
\frac{dv^i}{dt} & = & \frac{1}{\rho}\pder{S^{ij}}{x^j} -
\frac{B^i}{\rho}\pder{B^j}{x^j}, \label{eq:mompowell} \\
\frac{de}{dt} & = & \frac{1}{\rho}\pder{(v_i S^{ij})}{x^j} -
\frac{v_i B^i}{\rho}\pder{B^j}{x^j},
\label{eq:enerpowell}
\end{eqnarray}
 However, \citet{janhunen00} and \citet{dellar01} have strongly argued that a consistent formulation in the presence of monopoles, whilst adopting (\ref{eq:dBdt}) instead of (\ref{eq:induction}), should nevertheless still conserve linear (but not angular) momentum and energy. For SPMHD we have already proved that this is the case because we derived the momentum-conserving force (\ref{eq:spmhdmomtensor}) from the surface-flux-and-monopole conserving induction equation (\ref{eq:mhddBrhodt}). This is a somewhat moot point in practice though, since we require the subtraction of the source term in the momentum equation anyway in order to stabilise the tensile instability (Sec.~\ref{sec:borve}), giving essentially the 8-wave formulation.

\subsection{$\nabla\cdot{\bf B} \approx 0$: The ``ignore'' approach}
 The simplest approach to the divergence constraint is to do nothing at all. That is, simply monitor the divergence error and ``trust'' the simulations if it remains small. In SPMHD it is usual to monitor the divergence error using the dimensionless quantity
\begin{equation}
\frac{h \nabla\cdot{\bf B}}{\vert B \vert}, \label{eq:divBerr}
\end{equation}
typically hoping that it remains smaller than a few percent. For a number of problems, including many of the test problems shown in this paper, this approach works surprisingly well. There is also the question of what one means by the ``divergence error'', since (\ref{eq:divBerr}) can be computed using a variety of operators (Sec.~\ref{sec:gradients}) and the `stencil' in which one measures divergence changes in SPH as soon as one changes the particle positions. However there are clearly many problems where this approach is not sufficient.

\subsection{$\nabla\cdot{\bf B} \to 0$: The ``clean'' approach}
 Divergence cleaning in SPMHD also faces the problem of defining what $\nabla\cdot{\bf B} = 0$ actually means. Clearly from a stability point of view it would be desirable for the divergence to be zero in the discretisation in which it enters the force equation, but we have already seen that this is not possible even in 1D because the conservative SPH derivative does not vanish for constant functions. However, it may be hoped that cleaning in a particular discretisation will at least produce solenoidal fields to the order of the truncation error [i.e., $\mathcal{O}(h^{2})$]. Approaches to divergence cleaning fall into three general categories based on the solution to an elliptic, parabolic or hyperbolic partial differential equation.

\subsubsection{Elliptic cleaning}
 Elliptic cleaning corresponds to using projection methods: Solving either for a correction term,
\begin{equation}
{\bf B} = {\bf B}^{*} - \nabla\phi; \hspace{1cm} \nabla^{2} \phi = \nabla\cdot{\bf B}^{*},
\label{eq:phiproject}
\end{equation}
or alternatively for the physical (solenoidal) component of the field,
\begin{equation}
{\bf B} = \nabla\times {\bf A}; \hspace{1cm} \nabla^{2} {\bf A} = -\nabla\times{\bf B}^{*}.
\label{eq:biotsavart}
\end{equation}
SPH methods for both of the above have been discussed by \citet{pm05}. The main issue is that, if one simply uses the Green's function solution -- analogous to the computation of gravitational forces -- with the source function computed with a chosen div or curl operator, the projection is only approximate (that is, the Poisson equation is not discretely satisfied\footnote{The approximate nature of the projection is related to the question of how one should ``soften'' the Green's function solution, similar to the case for gravity. What is required is a generalisation of the softening formulation given by \citet{pm07} which shows how the gravitational Poisson equation can be discretely satisfied if the SPH density is used as the source term, so this would be an approach worth pursuing.}). For 3D simulations with particles on individual timesteps, it is also impractical and highly inefficient to compute a global projection every $n$ timesteps.

\subsubsection{Parabolic cleaning}
 A parabolic approach corresponds to adding a $\nabla\cdot{\bf B}$ diffusion term to the induction equation, i.e.,
\begin{equation}
\left( \frac{d{\bf B}}{dt}\right)_{clean} = \nabla \cdot (\eta_{c} \nabla\cdot{\bf B}).
\end{equation}
Having formulated our artificial dissipative terms using the jump in total magnetic energy (\ref{eq:mhdestar}) means that the artificial resistivity already contains a term of this form (see \ref{eq:artresistinterpret}), so applying artificial resistivity (with a switch that responds to the divergence, Eq.~\ref{eq:alphaBsource}) corresponds to a form of parabolic divergence cleaning. This is partly why the ``ignore'' approach is not as bad as it sounds, and has successfully used for a number of problems \citep[e.g.][]{ds09,kotarbaetal10,buerzleetal10}.

\subsubsection{Hyperbolic/parabolic cleaning}
\citet{dea02} suggested a hyperbolic cleaning method for the MHD equations, given by
\begin{equation}
\frac{d{\bf B}}{dt} = -{\bf B}(\nabla\cdot{\bf v}) + ({\bf B}\cdot\nabla){\bf v} -\nabla\psi,
\label{eq:dBdtpsi}
\end{equation}
where $\psi$ is an additional variable that evolves according to \citep{pm05}
\begin{equation}
\frac{d\psi}{dt} = -c_{h}^2 (\nabla\cdot{\bf B}) - \frac{\psi}{\tau},
\label{eq:dpsidt}
\end{equation}
where for SPH it is natural to use $c_{h} \equiv v_{sig}$ and $\tau \equiv \alpha_{h} h/v_{sig}$ where $\alpha_{h} \in [0,1]$ is a dimensionless parameter. The first term in (\ref{eq:dpsidt}) corresponds to hyperbolic propagation of the ``divergence wave'' (at speed $c_{h}$), whilst $\psi/\tau$ is a parabolic decay term (with decay time $\tau$). This can be shown combining the equations (assuming ${\bf v}=const$) to yield a wave equation,
\begin{equation}
\frac{1}{c_{h}^{2}}\pder{^{2} \psi}{t^{2}} - \nabla^{2} \psi + \tau \pder{\psi}{t} = 0.
\label{eq:psiwave}
\end{equation}
As reported by \citet{pm05} and others \citep[e.g.][]{mt10}, the original \citet{dea02} formulation contained a parameter that is \emph{not} dimensionless, but that is equivalent to the wavelength of critical damping $\lambda_{c}$ in the parabolic decay term (evident by writing $\tau\equiv \lambda_{c}/c_{h}$ in the above). Indeed, \citet{pm05} found that whilst using (\ref{eq:dBdtpsi})-(\ref{eq:dpsidt}) could be effective at cleaning well-resolved divergence errors ($\lambda_{c} \gg h$), it was found to have little or no effect when used in ``real'' problems (where $\lambda_{c} \sim h$), giving at most a factor of $\sim 2$ reduction in $\nabla\cdot{\bf B}_{\rm max}$. It was also found that a poor choice of parameters could \emph{increase} the divergence error substantially via constructive interference of divergence waves. 
 
  More recently Stasyszyn \& Dolag \citep[following][]{ds09} report better success using this method, by two adaptations: i) accounting for the magnetic energy dissipation in the energy equation, and ii) implementation of a limiter to control the amount by which the magnetic field can change in a single timestep (Dolag 2010, priv. commun.). Nevertheless, it does not appear that the method is being used on real problems \citep[e.g.][]{kotarbaetal10}, suggesting that it remains ineffective.

\subsection{$\nabla\cdot{\bf B} = 0$: The ``prevent'' approach}
 The third approach is to try to `prevent' divergence errors altogether by formulating the MHD equations such that $\nabla\cdot{\bf B} = 0$ is satisfied by construction. Unfortunately, the approach most successfully adopted in grid based schemes -- the constrained transport method \citep{eh88,gs05} -- cannot be directly applied in an SPH context because it requires the computation of surface rather than volume integrals.

\subsubsection{Euler potentials}
  Perhaps the closest to a `constrained transport' approach in SPMHD is the formulation in terms of Euler potentials \citep{stern70,stern76} (also referred to as ``Clebsch variables'', e.g. by \citealt{pm85}), $\alpha_{E}$ and $\beta_{E}$, where the magnetic field is expressed as
\begin{equation}
{\bf B} = \nabla\alpha_{E} \times \nabla\beta_{E}.
\label{eq:Beuler}
\end{equation}
Taking the divergence shows that $\nabla\cdot{\bf B} = 0$ is satisfied by construction. A further advantage is that for ideal MHD, the induction equation (\ref{eq:mhddBrhodt}) becomes simply
\begin{equation}
\frac{d\alpha_{E}}{dt} = 0, \hspace{5mm} \frac{d\beta_{E}}{dt} = 0,
\label{eq:eulerevol}
\end{equation}
which corresponds physically to the advection of magnetic field lines by Lagrangian particles \citep{stern66}. That is, the Euler potentials reflect the physical fact that the field lines are `frozen' to the fluid in ideal MHD and are therefore a natural description for a Lagrangian scheme. For SPH it is straightforward to write down expressions for (\ref{eq:Beuler}) using any of our standard first derivative operators (Sec.~\ref{sec:gradients}) and to simply use the resultant ${\bf B}$ in the SPMHD equations of motion -- though this will give only approximate conservation of energy, see \citealt{price10}. However, in order to capture shocks, some dissipation in the magnetic field is required. For this reason \citet{pb07} and \citet{rp07} introduced dissipation terms of the form
\begin{equation}
\left(\frac{d\alpha_{E}^{a}}{dt}\right)_{diss} = \sum_{b} \frac{m_{b}}{\bar{\rho}_{ab}} \alpha_{B} v^{B}_{sig} \left( \alpha_{E}^{a} - \alpha_{E}^{b} \right)\overline{F_{ab}}; \hspace{0.5cm}
\left(\frac{d\beta_{E}^{a}}{dt}\right)_{diss} = \sum_{b} \frac{m_{b}}{\bar{\rho}_{ab}} \alpha_{B} v^{B}_{sig} \left( \beta_{E}^{a} - \beta_{E}^{b} \right)\overline{F_{ab}},
\label{eq:dAdtdisswrong}
\end{equation}
corresponding to the continuum equations (assuming $\eta_{\rm M}\approx const$)
\begin{equation}
\frac{d\alpha_{E}}{dt} \approx \eta_{\rm M} \nabla^{2} \alpha; \hspace{1cm} \frac{d\beta_{E}}{dt} = \eta_{\rm M} \nabla^{2} \beta,
\label{eq:Eulerdiss}
\end{equation}
where $\eta_{\rm M}$ is given by (\ref{eq:artresistinterpret}). However it is clear that this is \emph{not} a consistent formulation of (physical) resistivity for the Euler potentials, since we have neglected mixed terms that are of comparable magnitude to those in (\ref{eq:Eulerdiss}) \citep[e.g.][]{brandenburg10}. It is also difficult to ensure that the contribution to the entropy is positive definite\footnote{\citet{rp07} simply used Eq~\ref{eq:mhddudtdiss} with ${\bf B}$ calculated from (\ref{eq:Beuler}) such that the contribution is guaranteed to be positive. However, the total energy is only approximately conserved by this procedure. See \citet{price10}.}, although \citet{price10} shows how the latter requirement can be achieved for the vector potential, and similar formulations should be used in place of (\ref{eq:dAdtdisswrong}) for the Euler potentials.

  Nevertheless, the control of the divergence errors means that the Euler potentials have found successful application to a range of problems \citep[e.g.][]{pb07,dp08,kotarbaetal09}, most notably in star formation \citep[e.g.][]{pb07,pb08,pb09} since one is able to form stars in the presence of magnetic fields without blow-up of the numerical solution due to divergence errors\footnote{\citet{pr06} also employed the Euler potentials to simulate magnetic field growth during Neutron star mergers. Whilst similar results were found using a ${\bf B}$-based formulation, the simulations with Euler potentials were later found to be incorrect, showing exaggerated field growth because the boundary conditions on the potentials for the stars were not correctly accounted for.}. The caveat is that the Euler potentials do not capture the full physics of MHD, since the Helicity ${\bf A}\cdot{\bf B}$ is identically zero, meaning that topologically non-trivial fields cannot be represented (or by corollary, created during a simulation) using (\ref{eq:Beuler})--(\ref{eq:eulerevol}). For example, the Euler potentials cannot be used to follow multiple complete windings of a magnetic field in a rotating fluid, a corollary of the fact that the potentials are simply advected by the particles (Eq.~\ref{eq:eulerevol}), so reconstruction of the field via (\ref{eq:Beuler}) requires a one-to-one mapping between the initial and final particle positions. These restrictions are easily demonstrated by simple test problems \citep{brandenburg10} and mean in practice that one is limited to studying problems with initially simple field geometries, and that important field winding processes (i.e., any dynamo action) are missed. On the other hand, it may be possible to formulate a more general Euler potentials description without such limitations.

\subsubsection{The vector potential}
 Another possibility for a divergence-free formulation without the restrictions of the Euler potentials is to simply employ the vector potential ${\bf B} = \nabla\times {\bf A}$ directly. Indeed in 2D they are exactly equivalent, since for a 2D field $\alpha_{E} \equiv A_{z} (x,y)$ and $\beta_{E} \equiv z$. The main disadvantage of the vector potential in 3D is that the evolution equation is complicated and requires an appropriate gauge choice. For example, setting the scalar potential $\phi = {\bf v}\cdot{\bf A}$ one can obtain a Galilean-invariant evolution equation for ${\bf A}$ of the form \citep{price10}
\begin{equation}
\frac{d{\bf A}}{dt} = - {\bf A}\times (\nabla\times {\bf v}) - ({\bf A}\cdot\nabla){\bf v} + {\bf v}\times{\bf B}_{ext} = -A^{j}\nabla v^{j} + {\bf v}\times{\bf B}_{ext}.
\label{eq:dAdt}
\end{equation}
 Using this, one can proceed to derive an SPMHD algorithm for the vector potential from the Lagrangian (\ref{eq:Lmhd}). In particular, \citet{price10} shows that by writing the curl according to
\begin{equation}
{\bf B}_{a} = (\nabla\times{\bf A})_{a} + {\bf B}_{ext} = \frac{1}{\rho_{a}} \sum_{b} m_{b} ({\bf A}_{a} - {\bf A}_{b}) \times \nabla_{a} W_{ab} + {\bf B}_{ext},
\label{eq:curlA}
\end{equation}
the effective evolution equation for ${\bf B}$ is given by
\begin{eqnarray}
\frac{d{\bf B}_{a}}{dt} &= & \frac{1}{\rho_{a}} \sum_{b} m_{b} \left({\bf A}_{a} - {\bf A}_{b} \right) \times [({\bf v}_{a} - {\bf v}_{b})\cdot\nabla] \nabla_{a} W_{ab} \nonumber \\
& + & \frac{1}{\rho_{a}} \sum_{b} m_{b} \left(\frac{d{\bf A}_{a}}{dt} - \frac{d{\bf A}_{b}}{dt} \right) \times \nabla_{a} W_{ab}
 - \frac{{\bf B}_{int}}{\rho_{a}} \frac{d\rho_{a}}{dt}. \label{eq:dBdtvecp}
\end{eqnarray}
Thus, writing the evolution equation for ${\bf A}$, (\ref{eq:dAdt}), in the form
\begin{equation}
\frac{d{\bf A}_{a}}{dt} = \frac{A^{a}_{j}}{\rho_{a}} \sum_{b} m_{b} (v^{j}_{a} - v^{j}_{b}) \nabla_{a} W_{ab} + ({\bf v}\times{\bf B}_{ext})_{a},
\label{eq:dAdtSPH}
\end{equation}
one can self-consistently derive equations of motion for the vector potential from (\ref{eq:deltaL}) by writing the perturbation to $\delta {\bf B}$ in terms of $\delta{\bf A}$ using (\ref{eq:dBdtvecp}) and $\delta{\bf A}$ in terms of $\delta{\bf r}$ using (\ref{eq:dAdtSPH}), giving
\begin{eqnarray}
\frac{d{\bf v}_{a}}{dt} & = & -\sum_{b} m_{b} \left( \frac{P_{a} - \frac{3}{2\mu_{0}} B_{a}^{2}}{\rho_{a}^{2}} + \frac{P_{b} - \frac{3}{2\mu_{0}} B_{b}^{2}}{\rho_{b}^{2}} \right) \nabla_{a} W_{ab} \nonumber \\
& - & \frac{1}{\mu_{0}} \sum_{b} m_{b} \left\{ \left( \frac{{\bf B}_{a}}{\rho_{a}^{2}} +  \frac{{\bf B}_{b}}{\rho_{b}^{2}} \right)\cdot \left[({\bf A}_{a} - {\bf A}_{b}) \times \nabla \right] \right\} \nabla_{a} W_{ab} \nonumber \\
& - & \frac{2}{\mu_{0}}\sum_{b} m_{b} \left( \frac{{\bf B}_a}{\rho_{a}^{2}} + \frac{{\bf B}_b}{\rho_{b}^{2}} \right)\cdot {\bf B}_{ext} \nabla_{a} W_{ab} + \frac{1}{\mu_{0}}\sum_{b} m_{b} \left( \frac{{\bf B}_a}{\rho_{a}^{2}} + \frac{{\bf B}_b}{\rho_{b}^{2}} \right){\bf B}_{ext} \cdot \nabla_{a} W_{ab} \nonumber \\ 
& - & \sum_{b} m_{b} \left[ \frac{{\bf A}_{a}}{\rho_{a}^{2}} {\bf J}_{a}\cdot\nabla_{a} W_{ab}  +   \frac{{\bf A}_{b}}{\rho_{b}^{2}}{\bf J}_{b}\cdot\nabla_{a} W_{ab} \right],
\label{eq:eomfixedh}
\end{eqnarray}
where ${\bf J}$ the magnetic current is computed using the symmetric curl operator (the conjugate to \ref{eq:curlA}),
\begin{equation}
{\bf J}_{a} \equiv \frac{\left(\nabla\times{\bf B}\right)_{a}}{\mu_{0}} \equiv -\frac{\rho_{a}}{\mu_{0}} \sum_{b} m_{b} \left[ \frac{{\bf B}_{a}}{\rho_{a}^{2}} +  \frac{{\bf B}_{b}}{\rho_{b}^{2}} \right] \times \nabla_{a} W_{ab}.
\label{eq:curlBfixedh}
\end{equation}
The proof that the above equations do indeed translate into an continuum expression of the conservative MHD equations, as well as a more general version taking full account of smoothing length gradient terms is given by \citet{price10}. However, the major flaw, from simple inspection of (\ref{eq:eomfixedh}), is that the isotropic (first) term will give a negative pressure -- and thus the tensile instability -- whenever $3 B^{2}/(2\mu_{0}) > P$, a much more restrictive regime than for the usual SPMHD equations (see Sec.~\ref{sec:mhdtensile}). This is verified in test problems, but unlike the standard SPMHD instability, proves difficult to stabilise without significantly affecting the solution. A second instability was also found to arise purely due to the evolution of ${\bf A}$ according to (\ref{eq:dAdt}), resulting in unconstrained growth of vector potential components, most likely related to the need to explicitly enforce a gauge condition alongside the evolution of the vector potential. \citet{price10} therefore concluded that using the vector potential was not a viable approach for SPMHD.
%
%
\vspace{-0.4cm}
\section{Summary}
 In summary, we have given an overview of SPH methodology, starting with the density estimate as the basis of all SPH formulations (Sec.~\ref{sec:calculatingdensity}) and showing how the equations of motion and energy can be self-consistently derived from the density estimate using a variational principle (Sec.~\ref{sec:densitytoequationsofmotion}). Kernel interpolation theory has been introduced mainly as a way of interpreting the SPH equations, and we have discussed how linear error analysis can be used to construct more accurate and very general derivative operators (Sec.~\ref{sec:interpolation}). In Sec.~\ref{sec:localconservation} the importance of local conservation was highlighted with respect to maintaining a regular particle distribution and thus accurate derivatives, giving us an understanding of how the tensile instability arises in SPMHD and why one should be careful in setting the ratio of smoothing length to particle spacing. Second derivatives in SPH were discussed in Sec.~\ref{sec:2ndderivs}, mainly as a way of formulating dissipative terms necessary to capture shocks and other kinds of discontinuities. In the second half of the paper, we have shown how SPMHD, like SPH, can also be formulated from a variational principle (Sec.~\ref{sec:spmhdfromL}) and have addressed the three main issues with regards to the accuracy of SPMHD: The tensile instability (Sec.~\ref{sec:mhdtensile}), the formulation of shock-capturing dissipation terms (Sec.~\ref{sec:mhddiss}) and the enforcement of the divergence-free condition on the magnetic field (Sec.~\ref{sec:divB}). Finally, this paper marks the public release of the \textsc{ndspmhd} SPH/SPMHD code that can be used to test and verify all of the ideas and methods  that have been discussed.

\vspace{-0.3cm}
\section*{Acknowledgments}
My knowledge and understanding of SPH would be nothing without the endless insight derived from my long-time mentor and colleague (and godfather-of-SPH), Joe Monaghan. In gathering my thoughts for this paper I would like to acknowledge particularly useful discussions with Klaus Dolag, Evghenii Gaburov, Volker Springel and Guillaume Laibe as well as many of the students at the ASTROSIM summer school -- thanks in particular go to Micha{\l} Hanasz for the invitation to present this material and the excellent organisation of the school. Figures were produced using \textsc{splash} \citep{splashpaper}, a freely available visualisation tool for SPH that is downloadable from \url{http://users.monash.edu.au/~dprice/splash/}. Thanks to James Wetter for the excellent work on the new giza backend. Finally, thanks to the (anonymous) referees for helpful comments on the manuscript.

\bibliography{sph,mhd,starformation}

\begin{thebibliography}{110}
\expandafter\ifx\csname natexlab\endcsname\relax\def\natexlab#1{#1}\fi
\expandafter\ifx\csname url\endcsname\relax
  \def\url#1{\texttt{#1}}\fi
\expandafter\ifx\csname urlprefix\endcsname\relax\def\urlprefix{URL }\fi

\bibitem[{{Abel}(2010)}]{abel10}
{Abel}, T., Mar. 2010. {rpSPH: a much improved Smoothed Particle Hydrodynamics
  Algorithm}. arXiv:1003.0937.

\bibitem[{{Agertz} et~al.(2007){Agertz}, {Moore}, {Stadel}, {Potter},
  {Miniati}, {Read}, {Mayer}, {Gawryszczak}, {Kravtsov}, {Nordlund}, {Pearce},
  {Quilis}, {Rudd}, {Springel}, {Stone}, {Tasker}, {Teyssier}, {Wadsley}, and
  {Walder}}]{agertzetal}
{Agertz}, O., {Moore}, B., {Stadel}, J., {Potter}, D., {Miniati}, F., {Read},
  J., {Mayer}, L., {Gawryszczak}, A., {Kravtsov}, A., {Nordlund}, {\AA}.,
  {Pearce}, F., {Quilis}, V., {Rudd}, D., {Springel}, V., {Stone}, J.,
  {Tasker}, E., {Teyssier}, R., {Wadsley}, J., {Walder}, R., Sep. 2007.
  {Fundamental differences between SPH and grid methods}. MNRAS 380, 963--978.

\bibitem[{{Artymowicz} and {Lubow}(1994)}]{al94}
{Artymowicz}, P., {Lubow}, S.~H., Feb. 1994. {Dynamics of binary-disk
  interaction. 1: Resonances and disk gap sizes}. ApJ 421, 651--667.

\bibitem[{{Balsara}(1998)}]{balsara98}
{Balsara}, D.~S., May 1998. {Total Variation Diminishing Scheme for Adiabatic
  and Isothermal Magnetohydrodynamics}. ApJS 116, 133.

\bibitem[{{Benz}(1990)}]{benz90}
{Benz}, W., 1990. {Smoothed Particle Hydrodynamics - A review}. In: {Buchler},
  J.~R. (Ed.), {The numerical modelling of nonlinear stellar pulsations}.
  Kluwer, pp. 269--288.

\bibitem[{{B{\o}rve} et~al.(2001){B{\o}rve}, {Omang}, and {Trulsen}}]{bot01}
{B{\o}rve}, S., {Omang}, M., {Trulsen}, J., Nov. 2001. {Regularized Smoothed
  Particle Hydrodynamics: A New Approach to Simulating Magnetohydrodynamic
  Shocks}. ApJ 561, 82--93.

\bibitem[{{B{\o}rve} et~al.(2004){B{\o}rve}, {Omang}, and {Trulsen}}]{bot04}
{B{\o}rve}, S., {Omang}, M., {Trulsen}, J., Aug. 2004. {Two-dimensional MHD
  Smoothed Particle Hydrodynamics Stability Analysis}. ApJS 153, 447--462.

\bibitem[{{B{\o}rve} et~al.(2006){B{\o}rve}, {Omang}, and {Trulsen}}]{bot06}
{B{\o}rve}, S., {Omang}, M., {Trulsen}, J., Dec. 2006. {Multidimensional MHD
  Shock Tests of Regularized Smoothed Particle Hydrodynamics}. ApJ 652,
  1306--1317.

\bibitem[{{Brandenburg}(2010)}]{brandenburg10}
{Brandenburg}, A., Jan. 2010. {Magnetic field evolution in simulations with
  Euler potentials}. MNRAS 401, 347--354.

\bibitem[{{Brio} and {Wu}(1988)}]{bw88}
{Brio}, M., {Wu}, C.~C., Apr. 1988. {An upwind differencing scheme for the
  equations of ideal magnetohydrodynamics}. J. Comp. Phys. 75, 400--422.

\bibitem[{{Brookshaw}(1985)}]{brookshaw85}
{Brookshaw}, L., 1985. {A method of calculating radiative heat diffusion in
  particle simulations}. Proceedings of the Astronomical Society of Australia
  6, 207--210.

\bibitem[{{B{\"u}rzle} et~al.(2010){B{\"u}rzle}, {Clark}, {Stasyszyn}, {Greif},
  {Dolag}, {Klessen}, and {Nielaba}}]{buerzleetal10}
{B{\"u}rzle}, F., {Clark}, P.~C., {Stasyszyn}, F., {Greif}, T., {Dolag}, K.,
  {Klessen}, R.~S., {Nielaba}, P., Aug. 2010. {Protostellar collapse and
  fragmentation using an MHD GADGET}. arXiv:1008.3790.

\bibitem[{{Byleveld} and {Pongracic}(1996)}]{bp96}
{Byleveld}, S.~E., {Pongracic}, H., Jan. 1996. {The influence of magnetic
  fields on star formation}. PASA 13, 71--74.

\bibitem[{{Cerqueira} and {de Gouveia Dal Pino}(2001)}]{cg01}
{Cerqueira}, A.~H., {de Gouveia Dal Pino}, E.~M., Oct. 2001. {Three-dimensional
  Magnetohydrodynamic Simulations of Radiatively Cooling, Pulsed Jets}. ApJ
  560, 779--791.

\bibitem[{{Cha} and {Whitworth}(2003)}]{cw03}
{Cha}, S.-H., {Whitworth}, A.~P., Mar. 2003. {Implementations and tests of
  Godunov-type particle hydrodynamics}. MNRAS 340, 73--90.

\bibitem[{{Chow} and {Monaghan}(1997)}]{cm97}
{Chow}, E., {Monaghan}, J.~J., 1997. {Ultrarelativistic SPH}. J. Comp. Phys.
  134, 296--305.

\bibitem[{{Cleary} et~al.(2002){Cleary}, {Ha}, {Alguine}, and
  {Nguyen}}]{clearyha02}
{Cleary}, P., {Ha}, J., {Alguine}, V., {Nguyen}, T., 2002. {Flow modelling in
  casting processes}. Appl. Math. Modelling 26~(2), 171 -- 190.

\bibitem[{{Cleary} and {Monaghan}(1999)}]{cm99}
{Cleary}, P.~W., {Monaghan}, J.~J., Jan. 1999. {Conduction Modelling Using
  Smoothed Particle Hydrodynamics}. J. Comp. Phys. 148, 227--264.

\bibitem[{{Couchman}(1991)}]{couchman91}
{Couchman}, H.~M.~P., Feb. 1991. {Mesh-refined P3M - A fast adaptive N-body
  algorithm}. ApJL 368, L23--L26.

\bibitem[{{Cullen} and {Dehnen}(2010)}]{cd10}
{Cullen}, L., {Dehnen}, W., Jul. 2010. {Inviscid smoothed particle
  hydrodynamics}. MNRAS, 1126.

\bibitem[{{Cummins} and {Rudman}(1999)}]{cr99}
{Cummins}, S.~J., {Rudman}, M., Jul. 1999. {An SPH Projection Method}. J. Comp.
  Phys. 152, 584--607.

\bibitem[{{Dai} and {Woodward}(1994)}]{dw94}
{Dai}, W., {Woodward}, P.~R., Dec. 1994. {Extension of the Piecewise Parabolic
  Method to Multidimensional Ideal Magnetohydrodynamics}. J. Comp. Phys. 115,
  485--514.

\bibitem[{{Dedner} et~al.(2002){Dedner}, {Kemm}, {Kr{\" o}ner}, {Munz},
  {Schnitzer}, and {Wesenberg}}]{dea02}
{Dedner}, A., {Kemm}, F., {Kr{\" o}ner}, D., {Munz}, C.-D., {Schnitzer}, T.,
  {Wesenberg}, M., Jan. 2002. {Hyperbolic Divergence Cleaning for the MHD
  Equations}. J. Comp. Phys. 175, 645--673.

\bibitem[{{Dehnen}(2001)}]{dehnen01}
{Dehnen}, W., Jun. 2001. {Towards optimal softening in three-dimensional N-body
  codes - I. Minimizing the force error}. MNRAS 324, 273--291.

\bibitem[{{Dellar}(2001)}]{dellar01}
{Dellar}, P.~J., Sep. 2001. {A Note on Magnetic Monopoles and the
  One-Dimensional MHD Riemann Problem}. J. Comp. Phys. 172, 392--398.

\bibitem[{Dilts(1999)}]{dilts99}
Dilts, G., 1999. Moving-least-squares-particle hydrodynamics - i. consistency
  and stability. Int. J. Num. Meth. Eng. 44~(8), 1115--1155.

\bibitem[{{Dobbs} and {Price}(2008)}]{dp08}
{Dobbs}, C.~L., {Price}, D.~J., Jan. 2008. {Magnetic fields and the dynamics of
  spiral galaxies}. MNRAS 383, 497--512.

\bibitem[{{Dolag} et~al.(1999){Dolag}, {Bartelmann}, and {Lesch}}]{dbl99}
{Dolag}, K., {Bartelmann}, M., {Lesch}, H., Aug. 1999. {SPH simulations of
  magnetic fields in galaxy clusters}. A\&A 348, 351--363.

\bibitem[{{Dolag} and {Stasyszyn}(2009)}]{ds09}
{Dolag}, K., {Stasyszyn}, F., Oct. 2009. {An MHD GADGET for cosmological
  simulations}. MNRAS 398, 1678--1697.

\bibitem[{{Eckart}(1960)}]{eckart60}
{Eckart}, C., May 1960. {Variation Principles of Hydrodynamics}. Physics of
  Fluids 3, 421--427.

\bibitem[{{Espa{\~ n}ol} and {Revenga}(2003)}]{er03}
{Espa{\~ n}ol}, P., {Revenga}, M., Feb. 2003. {Smoothed dissipative particle
  dynamics}. Phys. Rev. E 67~(2), 026705.

\bibitem[{{Evans} and {Hawley}(1988)}]{eh88}
{Evans}, C.~R., {Hawley}, J.~F., Sep. 1988. {Simulation of magnetohydrodynamic
  flows - A constrained transport method}. ApJ 332, 659--677.

\bibitem[{{Field}(1986)}]{field86}
{Field}, G., 1986. {Magnetic helicity in astrophysics}. In: {Epstein}, R.,
  {Feldman}, W. (Eds.), AIP Conf. Proc. 144: Magnetospheric Phenomena in
  Astrophysics. pp. 324--341.

\bibitem[{{Flebbe} et~al.(1994){Flebbe}, {Muenzel}, {Herold}, {Riffert}, and
  {Ruder}}]{flebbeetal94}
{Flebbe}, O., {Muenzel}, S., {Herold}, H., {Riffert}, H., {Ruder}, H., Aug.
  1994. {Smoothed Particle Hydrodynamics: Physical viscosity and the simulation
  of accretion disks}. ApJ 431, 754--760.

\bibitem[{{Fulk} and {Quinn}(1996)}]{fq96}
{Fulk}, D.~A., {Quinn}, D.~W., Jun. 1996. {An Analysis of 1-D Smoothed Particle
  Hydrodynamics Kernels}. J. Comp. Phys. 126, 165--180.

\bibitem[{{Gardiner} and {Stone}(2005)}]{gs05}
{Gardiner}, T.~A., {Stone}, J.~M., May 2005. {An unsplit Godunov method for
  ideal MHD via constrained transport}. J. Comp. Phys. 205, 509--539.

\bibitem[{{Gingold} and {Monaghan}(1977)}]{gm77}
{Gingold}, R.~A., {Monaghan}, J.~J., Nov. 1977. {Smoothed particle
  hydrodynamics - Theory and application to non-spherical stars}. MNRAS 181,
  375--389.

\bibitem[{{Gray} et~al.(2001){Gray}, {Monaghan}, and {Swift}}]{gms01}
{Gray}, J., {Monaghan}, J.~J., {Swift}, R.~P., Oct. 2001. {SPH elastic
  dynamics}. Computer methods in applied mechanics and engineering 190,
  6641--6662.

\bibitem[{{Harlow} and {Welch}(1965)}]{hw65}
{Harlow}, F.~H., {Welch}, J.~E., Dec. 1965. {Numerical Calculation of
  Time-Dependent Viscous Incompressible Flow of Fluid with Free Surface}.
  Physics of Fluids 8, 2182--2189.

\bibitem[{{Hernquist} and {Katz}(1989)}]{hk89}
{Hernquist}, L., {Katz}, N., Jun. 1989. {TREESPH - A unification of SPH with
  the hierarchical tree method}. ApJS 70, 419--446.

\bibitem[{{He{\ss}} and {Springel}(2010)}]{hs10}
{He{\ss}}, S., {Springel}, V., Aug. 2010. {Particle hydrodynamics with
  tessellation techniques}. MNRAS 406, 2289--2311.

\bibitem[{{Hockney} and {Eastwood}(1981)}]{he81}
{Hockney}, R.~W., {Eastwood}, J.~W., 1981. {Computer Simulation Using
  Particles}. New York: McGraw-Hill, 1981.

\bibitem[{{Hosking} and {Whitworth}(2004)}]{hw04a}
{Hosking}, J.~G., {Whitworth}, A.~P., Jan. 2004. {Modelling ambipolar diffusion
  with two-fluid smoothed particle hydrodynamics}. MNRAS 347, 994--1000.

\bibitem[{{Hu} and {Adams}(2006)}]{huadams06}
{Hu}, X.~Y., {Adams}, N.~A., Apr. 2006. {A multi-phase SPH method for
  macroscopic and mesoscopic flows}. J. Comp. Phys. 213, 844--861.

\bibitem[{{Inutsuka}(2002)}]{inutsuka02}
{Inutsuka}, S., Jun. 2002. {Reformulation of Smoothed Particle Hydrodynamics
  with Riemann Solver}. J. Comp. Phys. 179, 238--267.

\bibitem[{{Janhunen}(2000)}]{janhunen00}
{Janhunen}, P., May 2000. {A Positive Conservative Method for
  Magnetohydrodynamics Based on HLL and Roe Methods}. J. Comp. Phys. 160,
  649--661.

\bibitem[{{Jubelgas} et~al.(2004){Jubelgas}, {Springel}, and {Dolag}}]{jsd04}
{Jubelgas}, M., {Springel}, V., {Dolag}, K., Apr. 2004. {Thermal conduction in
  cosmological SPH simulations}. MNRAS preprint, 000.

\bibitem[{{Kats}(2001)}]{kats01}
{Kats}, A.~V., May 2001. {Variational principle and canonical variables in
  hydrodynamics with discontinuities}. Physica D, 459--474.

\bibitem[{{Kotarba} et~al.(2010){Kotarba}, {Karl}, {Naab}, {Johansson},
  {Dolag}, {Lesch}, and {Stasyszyn}}]{kotarbaetal10}
{Kotarba}, H., {Karl}, S.~J., {Naab}, T., {Johansson}, P.~H., {Dolag}, K.,
  {Lesch}, H., {Stasyszyn}, F.~A., Jun. 2010. {Simulating Magnetic Fields in
  the Antennae Galaxies}. ApJ 716, 1438--1452.

\bibitem[{{Kotarba} et~al.(2009){Kotarba}, {Lesch}, {Dolag}, {Naab},
  {Johansson}, and {Stasyszyn}}]{kotarbaetal09}
{Kotarba}, H., {Lesch}, H., {Dolag}, K., {Naab}, T., {Johansson}, P.~H.,
  {Stasyszyn}, F.~A., Aug. 2009. {Magnetic field structure due to the global
  velocity field in spiral galaxies}. MNRAS 397, 733--747.

\bibitem[{{Kulasegaram} et~al.(2004){Kulasegaram}, {Bonet}, {Lewis}, and
  {Profit}}]{kulasegarametal04}
{Kulasegaram}, S., {Bonet}, J., {Lewis}, R.~W., {Profit}, M., 2004. {A
  variational formulation based contact algorithm for rigid boundaries in
  two-dimensional SPH applications}. Comp. Mech. 33, 316--325.

\bibitem[{{Lodato} and {Price}(2010)}]{lp10}
{Lodato}, G., {Price}, D.~J., Jun. 2010. {On the diffusive propagation of warps
  in thin accretion discs}. MNRAS 405, 1212--1226.

\bibitem[{{Lucy}(1977)}]{lucy77}
{Lucy}, L.~B., Dec. 1977. {A numerical approach to the testing of the fission
  hypothesis}. Astron. J. 82, 1013--1024.

\bibitem[{{Maron} and {Howes}(2003)}]{mh03}
{Maron}, J.~L., {Howes}, G.~G., Sep. 2003. {Gradient Particle
  Magnetohydrodynamics: A Lagrangian Particle Code for Astrophysical
  Magnetohydrodynamics}. ApJ 595, 564--572.

\bibitem[{{Marri} and {White}(2003)}]{mw03}
{Marri}, S., {White}, S.~D.~M., Oct. 2003. {Smoothed particle hydrodynamics for
  galaxy-formation simulations: improved treatments of multiphase gas, of star
  formation and of supernovae feedback}. MNRAS 345, 561--574.

\bibitem[{{Meglicki}(1995)}]{meglicki95}
{Meglicki}, Z., 1995. {Analysis and applications of Smoothed Particle
  Magnetohydrodynamics}. Ph.D. thesis, {Australian National University}.

\bibitem[{{Mignone} and {Tzeferacos}(2010)}]{mt10}
{Mignone}, A., {Tzeferacos}, P., Mar. 2010. {A second-order unsplit Godunov
  scheme for cell-centered MHD: The CTU-GLM scheme}. J. Comp. Phys. 229,
  2117--2138.

\bibitem[{{Monaghan}(1985)}]{monaghan85}
{Monaghan}, J.~J., Sep. 1985. {Extrapolating B. Splines for Interpolation}. J.
  Comp. Phys. 60, 253.

\bibitem[{{Monaghan}(1992)}]{monaghan92}
{Monaghan}, J.~J., 1992. {Smoothed particle hydrodynamics}. Ann. Rev. Astron.
  Astrophys. 30, 543--574.

\bibitem[{{Monaghan}(1997)}]{monaghan97}
{Monaghan}, J.~J., Sep. 1997. {SPH and Riemann Solvers}. J. Comp. Phys. 136,
  298--307.

\bibitem[{{Monaghan}(2000)}]{monaghan00}
{Monaghan}, J.~J., Apr. 2000. {SPH without a Tensile Instability}. J. Comp.
  Phys. 159, 290--311.

\bibitem[{{Monaghan}(2002)}]{monaghan02}
{Monaghan}, J.~J., Sep. 2002. {SPH compressible turbulence}. MNRAS 335,
  843--852.

\bibitem[{{Monaghan}(2004)}]{monaghan04}
{Monaghan}, J.~J., Mar. 2004. {Energy transfer in a particle {$\alpha$} model}.
  J. Turb. 5, 12.

\bibitem[{{Monaghan}(2005)}]{monaghan05}
{Monaghan}, J.~J., 2005. Smoothed particle hydrodynamics. Rep. Prog. Phys.
  68~(8), 1703--1759.

\bibitem[{{Monaghan} et~al.(2005){Monaghan}, {Huppert}, and
  {Worster}}]{monaghanetal05}
{Monaghan}, J.~J., {Huppert}, H.~E., {Worster}, M.~G., Jul. 2005.
  {Solidification using smoothed particle hydrodynamics}. J. Comp. Phys. 206,
  684--705.

\bibitem[{{Monaghan} and {Lattanzio}(1985)}]{ml85}
{Monaghan}, J.~J., {Lattanzio}, J.~C., Aug. 1985. {A refined particle method
  for astrophysical problems}. A\&A 149, 135--143.

\bibitem[{{Monaghan} and {Price}(2001)}]{mp01}
{Monaghan}, J.~J., {Price}, D.~J., Dec. 2001. {Variational principles for
  relativistic smoothed particle hydrodynamics}. MNRAS 328, 381--392.

\bibitem[{{Morris}(1996{\natexlab{a}})}]{morris96}
{Morris}, J.~P., Jan. 1996{\natexlab{a}}. {A study of the stability properties
  of smooth particle hydrodynamics}. PASA 13, 97--102.

\bibitem[{{Morris}(1996{\natexlab{b}})}]{morrisphd}
{Morris}, J.~P., 1996{\natexlab{b}}. {Analysis of smoothed particle
  hydrodynamics with applications}. Ph.D. thesis, {Monash University,
  Melbourne, Australia}.

\bibitem[{{Morris} and {Monaghan}(1997)}]{mm97}
{Morris}, J.~P., {Monaghan}, J.~J., 1997. {A switch to reduce SPH viscosity}.
  J. Comp. Phys. 136, 41--50.

\bibitem[{{Morrison}(1998)}]{morrison98}
{Morrison}, P.~J., Apr. 1998. {Hamiltonian description of the ideal fluid}.
  Rev. Mod. Phys. 70, 467--521.

\bibitem[{{Murray}(1996)}]{murray96}
{Murray}, J.~R., Mar. 1996. {SPH simulations of tidally unstable accretion
  discs in cataclysmic variables}. MNRAS 279, 402--414.

\bibitem[{{Nelson} and {Papaloizou}(1994)}]{np94}
{Nelson}, R.~P., {Papaloizou}, J.~C.~B., Sep. 1994. {Variable Smoothing Lengths
  and Energy Conservation in Smoothed Particle Hydrodynamics}. MNRAS 270, 1.

\bibitem[{{Newcomb}(1962)}]{newcomb62}
{Newcomb}, W.~A., 1962. {Lagrangian and Hamiltonian methods in
  magnetohydrodynamics}. Nucl. Fusion Suppl. 2, 451--463.

\bibitem[{{Pelupessy} et~al.(2003){Pelupessy}, {Schaap}, and {van de
  Weygaert}}]{pelupessyetal03}
{Pelupessy}, F.~I., {Schaap}, W.~E., {van de Weygaert}, R., May 2003. {Density
  estimators in particle hydrodynamics. DTFE versus regular SPH}. A\&A 403,
  389--398.

\bibitem[{{Phillips} and {Monaghan}(1985)}]{pm85}
{Phillips}, G.~J., {Monaghan}, J.~J., Oct. 1985. {A numerical method for
  three-dimensional simulations of collapsing, isothermal, magnetic gas
  clouds}. MNRAS 216, 883--895.

\bibitem[{{Powell} et~al.(1999){Powell}, {Roe}, {Linde}, {Gombosi}, and {de
  Zeeuw}}]{pea99}
{Powell}, K.~G., {Roe}, P.~L., {Linde}, T.~J., {Gombosi}, T.~I., {de Zeeuw},
  D.~L., Sep. 1999. {A Solution-Adaptive Upwind Scheme for Ideal
  Magnetohydrodynamics}. J. Comp. Phys. 154, 284--309.

\bibitem[{{Price}(2004)}]{price04}
{Price}, D.~J., 2004. {Magnetic fields in Astrophysics}. Ph.D. thesis,
  University of Cambridge, Cambridge, UK. astro-ph/0507472.

\bibitem[{{Price}(2007)}]{splashpaper}
{Price}, D.~J., Oct. 2007. {SPLASH: An Interactive Visualisation Tool for
  Smoothed Particle Hydrodynamics Simulations}. Publ. Astron. Soc. Aust. 24,
  159--173.

\bibitem[{{Price}(2008)}]{price08}
{Price}, D.~J., 2008. {Modelling discontinuities and Kelvin-Helmholtz
  instabilities in SPH}. J. Comp. Phys. 227, 10040--10057.

\bibitem[{{Price}(2010)}]{price10}
{Price}, D.~J., Nov. 2010. {Smoothed Particle Magnetohydrodynamics - IV. Using
  the vector potential}. MNRAS 401, 1475--1499.

\bibitem[{{Price} and {Bate}(2007)}]{pb07}
{Price}, D.~J., {Bate}, M.~R., May 2007. {The impact of magnetic fields on
  single and binary star formation}. MNRAS 377, 77--90.

\bibitem[{{Price} and {Bate}(2008)}]{pb08}
{Price}, D.~J., {Bate}, M.~R., Apr. 2008. {The effect of magnetic fields on
  star cluster formation}. MNRAS 385, 1820--1834.

\bibitem[{{Price} and {Bate}(2009)}]{pb09}
{Price}, D.~J., {Bate}, M.~R., 2009. {Inefficient star formation: The combined
  effects of magnetic fields and radiative feedback}. MNRAS 398, 33--46.

\bibitem[{{Price} and {Federrath}(2010)}]{pf10}
{Price}, D.~J., {Federrath}, C., Jun. 2010. {A comparison between grid and
  particle methods on the statistics of driven, supersonic, isothermal
  turbulence}. MNRAS, 960.

\bibitem[{{Price} and {Monaghan}(2004{\natexlab{a}})}]{pm04a}
{Price}, D.~J., {Monaghan}, J.~J., Feb. 2004{\natexlab{a}}. {Smoothed Particle
  Magnetohydrodynamics - I. Algorithm and tests in one dimension}. MNRAS 348,
  123--138.

\bibitem[{{Price} and {Monaghan}(2004{\natexlab{b}})}]{pm04b}
{Price}, D.~J., {Monaghan}, J.~J., Feb. 2004{\natexlab{b}}. {Smoothed Particle
  Magnetohydrodynamics - II. Variational principles and variable
  smoothing-length terms}. MNRAS 348, 139--152.

\bibitem[{{Price} and {Monaghan}(2005)}]{pm05}
{Price}, D.~J., {Monaghan}, J.~J., Dec. 2005. {Smoothed Particle
  Magnetohydrodynamics - III. Multidimensional tests and the $\nabla\cdot{\bf
  B}= 0$ constraint}. MNRAS 364, 384--406.

\bibitem[{{Price} and {Monaghan}(2007)}]{pm07}
{Price}, D.~J., {Monaghan}, J.~J., Feb. 2007. {An energy-conserving formalism
  for adaptive gravitational force softening in smoothed particle hydrodynamics
  and N-body codes}. MNRAS 374, 1347--1358.

\bibitem[{{Price} and {Rosswog}(2006)}]{pr06}
{Price}, D.~J., {Rosswog}, S., May 2006. {Producing Ultrastrong Magnetic Fields
  in Neutron Star Mergers}. Science 312, 719--722.

\bibitem[{{Ritchie} and {Thomas}(2001)}]{rt01}
{Ritchie}, B.~W., {Thomas}, P.~A., May 2001. {Multiphase smoothed-particle
  hydrodynamics}. MNRAS 323, 743--756.

\bibitem[{{Rosswog}(2009)}]{rosswog09}
{Rosswog}, S., Apr. 2009. {Astrophysical smooth particle hydrodynamics}. New
  Astron. Reviews 53, 78--104.

\bibitem[{{Rosswog} and {Price}(2007)}]{rp07}
{Rosswog}, S., {Price}, D., Aug. 2007. {MAGMA: a three-dimensional, Lagrangian
  magnetohydrodynamics code for merger applications}. MNRAS 379, 915--931.

\bibitem[{{Ryu} and {Jones}(1995)}]{rj95}
{Ryu}, D., {Jones}, T.~W., Mar. 1995. {Numerical magetohydrodynamics in
  astrophysics: Algorithm and tests for one-dimensional flow`}. ApJ 442,
  228--258.

\bibitem[{{Salmon}(1988)}]{salmon88}
{Salmon}, R., 1988. {Hamiltonian fluid mechanics}. Annual Review of Fluid
  Mechanics 20, 225--256.

\bibitem[{{Schoenberg}(1946)}]{schoenberg46a}
{Schoenberg}, I.~J., 1946. {Contributions to the problem of approximation of
  equidistant data by analytic functions. A. On the problem of smoothing or
  graduation -- a 1st class of analytic approximation formulae}. Q. Appl. Math.
  4, 45.

\bibitem[{{Serrano} et~al.(2005){Serrano}, {Espa{\~n}ol}, and
  {Z{\'u}{\~n}iga}}]{sez05}
{Serrano}, M., {Espa{\~n}ol}, P., {Z{\'u}{\~n}iga}, I., Oct. 2005. {Voronoi
  Fluid Particle Model for Euler Equations}. J. Stat. Phys. 121, 133--147.

\bibitem[{{Springel}(2005)}]{springel05}
{Springel}, V., Dec. 2005. {The cosmological simulation code GADGET-2}. MNRAS
  364, 1105--1134.

\bibitem[{{Springel} and {Hernquist}(2002)}]{sh02}
{Springel}, V., {Hernquist}, L., Jul. 2002. {Cosmological smoothed particle
  hydrodynamics simulations: the entropy equation}. MNRAS 333, 649--664.

\bibitem[{{Stern}(1966)}]{stern66}
{Stern}, D.~P., 1966. {The Motion of Magnetic Field Lines}. Space Science
  Reviews 6, 147.

\bibitem[{{Stern}(1970)}]{stern70}
{Stern}, D.~P., Apr. 1970. {Euler Potentials}. Am. J. Phys. 38, 494--501.

\bibitem[{{Stern}(1976)}]{stern76}
{Stern}, D.~P., May 1976. {Representation of magnetic fields in space}. Rev.
  Geophys. \& Space Phys. 14, 199--214.

\bibitem[{{Stone} et~al.(1992){Stone}, {Hawley}, {Evans}, and {Norman}}]{sea92}
{Stone}, J.~M., {Hawley}, J.~F., {Evans}, C.~R., {Norman}, M.~L., Apr. 1992. {A
  test suite for magnetohydrodynamical simulations}. ApJ 388, 415--437.

\bibitem[{{T{\' o}th}(2000)}]{toth00}
{T{\' o}th}, G., Jul. 2000. {The {$\nabla$}.B=0 Constraint in Shock-Capturing
  Magnetohydrodynamics Codes}. J. Comp. Phys. 161, 605--652.

\bibitem[{{T{\' o}th}(2002)}]{toth02}
{T{\' o}th}, G., Oct. 2002. {Conservative and Orthogonal Discretization for the
  Lorentz Force}. J. Comp. Phys. 182, 346--354.

\bibitem[{{Thomas} and {Couchman}(1992)}]{tc92}
{Thomas}, P.~A., {Couchman}, H.~M.~P., Jul. 1992. {Simulating the formation of
  a cluster of galaxies}. MNRAS 257, 11--31.

\bibitem[{{Wadsley} et~al.(2008){Wadsley}, {Veeravalli}, and
  {Couchman}}]{wadsleyetal08}
{Wadsley}, J.~W., {Veeravalli}, G., {Couchman}, H.~M.~P., Jun. 2008. {On the
  treatment of entropy mixing in numerical cosmology}. MNRAS 387, 427--438.

\bibitem[{{Watkins} et~al.(1996){Watkins}, {Bhattal}, {Francis}, {Turner}, and
  {Whitworth}}]{watkins96}
{Watkins}, S.~J., {Bhattal}, A.~S., {Francis}, N., {Turner}, J.~A.,
  {Whitworth}, A.~P., Oct. 1996. {A new prescription for viscosity in smoothed
  particle hydrodynamics.} A\&AS 119, 177--187.

\bibitem[{{Whitehouse} and {Bate}(2004)}]{wb04}
{Whitehouse}, S.~C., {Bate}, M.~R., Oct. 2004. {Smoothed particle hydrodynamics
  with radiative transfer in the flux-limited diffusion approximation}. MNRAS
  353, 1078--1094.

\bibitem[{{Whitehouse} et~al.(2005){Whitehouse}, {Bate}, and {Monaghan}}]{wb05}
{Whitehouse}, S.~C., {Bate}, M.~R., {Monaghan}, J.~J., Dec. 2005. {A faster
  algorithm for smoothed particle hydrodynamics with radiative transfer in the
  flux-limited diffusion approximation}. MNRAS 364, 1367--1377.

\end{thebibliography}

\end{document}